\documentclass[11pt,reqno]{article}
\usepackage{amsmath,hyperref,amssymb, amsfonts,amsthm}
\usepackage{mathrsfs}
\usepackage{authblk}
\usepackage[all]{xy}
\usepackage{graphicx,url}
\usepackage{color}
\usepackage{slashed}

\newcommand{\be}{\begin{eqnarray}}
\newcommand{\ee}{\end{eqnarray}}
\newcommand{\eeq}{\end{equation}}
\newcommand{\beq}{\begin{equation}}

\usepackage{simplewick}

\allowdisplaybreaks \numberwithin{equation}{section}

\DeclareSymbolFont{AMSa}{U}{msa}{m}{n}
\DeclareSymbolFont{AMSb}{U}{msb}{m}{n}
\DeclareMathSymbol{\fieldR}{\mathalpha}{AMSb}{"52}

\renewcommand{\Im}{\imag}
\DeclareMathOperator{\imag}{Im}
\DeclareMathOperator{\rank}{rank}

\newcommand{\CH}{\mathcal{H}}

\newcommand{\CN}{\mathcal{N}}
\newcommand{\CM}{\mathcal{M}}

\newcommand{\CQ}{\mathcal{Q}}
\newcommand{\CV}{\mathcal{V}}
\newcommand{\Cc}{\mathsf{C}}
\newcommand{\Hh}{\mathsf{H}}
\DeclareMathOperator{\Tr}{Tr}
\DeclareMathOperator{\sTr}{sTr}

\newcommand{\qsf}{\mathsf{q}}
\newcommand{\usf}{\upsilon}
\newcommand{\vsf}{\nu}
\newcommand{\bF}{\mathbf{F}}

\newcommand{\ZZ}{\mathbb{Z}}
\newcommand{\RR}{\mathbb{R}}
\newcommand{\CC}{\mathbb{C}}
\newcommand{\g}{\mathfrak{g}}

\def\t{\tau}

\def\beq{\begin{equation}}
\def\eeq{\end{equation}}
\def\bea{\begin{eqnarray}}
\def\eea{\end{eqnarray}}

\def\<{\langle}

\newcommand\nn{\nonumber}

\title{BPS Algebras in 2D String Theory}
\author[1, 2]{Sarah M. Harrison\thanks{sarharr@physics.mcgill.ca}}
\author[3]{Natalie M. Paquette\thanks{paquette@ias.edu}}
\author[4]{Daniel Persson\thanks{daniel.persson@chalmers.se}}
\author[5]{Roberto Volpato\thanks{volpato@pd.infn.it}}

{\small \affil[1]{\small Department of Mathematics and Statistics, McGill University, Montreal, QC, Canada}
\affil[2]{\small Department of Physics, McGill University, Montreal, QC, Canada}
\affil[3]{\small Institute for Advanced Study, School of Natural Sciences, Princeton NJ, 08540, USA}
\affil[4]{Department of Mathematical Sciences, Chalmers University of Technology, Gothenburg, Sweden}
\affil[5]{\small Dipartimento di Fisica e Astronomia `Galileo Galilei', Universit\`a di Padova \& INFN, sez. di Padova, Via Marzolo 8, 35131, Padova, Italy}}
\date{}

\begin{document}
\maketitle
\abstract{We discuss a set of heterotic and type II string theory compactifications to 1+1 dimensions that are characterized by factorized internal worldsheet CFTs of the form $V_1\otimes \bar V_2$, where $V_1, V_2$ are self-dual (super) vertex operator algebras. In the cases with spacetime supersymmetry, we show that the BPS states form a module for a Borcherds-Kac-Moody (BKM) (super)algebra, and we prove that for each model the BKM (super)algebra is a symmetry of genus zero BPS string amplitudes. We compute the supersymmetric indices of these models using both Hamiltonian and path integral formalisms. The path integrals are manifestly automorphic forms closely related to the Borcherds-Weyl-Kac denominator. Along the way, we comment on various subtleties inherent to these low-dimensional string compactifications.}

\vspace{2cm}
\newpage
\tableofcontents

\section{Introduction}
Critical string theories compactified to low dimensions provide a useful formal playground for exploring various theoretical phenomena. They enjoy enhanced (nonperturbative) duality groups, uncommon supersymmetry algebras, exhibit relatively explicit but rich mathematical structures in their spacetime physics, and may still possess an interesting family of moduli enabling one to decompactify back to more physically relevant dimensions. Distinguished points in such low dimensional moduli spaces exhibit maximal symmetry groups that can usefully organize or classify (broken) symmetries in higher dimensional models upon decompactification.  Alternatively, one can use the enhanced symmetries at special points to construct orbifolds of the original theory; such orbifolds with respect to large symmetry groups can produce exotic models with few moduli. These models may even be completely rigid, sitting at isolated points in moduli space, and are hence distinguished in their own right. It is also interesting and nontrivial to determine the duality webs among these exceptional theories.

It is known that some of the points of maximal symmetry are given by string models whose internal worldsheet theories factorize into holomorphic and anti-holomorphic factors.  Such holomorphically factorized theories must be built from consistent (self-dual)\footnote{Throughout this paper we use the same definition of self-dual as the authors of \cite{CDR}: essentially, that an SVOA $V$ is self-dual if $V$ is the only irreducible, admissible $V$-module up to isomorphism.} vertex operator algebras (VOAs) of appropriate central charge. The latter are quite rare for low values of the central charge and have been determined at several key values of interest, including $c=24$ for bosonic VOAs and $c=12$ for super VOAs (SVOAs). In this work, we will focus on compactifications of IIA string theory to two spacetime dimensions built from holomorphically factorized worldsheet theories of the form $V_1 \otimes \bar{V_2}$, where $V_1, V_2$ are self-dual SVOAs with $c=12$. Much of our analysis also carries through for heterotic string theories built from $V_1 \otimes \bar{V_2}$, where $V_1$ is now a self-dual VOA with $c=24$.

One nice feature of these factorized models is that BPS states are closely related to  infinite-dimensional Lie (super)algebras called Borcherds-Kac-Moody algebras \cite{BorcherdsMM}. For a variety of instances of BKMs in string theory see, e.g, \cite{HM1, HM2, Neumann, KP, Galakhov, Cheng:2008fc, Cheng:2008kt}. Such algebras can be constructed by applying a certain functor to a self-dual VOA that looks analogous to constructing a ``chiral" string theory: that is, tensor in the vertex algebra corresponding to the even self-dual lattice of signature $(1,1)$ (a chiral analogue of the lightcone directions) to obtain a theory with critical central charge, tensor in the vertex algebra corresponding to ghosts, and compute the BRST cohomology of the resulting complex. The resulting cohomology classes generate a BKM. 

BKM algebras were originally defined by Borcherds in the course of his proof of the monstrous moonshine conjectures \cite{borcherds1988generalized,BorcherdsFake,BorcherdsMM}. Harvey and Moore later proposed that BKM algebras should play a role as the underlying organizing structure of BPS states in string compactifications \cite{HM1,HM2}. While a complete understanding of the elusive ``algebra of BPS states'' is still lacking (though see, e.g, \cite{KS} for a mathematical perspective), the fact that BKM algebras are intimately connected with BPS states is indisputable. 

Let us explain this connection in some more detail. Consider heterotic string theory compactified to two dimensions with an internal CFT of the form $V^\natural\otimes \bar{V}^{f\natural}$. Here, $V^\natural$ is the monster module, constructed by Frenkel, Lepowsky and Meurman \cite{FLM0,FLM}, and ${V}^{f\natural}$ is Duncan's super-moonshine module for the Conway group \cite{Duncan}.  This theory was used in \cite{PPV1,PPV2} to provide a physical interpretation of the genus zero property of monstrous moonshine. A key aspect is that it gives a \emph{spacetime} description of the McKay-Thompson series, which appear in certain BPS indices. In particular, after compactifying the space direction on a circle $S^1$, one can consider the following index
\begin{equation}
\text{Tr}_{\mathcal{H}_{\text{BPS}}}\Big((-1)^F e^{2\pi i TW}e^{2\pi i UM}\Big),
\end{equation}
where $\mathcal{H}_{\text{BPS}}$ is a second-quantized space of BPS string states, $F$ is the fermion number and $(W, M)$ are the winding and momenta along the circle $S^1$. In addition, $(T, U)$ are the associated chemical potentials which involve the spacetime radius $R$ and inverse temperature $\beta$. It was shown in \cite{PPV1} that this index can be written as an infinite product
\begin{equation}
\Big(e^{2\pi i (w_0T+m_0U)}\prod_{m>0\atop w\in \mathbb{Z}}(1-e^{2\pi i mU}e^{2\pi i wT})^{c(mw)}\Big)^{24},
\end{equation}
where $c(n)$ are the Fourier coefficients of the modular invariant $J$-function. This is recognized as the (24th power of) the Borcherds--Weyl--Kac denominator formula of the Monster Lie algebra $\mathfrak{m}$. The roots $\alpha$ of the Monster Lie algebra are labelled by pairs of integers $(m,w)$ and the Fourier coefficients $c(mw)$ encode the root multiplicities $\text{mult}\, \alpha$. This implies that the denominator formula of the Monster Lie algebra can be viewed as a generating function of BPS states in this model. The denominator function turns out to be a special type of automorphic form on $SO(2,2)/(SO(2)\times SO(2))$. 

One can further show that the space of BPS states is actually a \emph{module} for the monster Lie algebra $\mathfrak{m}$. This gives $\mathfrak{m}$ a natural interpretation as a BPS algebra \cite{PPV1}. This analysis extends to all CHL orbifolds \cite{CHL} of  $V^\natural\otimes \bar{V}^{f\natural}$  in which case the relevant algebras are the $\mathfrak{m}_g$'s constructed by Carnahan in his proof of generalised monstrous moonshine \cite{Carnahan0,Carnahan,Carnahan2}. 

The VOA $V^{f\natural}$  was further studied in \cite{Harrison:2018joy}, where a BKM algebra was constructed upon which the Conway group $Co_0$ acts faithfully. This is a candidate for an algebra of BPS states in a certain compactification of type IIA string theory. In a similar spirit, in reference \cite{Harrison:2020wxl}, we constructed a family of BKM algebras associated with $F_{24}$, the holomorphic $c=12$ SCFT based on 24 free fermions. More precisely, we obtained one BKM for every choice of $\mathcal{N}=1$ superconformal structure on $F_{24}$. Additionally, for these BKM algebras one can identify the denominator formulas with certain higher rank automorphic forms. A similar story unfolds if one instead starts with the self-dual SVOA based on the $E_8$ lattice, $V^{fE_8}$\cite{Sch1, Sch2}.

It is the purpose of the present work to provide a spacetime interpretation of the aforementioned BKM (super)algebras, along the lines of \cite{PPV1,PPV2}.  The models we consider are now full (non-chiral) string theory, whose spacetime BPS states moreover still enjoy a close relationship with these BKMs, in the sense that 1) they form a module for a BKM (super)algebra and 2) the BKM (super)algebras act as symmetries of amplitudes which contain insertions only of BPS vertex operators. This provides a  realization of these BKM algebras  as algebras of BPS states in type II string theories, such that their denominator formulas correspond to BPS indices. 

\bigskip
\noindent \emph{Summary and outline}
\medskip

Now we come to highlight the key results in the present work. Let us consider 2d type IIA and heterotic string compactifications with holomorphically factorized worldsheet theories of the form $V_1 \otimes \bar V_2$, as introduced above. Let $\mathfrak g$ be the BKM (super)algebra associated to $V_1$ via the standard chiral construction due to Borcherds, briefly reviewed above. In this work, we show the following:
\begin{enumerate}
    \item The spacetime BPS states in the 2d compactification associated to $V_1 \otimes \bar V_2$ form a representation of the BKM algebra $\mathfrak g.$ (\S \ref{sec:algebras})
    \item  $\mathfrak g$ acts as a symmetry on certain genus zero BPS saturated amplitudes  in the theory. (\S \ref{sec:algebras})
    \item A suitably defined spacetime supersymmetric index in the theory reproduces the (super)denominator formula of $\mathfrak g$. (\S \ref{sec:secondquantized})
    \item This index can also be reproduced via a path integral computation on a 2d Euclidean spacetime torus. The path integral precisely reduces to a familiar theta lift in the theory of automorphic forms. (\S \ref{sec:pathintegrals})
\end{enumerate}

The plan of the rest of the paper is as follows. In Section \ref{sec:models}, we describe basic features of the 2d spacetime string theories we are interested in, including worldsheet SCFTs, massless field content, spacetime supersymmetry algebras, and gravitational anomalies. In Section \ref{sec:states}, we explicitly construct the cohomology of physical states in our holomorphically factorized models, leaving some technical details to appendix \ref{app:semirelative}. In Section \ref{sec:algebras}, we focus on the subspace of spacetime BPS states $\mathcal{H}_{BPS}$ and describe the action of the BKM algebra $\mathfrak{g}$ (constructed from the (S)VOAs $V_1$) on $\mathcal{H}_{BPS}$. We further show that $\mathfrak{g}$ is a symmetry of genus zero BPS amplitudes. In Section \ref{sec:secondquantized} we consider a second quantized version of our 2d string theories, where we allow for an arbitrary number of fundamental strings, and study the resulting Hilbert space. We compute a number of natural supersymmetric indices in the second quantized theories from the Hamiltonian point of view, and explain their relation to denominators of the BKM algebras from the previous section. In Section \ref{sec:pathintegrals} we revisit the computation of the indices from the path integral point of view using the formalism of theta lifts from the theory of automorphic forms, and clarify some subtleties that arise in these low dimensional models. Finally, we conclude in section~\ref{sec:discussion} with a discussion of several open questions that emerged during the course of this work. 

This paper also contains four appendices. In appendix~\ref{s:Bfield} we give a detailed analysis of gravitational anomalies and the B-field tadpole in our models. In appendix~\ref{app:semirelative} we provide a careful treatment of the semi-relative cohomology of physical states. In appendix \ref{app:0mom} we discuss some details related to the zero momentum R-R spectrum in one of our models, relevant for certain computations carried out in \S \ref{sec:secondquantized}. And finally, in appendix \ref{app:vectorvaluedtheta} we review some standard facts about theta lifts of vector-valued modular forms.

\bigskip
\noindent \emph{Conventions and notation}
\medskip

\noindent We use the two-dimensional metric $\eta^{\mu\nu}={\rm diag}(-,+)$. The gamma matrices $\Gamma^0=\left(\begin{smallmatrix} 0 & 1\\ -1 & 0\end{smallmatrix}\right)$, $\Gamma^1=\left(\begin{smallmatrix} 0 & 1\\ 1 & 0\end{smallmatrix}\right)$ obey the algebra relations $\{\Gamma^\mu,\Gamma^\nu\}=2\eta^{\mu\nu}$ and the chirality matrix is $\Gamma\equiv \Gamma^0\Gamma^1=\left(\begin{smallmatrix} 1 & 0\\ 0 & -1\end{smallmatrix}\right)$.  Majorana-Weyl spinors of positive and negative chirality have the form $\left(\begin{smallmatrix} u_+ \\ 0\end{smallmatrix}\right)$ and $\left(\begin{smallmatrix} 0 \\ u_-\end{smallmatrix}\right)$, respectively. The massless Dirac equation is
$$ 0=k_\mu \Gamma^0\Gamma^\mu u=\begin{pmatrix} -k_0+k_1 & 0\\ 0 & k_0+k_1
\end{pmatrix}\begin{pmatrix} u_+\\ u_-\end{pmatrix}=\begin{pmatrix} k^0+k^1 & 0\\ 0 & -k^0+k^1
\end{pmatrix}\begin{pmatrix} u_+\\ u_-\end{pmatrix},
$$ such that a positive chirality fermion exists only with $k^0=-k^1$ and a negative chirality one only for $k^0=k^1$. We say that a theory has $(\CN_+,\CN_-)$ supersymmetry when there are $\CN_+$ supercharges $\CQ^i_+$ of positive chirality  and $\CN_-$ supercharges $\CQ^i_-$ of negative chirality.   In the absence of central charges, the supersymmetry algebra in two uncompactified dimensions is
\be
\{\CQ^i_\alpha,\CQ^j_\beta\}=2\delta^{ij}(\Gamma^0\Gamma^\mu)_{\alpha\beta}P_\mu=2\delta^{ij}(P^0\delta_{\alpha\beta}+P^1\Gamma_{\alpha\beta});
\ee i.e.
\be
\{\CQ_+,\CQ_+\}=2\delta^{ij}(P^0+P^1)\ ,\qquad \{\CQ_-,\CQ_-\}=2\delta^{ij}(P^0-P^1)\ ,\qquad \{\CQ_+,\CQ_-\}=0\ .
\ee

\section{Two--dimensional models}\label{sec:models}
In this section, we discuss the class of models of interest in this paper. These models are string theory compactifications to 1+1 dimensions where the worldsheet CFT takes the holomorphically factorized form $V_1 \otimes \bar V_2.$ We will always take $V_2$ to be one of the three self-dual SVOAs of central charge 12 (see \S \ref{sec:VOAs}). In the case of type IIA compactifications, we will also take $V_1$ to be one of these three theories. For the heterotic models we consider, we will allow $V_1$ to be one of the 71 known self-dual VOAs of central charge 24 \cite{Schellekens:1992db}.

\subsection{The three $c=12$ SVOAs}\label{sec:VOAs}

First we briefly review the three $c=12$ SVOAs we will consider. In \cite{CDR} it was proven that there are exactly three so-called self-dual SVOAs with central charge 12. These theories are:
\begin{enumerate}
	\item $V^{fE_8}$: this is the supersymmetric vertex operator algebra based on the $E_8$ lattice.  It contains 8 chiral free bosons compactified on the $E_8$ root lattice, and their 8 fermionic superpartners. 
	\item $V^{f\natural}$: This is the unique self-dual SVOA with $c=12$ and no fields of conformal weight $1/2$ in the NS sector. In \cite{Duncan} it was proved that, up to automorphisms, it admits a unique choice of $\mathcal N=1$ superconformal current which is stabilized by Conway's group $Co_0$. (The $\ZZ_2$ center of $Co_0$ acts non-trivially only on the Ramond sector).
	\item $F_{24}$: This is a theory of 24 free chiral fermions. An $\mathcal N=1$ supercurrent can be defined by taking a linear combination of cubic Fermi terms, and the 8 inequivalent choices are in one-to one correspondence with semisimple Lie algebras of dimension 24. Given a choice of $\CN=1$ superconformal structure, the $24$ currents that are supersymmetric descendants of the free fermions generate an affine Kac-Moody algebra, of which there are eight possibilities \cite{Goddard:1984hg}:
	\begin{align}(\widehat{su}(2)_2)^{\oplus 8}\ ,\quad (\widehat{su}(3)_3)^{\oplus 3}\ ,\quad \widehat{su}(4)_4\oplus (\widehat{su}(2)_2)^{\oplus 3}\ ,\quad \widehat{su}(5)_5\ ,\quad  \widehat{so}(5)_3\oplus \hat g_{2,4}\ ,\notag\\
\widehat{so}(5)_3\oplus \widehat{su}(3)_3\oplus (\widehat{su}(2)_2)^{\oplus 2}\ ,\quad \widehat{so}(7)_5\oplus \widehat{su}(2)_2\ ,\quad \widehat{sp}(6)_4\oplus \widehat{su}(2)_2\ .\notag
\end{align}
	We discussed further aspects of this theory in \cite{Harrison:2020wxl}.
\end{enumerate}

Strictly speaking, a self-dual SVOA corresponds to the NS sector of a chiral superconformal field theory, while the Ramond sector is a (canonically) twisted module. We will be slightly imprecise in this respect, and use the word self-dual SVOA as a synonym of chiral superconformal field theory.

In the NS sector of each holomorphic SCFT there is a canonical $\ZZ_2$ symmetry---the fermion number operator $(-1)^F$---which acts trivially on states of integral $L_0$ eigenvalue and as multiplication by $-1$ on states with half-integral eigenvalue. The Ramond sector can be seen as a twisted module with respect to this symmetry. For all the theories we consider, the definition of the fermion number can be extended to the Ramond sector in such a way that it is a symmetry of the OPE of order $2$. However, there is an ambiguity by an overall sign in this definition. For $F_{24}$ and $V^{fE_8}$, the zero mode of any weight $1/2$ NS field provides an isomorphism between the Ramond eigenspaces of positive and negative fermion number, so the choice of the sign of $(-1)^F$ is immaterial. For $V^{f\natural}$ there is no such weight $1/2$ NS field, and indeed all the $24$ Ramond ground states (i.e. of conformal weight $1/2$) have the same $(-1)^F$ eigenvalue, with can be chosen to be $+1$ or $-1$. This choice will be important when we discuss the GSO projection in superstring compactifications in the next subsections. It is useful to introduce the notation $V^{f\natural}_+$ and $V^{f\natural}_-$ to denote the SVOA $V^{f\natural}$ with the choice of fermion number equal, respectively, to $+1$ and $-1$ on the Ramond ground states. For conformal weight strictly greater than $1/2$, the Ramond subspaces with positive and negative fermion number are again isomorphic, with the isomorphism given by the zero mode of the $\CN=1$ supercurrent (see below).

For each of these theories $V \in \{V^{fE_8},V^{f\natural},F_{24}\}$, we can define a set of four torus partition functions, corresponding to the four choices of boundary condition for the fermions,
\bea\label{eq:partfns}
\phi_{\rm NS}(\tau;V)&:=& \Tr_{\rm NS}(\qsf^{L_0 - 1/2}) = \frac{\eta^{48}(\t)}{\eta^{24}(\t/2)\eta^{24}(2\t)}-24+\chi^{\rm NS}(V)\\
\phi_{\widetilde{\rm NS}}(\tau;V)&:=& \sTr_{\rm NS}( \qsf^{L_0 - 1/2}) =  {\eta^{24}(\t/2)\over \eta^{24}(\t)}+24 -\chi^{\rm NS}(V)\\
\phi_{\rm R}(\tau;V)&:=& \Tr_{\rm R}(\qsf^{L_0 - 1/2})=  2^{12}{\eta^{24}(2\t)\over \eta^{24}(\t)}+ \left (\chi^{\rm R}_+(V)+\chi^{\rm R}_-(V) \right )
\\
\phi_{\tilde{\rm R}}(\tau;V)&:=& \sTr_{\rm R}(\qsf^{L_0 - 1/2})=\chi^{\rm R}_+(V)-\chi^{\rm R}_-(V)\ ,
\eea
where by $\sTr$ we denote the supertrace; i.e. the trace with the insertion of the fermion number operator $(-1)^F.$ 
The partition functions $\{\phi_{\rm NS},\phi_{\widetilde{\rm NS}},\phi_{\rm R}\}$ form a 3-component vector-valued representation of $SL(2,\mathbb Z)$, whereas $\phi_{\tilde{\rm R}}$ is a singlet under the action of $SL(2,\mathbb Z)$. We have expressed them explicitly in terms of eta quotients, where we have have used the Dedekind eta function, defined as
\be
\eta(\tau):=\qsf^{1/24}\prod_{n=1}^\infty (1-\qsf^n),
\ee
and throughout we use the definition $\qsf:= e^{2\pi i\tau}$.
The first several terms in the $\qsf$-expansion of each of these functions are
\bea\label{eq:qexp}
\phi_{\rm NS}(\tau; V)&=& \qsf^{-1/2} + \chi^{\rm NS}(V) + 276 \qsf^{1/2} + 2048 \qsf + 11202 \qsf^{3/2} + \ldots \\
\phi_{\widetilde{\rm NS}}(\tau; V)&=& \qsf^{-1/2} -\chi^{\rm NS}(V) + 276 \qsf^{1/2} - 2048 \qsf + 11202 \qsf^{3/2} + \ldots \\
\phi_{\rm R}(\tau; V)&=&  \left (\chi^{\rm R}_+(V)+\chi^{\rm R}_-(V) \right )+ 4096 \qsf + 98304 \qsf^2 + 1228800 \qsf^3 
+ \ldots\\
\phi_{\tilde{\rm R}}(\tau; V)&=&\left (\chi^{\rm R}_+(V)-\chi^{\rm R}_-(V)\right )\ .
\eea

The constant $\chi^{\rm NS}(V)$ captures the number of spin-$1/2$ fields of $V$ in the NS sector. All of these fields are odd under the fermion number operator. The constants  $\chi^{\rm R}_+(V)$ and $\chi^{\rm R}_-(V)$ capture the number of even and odd spin-$1/2$ fields in the Ramond sector, respectively. The values of these constants for each choice of $V \in \{V^{fE_8},V^{f\natural},F_{24}\}$ are listed in the table below.
\begin{center}
	\begin{tabular}{c|ccc}
		SVOA $V$ & $\chi^{\rm NS}(V)$ & $\chi^{\rm R}_+(V)$ & $\chi^{\rm R}_-(V)$\\ \hline
		$V^f_{E_8}$ & 8 & 8 & 8\\
		$V^{f\natural}_+$ & 0 & 24 & 0\\
		$V^{f\natural}_-$ & 0 & 0 & 24\\
		$F_{24}$ & 24 & 0 & 0.
	\end{tabular}
\end{center}
Note that each of these theories has a total of 24 spin-$1/2$ fields across both of the NS and R sectors.  

In order to construct a physical string model, one also needs a choice of $\mathcal N=1$ supercurrent in these theories. As shown in \cite{Duncan}, $V^{f\natural}$ admits a unique (up to isomorphism) $\mathcal N=1$ supercurrent, which is stabilized by the sporadic group $Co_0$, the group of automorphisms of the Leech lattice. In the theory $V^{fE_8}$, there is a standard choice of supercurrent of the form
$$G(z)\sim \sum_{i=1}^8 :\partial X^i \psi^i:(z),$$
 which is left invariant by the group of inner automorphisms $U(1)^8:W(E_8)$, where $W(E_8)$ is the Weyl group of $E_8$. Finally, the theory $F_{24}$ has $8$ inequivalent choices for a supercurrent, all of the form
 $$G(z) \sim \sum_{i,j,k}c_{ijk} :\lambda^i\lambda^j\lambda^k:(z),$$
 where the coefficients $c_{ijk}$ are the structure constants of a semisimple Lie algebra $g$ of total dimension 24 \cite{Goddard:1984hg}.
 
 As described in \cite{CDR}, the three theories $V^{fE_8},V^{f\natural},F_{24}$ can be related to each other via orbifolding by a cyclic group of symmetries, that preserve the $\mathcal N=1$ superconformal structure. This is described in detail in \S 3 of \cite{Harrison:2020wxl} for orbifolds relating $F_{24}$ and $V^{fE_8}$. When we turn to full string theory models where these SVOAs are components of the worldsheet theory, we expect these relations will be important for understanding string dualities relating these models.

\subsection{Superstring models} 

We can build 2-dimensional compactifications of type II string theory by taking worldsheet theories of the form $V_1\otimes \bar V_2$ with $V_1,V_2 \in \{ V^{fE_8},V^{f\natural},F_{24}\}$. For each such model, we list the number of fields of spin $0$ or $1/2$ in the massless spectrum (there are also fields with no propagating local degrees of freedom, namely containing only $G_{\mu\nu}$, $B_{\mu\nu}$, the dilaton $\phi$, and possibly gravitinos and vector fields) and the number of spacetime supersymmetries in table \ref{tbl:IIAtad}.\footnote{Note that $F_{24}\otimes \overline F_{24}$ has massive spacetime fermions, though it has no spacetime supersymmetries.}

\begin{table}
\begin{center}
	\begin{tabular}{c|ccccc}
		Theory & NS-NS & R-R & NS-R & R-NS & SUSY\\\hline
		$V^{fE_8}\otimes \bar V^{fE_8}$ & $8\times 8$ & $8\times 8$ & $8\times 8$ & $8\times 8$ & $(16,16)$ \\
		$V^{fE_8}\otimes \bar V^{f\natural}_-$ & $0$ & $8\times 24$ & $8\times 24$ & $0$ & $(8,32)$ \\
		$F_{24}\otimes \bar V^{fE_8}$ & $24\times 8$ & $0$ & $24\times 8$ & $0$ & $(8,8)$ \\
		$V^{f\natural}_+\otimes \bar V^{f\natural}_-$ & $0$ & $24\times 24$ & $0$ & $0$ &  $(0,48)$ \\
		$V^{f\natural}_-\otimes \bar V^{f\natural}_-$ & $0$ & $24\times 24$ & $0$ & $0$ & $(24,24)$ \\
		$F_{24}\otimes \bar V^{f\natural}_-$ & $0$ & $0$ & $24\times 24$ & $0$ & $(0,24)$ \\
				$F_{24}\otimes \bar F_{24}$ & $24\times 24$ & $0$ & $0$ & $0$ & $(0,0)$\\
	\end{tabular}\caption{The two-dimensional type IIA models we consider in this paper. We enumerate the number of massless fields from each sector in the first four columns, and identify the spacetime supersymmetry algebra in the fifth column.} \label{tbl:IIAtad}
	\end{center}
	\end{table}

The theory $V^{fE_8}\otimes \bar V^{fE_8}$ is equivalent to the type IIA string compactified on $T^8=R^8/\Lambda_{E_8}$ at the holomorphically factorized point in moduli space. As both $V^{f\natural}$ and $F_{24}$ can be realized as orbifolds of $V^{fE_8}$, all other theories in the table can be thought of as asymmetric orbifolds of the type IIA string on $T^8$ at the holomorphically factorized point. Some of these models were recently considered in \cite{Florakis:2017zep}.

Similarly, we will also consider 2-dimensional compactifications of heterotic string theory by considering worldsheet theories of the form $V_1\otimes \bar V_2$, with  $V_1$  a self-dual bosonic VOA with $c=24$ and $V_2 \in \{ V^{fE_8},V^{f\natural},F_{24}\}$. In table \ref{tbl:hettad}, we list the number of massless fields and spacetime supersymmetries in terms of $ N$, the number of  currents in $V_1$.

\begin{table}
\begin{center}
	\begin{tabular}{c|ccc}
		Theory & NS & R  & SUSY\\\hline
		$V_1\otimes \bar V^{fE_8}$ & $N\times 8$ & $N\times 8$ &  $(8,8)$ \\
		$V_1\otimes \bar V^{f\natural}_+$ & $0$ & $N\times 24$ &  $(0,24)$\\
		$V_1\otimes \bar F_{24}$ & $N\times 24$ & $0$ &  $(0,0)$\\
	\end{tabular}\caption{The two-dimensional heterotic models we consider in this paper. Here, $N$ is the number of currents in the VOA $V_1$. }\label{tbl:hettad}
\end{center}
\end{table}
When considering type IIA or heterotic string theory compactified to two dimensions, there is the possibility that there is a tadpole for the $B$-field \cite{Lerche:1987sg,Lerche:1987qk,Vafa:1995fj} (see also \cite{Sethi:1996es} and \cite{Sen:1996na}). Such a tadpole can be canceled by adding a certain number of spacetime--filling strings. In appendix \ref{s:Bfield}, we discuss the derivation of the tadpole for the theories we consider.

\section{Physical states in type II on $V_1\otimes \bar V_2$}\label{sec:states}

In this section, we will identify the physical states of type II string theory compactified on $V_1\otimes \bar V_2$, where $V_1$ and $V_2$ are each one of the self-dual $\CN=1$ SVOAs with $c=12$, i.e. $V^{f\natural}$, $V_{E_8}$, or $F_{24}$ with a choice of $\CN=1$ structure. The final outcome of this section is that, roughly speaking, the space of physical states  furnishes a tensor product of the physical states for the `chiral' superstring on $V_1$ times the physical states for the `antichiral' superstring on $V_2$.  

A similar factorization holds for the physical states in the heterotic string models on $V_1\otimes \bar V_2$, where $V_1$ is a self-dual bosonic VOA of central charge $c=24$ and $V_2$ a self-dual $\CN=1$ SVOA with $c=12$.

\subsection{Chiral compactification}
Let us first recall the main steps in the construction of the physical states for the `chiral' version of superstrings compactified on a self-dual $c=12$ $\CN=1$ SVOA $V$. This material is largely standard, and we closely follow the notation and treatment of \cite{Harrison:2020wxl}, to which we refer for further details. One starts from a product vertex algebra $\CH^m\otimes \CH^{gh}$. The `matter sector' $\CH^m$ is itself the product of the internal SVOA $V$ and the super vertex algebra (SVA) of $1+1$-spacetime chiral bosons $X^\mu$ and fermions $\psi^\mu$, $\mu=0,1$.  The `ghost' sector $\CH^{gh}$ is the product of a $bc$ ghost and a $\beta \gamma$ superghost system \footnote{\label{fn:ghosts}More precisely, we restrict as always to the relevant subalgebra of the $\eta \zeta \phi$-VA, where $\beta = \partial \zeta e^{-\phi}$, $\gamma= \eta e^{\phi}$.}.

 The product $\CH=\CH^m\otimes \CH^{gh}$ splits into four sectors $\CH_{NS+}$, $\CH_{NS-}$, $\CH_{R+}$, and $\CH_{R-}$, corresponding to the four spin structures on the torus. Here, $$\CH_{NS}=\CH^m_{NS}\otimes \CH_{NS}^{gh}$$ and $$\CH_{R}=\CH^m_{R}\otimes \CH_{R}^{gh}$$ denote the NS and Ramond sector, respectively, and   $\CH_{NS\pm}$ and $\CH_{R\pm}$ are the  projections onto states of positive or negative fermion number $(-1)^F$.\footnote{In the Ramond sector of the $\psi^\mu,X^\mu$ theory, the fermion number acts on the ground states $|\pm\rangle$ by $\pm i$ (i.e. $(-1)^F$ coincides with $i\Gamma$, where $\Gamma$ is the chirality matrix in $1+1$ dimensional spacetime), while in the $\beta,\gamma$ theory $(-1)^F$ acts by $e^{\pi i p}$ on the ground state with picture number $p$. Since $p$ is integral in the NS sector and half-integral in the Ramond sector, the fermion number $(-1)^F$ has order $2$ in the full matter+ghost theory.}
 Then one defines a GSO projected space of states
\be \CH_{GSO}:=\CH_{NS_+}\oplus \CH_{R_+},
\ee
 and further restricts to the $\ker b_0\cap \ker L_0$
\be C:=\CH_{GSO}\cap \ker b_0\cap \ker L_0\ .
\ee The theory contains the conserved currents $j^{gh}=-:\xi\eta:+:bc:$, $j^{p}=\partial\phi+:\xi\eta:$ and  $(P^0,P^1)=(i\partial X^0,i\partial X^1)$, whose zero modes eigenvalues correspond, respectively, to the ghost number $n$, the picture number $p$, and the winding-momenta $k:=(k^0,k^1)$. The space $C$ can be decomposed into components $C^{n}_{p}(k)$ with definite $n$, $p$, and $k$, as
\be C=\bigoplus_{\{k; n,p\}} C^{n}_{p}(k)\ .
\ee In \cite{Harrison:2018joy, Harrison:2020wxl}, the momenta $k$ were assumed to take values in the even unimodular lattice $\Gamma^{1,1}$; here we leave them unspecified, for the time being.

The nilpotent BRST charge $Q$ commutes with $L_0$, picture number and winding-momenta, and has ghost number $1$, so that for each momentum $k$ and picture number $p$ it defines a complex
\be \ldots \rightarrow C^{n-1}_{p}(k)\rightarrow C^{n}_{p}(k)\rightarrow C^{n+1}_{p}(k)\rightarrow \ldots
\ee graded by the ghost number $n$. From this complex, one builds the BRST cohomology $H^{n}_{p}(k)$. Note that the picture number takes integral values in the NS sector and half-integral values in the Ramond sector.

It is also useful to recall the picture-changing homomorphism
\be X:H^n_p(k)\to H_{p+1}^{n}(k),
\ee which is an isomorphism for $k\neq 0$. This allows us to focus on the `canonical pictures', which are $p=-1$ for the NS sector and $p=-1/2$ for the R sector. Furthermore, for $k\neq 0$ the chiral cohomology groups $H^{n}_{-1}(k)$ and $H^{n}_{-1/2}(k)$ are trivial unless $n= 1$. Given this, we define the Hilbert space of (chiral) physical states as
\be \mathcal H^{phys}:= \bigoplus_k H_{-1}^1(k)\oplus H_{-{1\over 2}}^1(k)=\bigoplus_k H^1(k)\ ,
\ee where
\be H^n(k):=H^n_{-1}(k)\oplus H^n_{-1/2}(k)\ .
\ee Note that the same construction could be implemented starting from the GSO projection $\CH_{GSO}=\CH_{NS_+}\oplus \CH_{R_-}$; the corresponding physical states are related by a spacetime parity transformation, which inverts the sign of $X^1, \psi^1$, and $k^1$. 

Analogously, starting from anti-holomorphic rather than holomorphic vertex algebras, one can define a space of physical states for an anti-chiral superstring, in terms of the cohomology groups $\bar H^{\bar n}_{\bar p}(k)$ of a complex $\bar C^\bullet_{\bar p}(k)$ with respect to a nilpotent BRST operator $Q_r$.

\subsection{Non-chiral compactification}
We are now ready to construct the non--chiral (left-- and right--moving) compactification of type IIA on $V_1\otimes \bar V_2$, where $V_1$ and $V_2$ are not necessarily the same SVOA. We define the spaces $\CH^m, \CH^{gh}$ and their right-moving analogues $\bar \CH^m, \bar \CH^{gh}$ as above. Similarly, we define  NS$_\pm$ and R$_\pm$ sectors for both the left-- and the right--moving spaces of states and the corresponding spaces $\CH_{NS_\pm}$, $\CH_{R_\pm}$, $\bar \CH_{NS_\pm}$, $\bar\CH_{R_\pm}$. The GSO projected space of states is
\be \CH_{GSO}:=(\CH_{NS_+}\otimes \bar \CH_{NS_+})\oplus (\CH_{NS_+}\otimes \bar \CH_{R_-})\oplus  (\CH_{R_+}\otimes \bar \CH_{NS_+})\oplus (\CH_{R_+}\otimes \bar \CH_{R_-})\ .
\ee 
Note that we are only considering type IIA GSO projection. We could also consider type IIB theories, by projecting the NS-R and R-R sectors on the subspace with positive, rather than negative right-moving fermion number. However, we would not obtain anything new in this way: as stressed in section \ref{sec:models}, there is a choice of sign for the fermion number in the Ramond sector of the internal SVOA $V_2$, and type IIB on $V_1\otimes \bar V_2$ with one choice of sign for $V_2$ is equivalent to type IIA with the opposite choice. In particular, type IIA on $V_1\otimes \bar V^{f\natural}_{\pm}$ is equivalent to type IIB on $V_1\otimes \bar V^{f\natural}_{\mp}$; when $V_2$ is either $V^{fE_8}$ or $F_{24}$, type IIA and type IIB GSO projections give equivalent theories.

We now restrict to the space
\be \Cc:=\CH_{GSO}\cap \ker b_0\cap \ker \bar b_0\cap \ker L_0\cap \ker\bar L_0\ ,
\ee which decomposes as
\be \Cc=\bigoplus_{\{k_l,k_r\}} \Cc(k_l,k_r)=\bigoplus_{\{k_l,k_r; n,\bar n,p,\bar p\}}\Cc^{n,\bar n}_{p;\bar p}(k_l,k_r)\ ,
\ee where $k_l := (k_l^0,k_l^1)$ and $k_r:= (k_r^0,k_r^1)$ are left-- and right--moving momenta, i.e. the eigenvalues of $(P^0_l,P^1_l)$ and $(P^0_r,P^1_r)$, respectively. Here, $P^\mu_l$ and $P^\mu_r$ are the zero modes of $i\partial X^\mu(z)$ and $i\bar\partial X^\mu(\bar z)$, respectively. We will consider two possibilities for the values of these momenta: in the `uncompactified case', where the target spacetime is $\RR^{1,1}$, one has that $k^0_l=k^0_r$ and $k^1_l=k^1_r$ and both take values in $\RR$; in the `compactified case', where a spacelike direction of spacetime is compactified on a circle of radius $R$, one has
\be k^1_l=\frac{1}{\sqrt{2}}\left (\frac{m}{R}-wR\right)\ ,\qquad k^1_r=\frac{1}{\sqrt{2}}\left(\frac{m}{R}+wR\right)\ ,\qquad m,w\in \ZZ\ ,
\ee and again $k^0_l=k^0_r\in \RR$.

Throughout this section, we will construct physical states in relative cohomology for simplicity, from the space $\Cc$ above. We will treat the physical states of the complete, non-chiral superstring more properly in semirelative cohomology (i.e. imposing the condition $b_0 - \bar{b}_0=0$ rather than $b_0=\bar{b}_0=0$) in Appendix \ref{app:semirelative}. These two cohomologies are isomorphic at nonzero momentum, and hence in particular are both supported in ghost number $(1, 1)$. The relative and semirelative cohomologies differ, however, at zero momentum. Moreover, zero momentum states in both relative and semirelative cohomologies are supported in various ghost numbers. At ghost number 2, which are the states on which we will construct a BKM algebra action in the next section, the semirelative cohomology only differs from the relative cohomology ($\ker b_0 \cap\ker \bar{b}_0$) by a single state at zero-momentum which is largely unimportant for our subsequent analyses.

 Now, if we denote by $C^{n}_{p}(k_l)$ and $\bar{C}^{\bar n}_{ \bar p}(k_r)$ the respective spaces that we would obtain by the chiral and anti-chiral construction described above for $V_1$ and $\bar V_2$ (with the appropriate GSO projections), one has
\be \Cc^{N}_{p,\bar p}(k_l,k_r)=\bigoplus_{n+\bar n=N} C^{n}_{p}(k_l)\otimes \bar{C}^{\bar n}_{ \bar p}(k_r)\ ,
\ee  where $N$ is the total (holomorphic plus anti-holomorphic) ghost number. Furthermore, if $Q_l$ and $Q_r$ are the chiral and antichiral BRST charges for the corresponding left-- and right--moving complexes, the operator $Q:=Q_l+Q_r\equiv Q_l\otimes 1+1\otimes Q_r$ is nilpotent (in particular, $\{Q_l,Q_r\}=0$) and has total ghost number $1$, so that it defines a complex 
\be \ldots \rightarrow \Cc^{N-1}_{p,\bar p}(k_l,k_r)\rightarrow \Cc^{N}_{p,\bar p}(k_l,k_r)\rightarrow \Cc^{N+1}_{p,\bar p}(k_l,k_r)\rightarrow \ldots
\ee
 This complex defines a $Q$-cohomology $\Hh^N_{p,\bar p}(k_l,k_r)$ such that
\be \Hh^N_{p,\bar p}(k_l,k_r)=\bigoplus_{n+\bar n=N} H^n_{p}(k_l)\otimes \bar H^{\bar n}_{\bar p}(k_r)\ ,
\ee 
where $H^n_{p}(k_l)$ and $\bar H^{\bar n}_{\bar p}(k_r)$ are the corresponding left-- and right--moving cohomologies of $Q_l$ and $Q_r$, respectively.

Finally, we restrict to the canonical pictures $p,\bar p\in \{-1/2,-1\}$, and   total ghost number $N=2$. If $k\neq 0$, then the cohomology group reduces to
\be \Hh^2_{p,\bar p}(k_l,k_r)=H^1_{p}(k_l)\otimes 
\bar{H}^1_{\bar p}(k_r)\ ,
\ee because the other components are zero.  Again, see Appendix \ref{app:semirelative} for a discussion of (unimportant for our purposes) subtleties at zero-momentum.  We define the Hilbert space of physical states as
\begin{align}  \CH^{phys}&:= \bigoplus_{\{k_l,k_r\}} \Hh^2(k_l,k_r)\nonumber \\
&=\bigoplus_{\{k_l,k_r\}} (\Hh^2_{-1,-1}(k_l,k_r)\oplus \Hh^2_{-1,-{1\over 2}}(k_l,k_r)\oplus \Hh^2_{-{1\over 2},-1}(k_l,k_r)\oplus \Hh^2_{-{1\over 2},-{1\over 2}}(k_l,k_r)),\nonumber \\
\end{align} where the four summands correspond, respectively, to the NS-NS, NS-R, R-NS and R-R sectors, and 
\be
\Hh^N(k_l,k_r):=\bigoplus_{p,\bar p\in \{-1,-1/2\}}\Hh^N_{p,\bar p}(k_l,k_r)\ .
\ee

\subsection{Heterotic strings}

Similar results hold for the physical states of heterotic strings compactified on $V_1\otimes \bar V_2$, where $V_1$ is a bosonic self-dual VOA of central charge $24$ and $V_2$ is one of the self-dual SVOA of central charge $12$. The spacetime matter SCFT in $1+1$ dimensions includes only right-moving fermions $\bar\psi^\mu(\bar z)$, in addition to the usual bosonic fields $X^\mu(z,\bar z)$. Similarly, the holomorphic ghost sector includes only the $bc$-system and not the $\beta\gamma$-system; the anti-holomorphic ghost sector is as in type II theories.

The GSO projected space of states is
\be \CH_{GSO}:=(\CH\otimes \bar \CH_{NS_+})\oplus   (\CH\otimes \bar \CH_{R_+})\ ,
\ee where $\CH$ is the space of states of the holomorphic bosonic matter and ghost CFT. We restrict to the space
\be \Cc:=\CH_{GSO}\cap \ker b_0\cap \ker \bar b_0\cap \ker L_0\cap \ker\bar L_0\ ,
\ee which decomposes as
\be \Cc=\bigoplus_{\{k_l,k_r\}} \Cc(k_l,k_r)=\bigoplus_{\{k_l,k_r; n,\bar n,\bar p\}}\Cc^{n,\bar n}_{\bar p}(k_l,k_r)\ ,
\ee where now only the anti-holomorphic picture number $\bar p$ is defined. The spaces $\Cc^{n,\bar n}_{\bar p}(k_l,k_r)$ decompose as
\be \Cc^{N}_{\bar p}(k_l,k_r)=\bigoplus_{n+\bar n=N} C^{n}(k_l)\otimes \bar{C}^{\bar n}_{ \bar p}(k_r)\ ,
\ee
where $\bar{C}^{\bar n}_{ \bar p}(k_r)$ is as in the previous section, and $C^{n}(k_l)$ is the space one would obtain by a chiral bosonic string construction starting from the internal VOA $V_1$, tensoring with the chiral bosonic spacetime matter in $1+1$ dimensions and $bc$-ghost system, and restricting to $\ker b_0\cap\ker L_0$. One can define a nilpotent BRST charge $Q$ of ghost number $1$ that is the sum $Q=Q_l+Q_r$ of BRST charges $Q_l$ and $Q_r$ acting only on the left- and right-moving sectors, respectively. For each $\bar p$ and $(k_l,k_r)$, one has a complex
\be \ldots \rightarrow \Cc^{N-1}_{\bar p}(k_l,k_r)\rightarrow \Cc^{N}_{\bar p}(k_l,k_r)\rightarrow \Cc^{N+1}_{\bar p}(k_l,k_r)\rightarrow \ldots
\ee and the corresponding cohomology $\Hh^N_{\bar p}(k_l,k_r)$ is given by
\be \Hh^N_{\bar p}(k_l,k_r)=\bigoplus_{n+\bar n=N} H^{n}(k_l)\otimes \bar{H}^{\bar n}_{ \bar p}(k_r)\ .
\ee Here, $\bar{H}^{\bar n}_{ \bar p}(k_r)$ is as in the type II case, while $H^{n}(k_l)$ is the cohomology of the complex $C^\bullet(k_l)$ with respect to the nilpotent operator $Q_l$. For $k_l,k_r\neq 0$, the cohomology group $\Hh^N_{\bar p}(k_l,k_r)$ is non-trivial only at ghost number $N=2$ and  factorizes as
\be \Hh^2_{\bar p}(k_l,k_r)= H^{1}(k_l)\otimes \bar{H}^{1}_{ \bar p}(k_r)\ .
\ee
Thanks to the picture changing isomorphism, one can focus on the canonical pictures $\bar p=-1 $ (NS sector) and $\bar p=-1/2$ (Ramond sector). The space of physical states of the heterotic strings is then defined as 
\begin{align}  \CH^{phys}&:= \bigoplus_{\{k_l,k_r\}} \Hh^2(k_l,k_r)
=\bigoplus_{\{k_l,k_r\}} (\Hh^2_{-1}(k_l,k_r)\oplus \Hh^2_{-{1\over 2}}(k_l,k_r)),
\end{align}
where the sum over momenta depends on whether the $1+1$-dimensional spacetime is compactified or not, as in the type II case.
\section{BKM algebras of BRST exact states}\label{sec:algebras}

Now that we have constructed the physical states of our model, we can define the corresponding action of a Borcherds-Kac-Moody (BKM) algebra  on a particular subsector of the space of physical states. A BKM algebra is a type of infinite-dimensional Lie algebra introduced by Borcherds \cite{BorcherdsMM} expanding the notion of a Kac-Moody algebra. The construction of a BKM superalgebra on a chiral superstring is well known \cite{Sch1, Sch2}, so at a point in string theory moduli space where the worldsheet theory holomorphically factorizes, one might naively expect two complementary BKMs acting on the left and right. However, a tensor product of two Lie algebras is not a Lie algebra, so a more complicated action of a BKM must arise in the physical superstring if it is to arise at all. Moreover, the chiral superstring construction of a BKM requires all spacetime dimensions to be compactified, and it would be far more interesting if the algebra structure arose in a standard non-chiral string theory where at least the time direction remains uncompactified. 

To set the notation we will begin this section with a brief overview of the basic structure theory of BKMs. Then we  proceed to discuss BPS states and show that they form a representation of a BKM algebra $\mathfrak{g}$. Finally we  demonstrate that $\mathfrak{g}$ is also a symmetry of BPS saturated genus zero amplitudes. We also offer some speculations on the physical interpretation of these results.

\subsection{Basic structure theory of BKM algebras}

Let us review the definition of a Borcherds-Kac-Moody superalgebra. For more details, see \cite{Ray}. Recall that, in general, a \emph{Lie superalgebra} is a $\mathbb{Z}_2$-graded vector space $\mathfrak{g}=\mathfrak{g}_0\oplus \mathfrak{g}_1$ equipped with a  Lie bracket satisfying the  $\ZZ_2$-graded version of the usual skew-symmetry properties and Jacobi identity. A BKM Lie superalgebra $\g$  is characterised 
by a Cartan matrix $A$, which is  allowed to have infinite rank and
is generically of indefinite signature \cite{borcherds1988generalized, Ray}. Let $I$ be a set  (finite or countably infinite) indexing the simple roots of $\g$ and let $S\subseteq I$ be a subset indexing the odd simple roots. Let $\mathfrak{h}_\mathbb{R}$ be a real vector space with generators $h_i, i\in I$, and equipped with a non-degenerate symmetric real-valued bilinear form $(\cdot |\cdot )$. The bilinear form satisfies the following three properties:
\begin{eqnarray}
& (1)& (h_i|h_j)\leq 0\quad  \text{if}\quad  i\neq j,\nonumber \\
& (2)& \text{If}\quad   (h_i|h_i)> 0\quad   \text{then}\quad  \frac{2(h_i|h_j)}{(h_i|h_i)}\in \mathbb{Z} \quad  \text{for all}\quad   j\in I,\nonumber \\
&(3) & \text{If}\quad  (h_i|h_i)>0\quad  \text{and} \quad  i\in S \quad  \text{then}\quad    \frac{(h_i|h_j)}{h_i|h_i)}\in \mathbb{Z} \quad  \text{for all}\quad   j\in I.
\end{eqnarray}

Now set $\mathfrak{h}=\mathfrak{h}_{\mathbb{R}}\otimes_{\mathbb{R}} \mathbb{C}$. The Cartan matrix is the symmetric matrix $A$ with entries $A_{ij}=(h_i|h_j)$. The BKM superalgebra  associated with the vector space $\mathfrak{h}$ and (generalized) Cartan matrix $A$ is the Lie superalgebra $\mathfrak{g}$ generated by the elements $h_i, e_i, f_i$, $i\in I$, subject to the relations:
\begin{eqnarray}
&& [e_i, f_j]=\delta_{ij}h_i,\quad [h_i,h_j]=0
\nonumber \\
&& [h, e_i]=(h|h_i)e_i, \quad [h, f_i]=-(h|h_i)f_i, \quad \forall h\in \mathfrak{h},
\nonumber \\
&& \text{deg}(e_i)=\text{deg}(f_i)=0 \quad \text{if } i\notin S\qquad \text{and}\qquad \text{deg}(e_i)=\text{deg}(f_i)=1,\quad \text{if } i\in S
\nonumber \\
&& \text{ad}_{e_i}^{1-2A_{ij}/A_{ii}}e_j=0, \quad \text{ad}_{f_i}^{1-2A_{ij}/A_{ii}}f_j=0 \quad \text{if} \quad A_{ii}>0 \quad \text{and} \quad i\neq j,
\nonumber \\
&& \text{ad}_{e_i}^{1-A_{ij}/A_{ii}}e_j=0, \quad \text{ad}_{f_i}^{1-A_{ij}/A_{ii}}f_j=0 \quad \text{if } i\in S,  \quad A_{ii}>0 \quad \text{and}\quad i\neq j,
\nonumber \\
&& [e_i, e_j]=[f_i, f_j]=0 \quad \text{if} \quad A_{ij}=0.
\end{eqnarray} Here, $\deg$ denotes the $\ZZ_2$-grading.
In the special case when $A_{ii}>0$ for all $i\in I$ the resulting algebra $\mathfrak{g}$ is a Kac-Moody Lie superalgebra.

Just like for ordinary Kac-Moody algebras, 
the diagonal elements 
$h_i$ generate the Cartan subalgebra $\mathfrak{h}$ and the 
$e_i$'s and $f_j$'s generate nilpotent subalgebras 
$\mathfrak{g}^+$ and $\mathfrak{g}^{-}$, respectively. This implies that $\mathfrak{g}$ has a  triangular decomposition 
\begin{align}
\mathfrak{g}=\mathfrak{g}^{-}\oplus \mathfrak{h}\oplus \mathfrak{g}^+\, \hspace{1cm} \text{(direct sums of vector spaces)}.
\end{align}

The bilinear form $( \cdot | \cdot )$ on $\mathfrak{h}_{\RR}$ induces a bilinear form on the dual space $\mathfrak{h}^*_{\RR}$. An important difference compared to standard Kac-Moody 
algebras is that the diagonal entries of the Cartan matrix $A$ are not required to
be positive. This implies that the simple roots of $\mathfrak{g}$ come in two classes: \emph{real} simple roots
satisfying 
$(\alpha_i|\alpha_i)>0$, and 
\emph{imaginary} simple roots satisfying
$(\alpha_i|\alpha_i)\leq 0$. 

We denote by $\Delta$ the set of all roots. We say that a  root is 
positive (resp.\ negative) if it is a non-negative (resp.\ non-positive) integer
linear combination 
of the simple roots. We therefore have a natural splitting into positive and negative roots
\begin{equation}
\Delta=\Delta^+\oplus \Delta^-.
\end{equation} Accordingly, the nilpotent subalgebras $\g^-$ and $\g^+$ can be written as direct sums
\be \g^\pm=\bigoplus_{\alpha\in\Delta^\pm}\g_\alpha\ ,
\ee  where the root spaces $\g_\alpha$ are finite dimensional ${\rm mult}(\alpha):=\dim(\g_\alpha)<\infty$. The bilinear form $(\cdot |\cdot)$ on $\mathfrak{h}$ extends uniquely to a (super-)symmetric, non-degenerate, invariant bilinear form on $\g$, such that $\g_\alpha$ and $\g_\beta$ are orthogonal unless $\beta=-\alpha$.
The integral span of all simple roots is the \emph{root lattice}
\begin{equation}
\Phi=\sum_{i=1}^{\rank\g}\mathbb{Z}\alpha_i 
\subset \mathfrak{h}^{*}.
\end{equation}

The Weyl group $\mathcal{W}(\mathfrak{g})$ is the group of reflections in 
$\Phi\otimes \mathbb{C}$ with respect to the even real simple roots $\alpha_i$, $i\in I\setminus S$. 
It is generated by the fundamental reflections 
\be
s_i\ : \alpha\ \longmapsto \ \alpha
-2\frac{(\alpha|\alpha_i)}{(\alpha_i|\alpha_i)} \alpha_i.
\ee
An additional important property of a BKM algebra is the existence of a Weyl
vector $\rho$, satisfying 
\begin{align}
(\rho|\alpha)\geq \frac{1}{2}(\alpha|\alpha)\ , \label{DefEquWeyl}  
\end{align}
with equality if and only if $\alpha$ is a simple root.

Let us focus on the case where $S=\emptyset$, so that $\g$ is a (purely even) BKM Lie algebra, rather than a Lie superalgebra. For any (integrable)
lowest weight 
representation $R(\lambda)$ of $\mathfrak{g}$ one has the
Weyl-Kac-Borcherds 
character formula
\be
\text{ch}\, R(\lambda)=\frac{\sum_{w\in\mathcal{W}} \epsilon(w)w(T)e^{\rho}}
{\prod_{\alpha\in \Delta^+} (1-e^{\alpha})^{\text{mult}(\alpha)}}\ ,
\ee
where $\epsilon(w)=(-1)^{\ell(w)}$ with $\ell(w)$  the length of the Weyl
element $w$. This  differs from the
standard Weyl-Kac character formula by the factor $w(T)$ which contains a
correction 
due to the imaginary simple roots
\be
T=e^{\lambda-\rho}\sum_{\mu} \xi(\mu)e^{\mu}
\ .
\ee
The sum is taken over all (unordered) sets of distinct mutually orthogonal imaginary simple roots, whose sum is denoted by $\mu$. Here $\xi(\mu)=(-1)^{m}$ if $\mu$ is a sum of $m$ distinct pairwise
orthogonal 
imaginary simple roots which are orthogonal to $\lambda$, and $\xi(\mu)=0$
otherwise. 
For our purposes we are mainly interested in the simplest case of the trivial
representation $\lambda=0$, 
for which $\text{ch}\, R(\lambda)=1$, and the character formula reduces to the
so called 
\emph{denominator formula}
\be
\sum_{w\in \mathcal{W}} \epsilon(w)w(T)=
e^{-\rho}\prod_{\alpha\in \Delta^+}\left(1-e^{\alpha}\right)^{\text{mult}(\alpha)}\ .
\label{denominatorformula}
\ee
This formula relates a sum over the Weyl group $\mathcal{W}$ to an
infinite product over all positive roots of $\g$.

Let us generalize the the (super-)denominator formula to the case where $\g$ is a BKM superalgebra. To this end we need a little bit of additional structure. Again, we refer to \cite{Ray} for details. Denote the set of roots of $\mathfrak{g}$ by $ \Delta$. This splits into even or odd roots    ${\Delta}_{0}, {\Delta}_{1}$, respectively. Define the corresponding even and odd root multiplicities as follows
\begin{equation}
m_0(\alpha)=\dim (\mathfrak{g}_\alpha\cap  \mathfrak{g}_0), \qquad m_1(\alpha)=\dim (\mathfrak{g}_\alpha \cap \mathfrak{g}_1)=\text{mult}(\alpha)-m_0(\alpha).
\end{equation}
 Any root $\alpha\in \Delta$ can be expanded as $\alpha=\sum_{i\in I} k_i\alpha_i$. Define the \emph{height} $\text{ht}(\alpha)$ and \emph{even height} $\text{ht}_0(\alpha)$ of $\alpha$ as follows
\begin{equation}
\text{ht}(\alpha)=\sum_{i\in I} k_i, \qquad \qquad ht_0(\alpha)=\sum_{i\in I \backslash S} k_i.
\end{equation}
Now introduce the following expressions where the sums are taken over all sums $\mu$ of distinct pairwise orthogonal imaginary simple roots:
\begin{equation}
T=e^{-{\rho}}\sum_{\mu}(-1)^{\text{ht}(\mu)} e^{\mu}, \qquad \qquad T'=e^{-{\rho}}\sum_{\mu}(-1)^{\text{ht}_0(\mu)} e^{\mu},
\end{equation}
where $\rho$ is the Weyl vector.

For any BKM superalgebra $\mathfrak{g}$ we have the {\it denominator formula} and {\it super-denominator formula}:
\begin{eqnarray} 
\sum_{w\in W}\epsilon(w)w(T)&=&\frac{e^{-{\rho}}\, \prod_{\alpha\in {\Delta}_0^+}(1-e^{\alpha})^{m_0(\alpha)}}{\prod_{\alpha\in {\Delta}_1^+}(1+e^{\alpha})^{m_1(\alpha)}},
\nonumber \\
\sum_{w\in W}\epsilon(w)w (T')&=&\frac{e^{-{\rho}}\, \prod_{\alpha\in {\Delta}_0^+}(1-e^{\alpha})^{m_0(\alpha)}}{\prod_{\alpha\in {\Delta}_1^+}(1-e^{\alpha})^{m_1(\alpha)}}.
\end{eqnarray}

\subsection{BPS states in superstring models}
\label{sec:BPSstates}
In \cite{PPV1, PPV2}, several of the authors conjectured the appearance of a BKM algebra in a string theory whose worldsheet matter CFT, prior to compactification on a spatial circle, was holomorphically factorized. In this section, we will expand upon and generalize the BKM action studied in \cite{PPV1, PPV2}.  

The basic idea of \cite{PPV1, PPV2} was as follows. We began with a heterotic string theory whose internal worldsheet CFT was given by the $c=24$ holomorphic Monster VOA, $V^{\natural}$, tensored with the antiholomorphic $c=12$ SVOA $V^{f\natural}$ (for the NS sector), or its ``canonically twisted module" (for the R sector). Completing the standard string theory voodoo (adding ghost sectors, GSO projecting, etc) resulted in a (1+1)-d theory with $\mathcal{N}=(0, 24)$ supersymmetry. In 2d string theory, the GSO projection relates the internal fermion number of the worldsheet CFT to fermion chirality in spacetime. Compactifying the remaining spatial direction on a circle $S^1$ of radius $R$ enables one to write the supersymmetry algebra as 
\be
\left\lbrace \mathcal{Q}^i, \mathcal{Q}^j \right\rbrace = 2\delta^{ij}(P^0_r - P^1_r), \ \ i,j=1,\ldots, 24,
\ee expressed in terms of the temporal and spatial components of the \textit{right-moving} momenta around the circle. BPS states in this model are annihilated by all supercharges and hence satisfy $k^0_r = k^1_r$, having equal eigenvalues of the $P^0_r, P^1_r$ operators. BPS states furnish a subspace of physical states in the (0+1)-d string theory, and the latter arise as usual from BRST cohomology classes with respect to a nilpotent operator $Q = Q_l + Q_r$, which can be expressed as a sum of left and right-moving pieces. When the right-moving momentum is nonzero, the space of physical states factorizes into a product of left and right-moving cohomology classes, graded by momenta, picture number, and ghost number. To the left-moving factor we can associate a BKM algebra following precisely the chiral construction; on the subspace of BPS states, the right-moving factor degenerates into 24 copies of the trivial representation of this algebra, contributing only a (24-fold) multiplicity to each BPS state. In other words, although we cannot tensor two algebras to obtain an algebra, a single copy of the BKM structure is preserved upon tensoring with (copies of) the trivial representation. It turns out that in this non-chiral string theory, for each choice of BRST-representative satisfying the BPS condition $k^0_r = k^1_r$, there is a corresponding nonzero BRST-exact state obtained from it by acting with a single supercharge $\mathcal{Q}^i$. The BRST-exact state is the $Q_r$-image of a $Q_l$-closed state. One can then obtain a correspondence between generators of a BKM algebra (in this case, the Monster Lie algebra) and the quotient space coming from the space of $Q_r$-exact $Q_l$-closed BPS states by the space of $Q_l$-exact states.

In the remainder of this section, we will expand and improve upon this algebra construction, generalize to the type II case, and clarify various subtleties and issues along the way. 

Let us consider either a heterotic or type II superstring theory model compactified on $(V_1\otimes \bar V_2)\otimes S^1$, where $V_2$ is  a self-dual SVOA of central charge $12$ and $V_1$ is either a self-dual VOA of central charge $24$ (in the heterotic case) or a self-dual SVOA of central charge $12$ (in the type II case). Following the analysis of the previous section, the space of physical states has the form
\be \CH^{phys}:=(\bigoplus_{\substack{k_l,k_r\neq 0\\ k^0_l=k^0_r}} H^1(k_l)\otimes \bar H^1(k_r))\oplus \Hh^{2}(0,0)\ ,
\ee where $H^1(k)$ and $\bar H^1(k)$ are the cohomology space for the chiral and anti-chiral models with respect to the left-and right-moving BRST charges $Q_l$ and $Q_r$. We have separated the nonzero momentum states from the zero-momentum states, which do not necessarily factorize (see Appendix \ref{app:semirelative}).  In the fermionic case, $H^1$ is a direct sum of $-1$ (NS) and $-1/2$ (R) picture components. 

We want to focus on the subspace 
\be \CH_{BPS}:=\left(\bigoplus_{\substack{k_l,k_r\neq 0\\ k^0_l=k^0_r=k^1_r}} H^1(k_l)\otimes \bar H^1(k_r)\right )\oplus \Hh^{2}(0,0)'\ ,
\ee
 of physical states satisfying \be\label{BPScond} k^0_r=k^1_r\ .\ee The zero momentum sector $\Hh^{2}(0,0)'$ is obtained from $\Hh^{2}(0,0)$ by taking a certain quotient of a suitable subspace; we will give the precise definition in subsection  \ref{s:zeromom}. If the model has spacetime supersymmetry, corresponding to gravitinos of positive chirality in the NS-R sector of the theory, then $\CH_{BPS}$ is the subspace of states annihilated by those supersymmetries, hence BPS states. We use the same notation $\CH_{BPS}$ even when there are no spacetime supersymmetries.  The possible momenta for states in $\CH_{BPS}$ are labeled by two integers $m,w\in \ZZ$ as follows
 \be k^0_l=k^0_r=k^1_r=\frac{1}{\sqrt{2}}(\frac{m}{R}+wR)\ ,\qquad k^1_l=\frac{1}{\sqrt{2}}(\frac{m}{R}-wR)\ ,
 \ee where $R$ is the radius of the circle.
 
Recall that with the self-dual VOA or SVOA $V_1$ is canonically associated, via the `chiral (super)string construction', a BKM algebra (if $V_1$ is bosonic) \cite{BorcherdsMM} or superalgebra (if $V_1$ is an SVOA) $\g$ \cite{Sch1, Sch2}. 
 
In the remainder of this section we will prove the following:
 \begin{enumerate}
 	\item There is a representation $\delta$
 	\begin{align}\delta: \ &\g\to \text{End}(\CH_{BPS})\\ 
 	&x\mapsto \delta_x\end{align}
 	 of the BKM (super)algebra $\g$ associated with $V_1$ on the space $\CH_{BPS}$ of physical BPS states. The elements $x \in \g$ are identified with certain $Q_l$-closed states $v_x$ of ghost number $(1,0)$ in the superstring theory satisfying the `BPS' condition $k^0_r=k^1_r$, or equivalently with their BRST variation $Qv_x$, which only has components with ghost numbers $(1,1)$. For the heterotic string, the BRST variations $Qv_x$ are exact `BPS' states in the NS sector; in the type II case, the even and odd components $\g_0$ and $\g_1$ correspond to exact `BPS' states in the NS-NS and R-NS sector, respectively. Here `BPS' is in quotes because these states satisfy the condition $k^0_r = k^1_r$, but they are not physical states since they are cohomologically trivial. Note that the algebra action does not mix the right-moving NS sector with the right-moving R sector of $\CH_{BPS}$, so that  the representation $\CH_{BPS}$ is, in general, the sum of two  representations corresponding to these sectors. This is  proven in section~\ref{sec:BPSrep} below. \\
 	\item The action of $\g$ on $\CH_{BPS}$ is a symmetry of tree-level string theory amplitudes involving only states in $\CH_{BPS}$ (\emph{purely BPS amplitudes}). Explicitly, for each $x \in \g$ and every $\varrho_1,\ldots, \varrho_n \in \CH_{BPS}$, we have
 	\be\label{BKMsymm} \delta_x\Bigl(\int_{\CM_{0,n}}  \langle \prod_{i=1}^n \CV_{\varrho_i}\rangle \Bigr)\equiv \sum_{j=1}^n \int_{\CM_{0,n}}  \langle \CV_{\delta_x(\varrho_j)}\prod_{i\neq j} \CV_{\varrho_i}\rangle=0\ ,
 	\ee where $\int_{\CM_{0,n}}$ denotes the integration over the appropriate (super)moduli space of genus $0$ with $n$ punctures. Here, $\CV_\varrho$ is the vertex operator corresponding to a state $\varrho\in \CH_{BPS}$. Roughly speaking, this result will be obtained as follows. One considers an $n+1$-point amplitude where the additional vertex operator corresponds to the insertion of the BRST exact state $Qv_x$ associated to $x\in \g$. By standard arguments, since this amplitude includes one BRST-exact and $n$ BRST-closed states, it is expected to vanish. One then integrates over the modulus parametrizing the position of the exact state.\footnote{When $x$ is an odd element of a superalgebra, so that $v_x$ is in the R-NS sector, this statement needs to be refined; see section \ref{s:symmBPS}.} The amplitude is a total derivative with respect to this modulus, so it gets contributions only from the boundary of the moduli space, i.e. from the limit where the insertion of the exact vertex operator coincides with one of the other $n$ punctures. One then just needs to prove that the sum over the boundary contributions has the form \eqref{BKMsymm}.  \end{enumerate}

Claim 1 above is proven in section~\ref{sec:BPSrep} and claim 2 is proven in section~\ref{s:symmBPS}.

We further conjecture that $\g$ is a symmetry of BPS amplitudes at all loops, by analogous reasoning to that outlined above. On the other hand, we do not expect the action of $\g$ to extend to the whole space of physical states $\CH^{phys}$.\footnote{Of course, the zero-momentum subalgebra of $\g$, which corresponds to the standard symmetries related to conservation of spacetime momentum and winding number, will act on all of $\CH^{phys}$.} 

Similarly, the super-algebra $\bar \g$ associated with the anti-holomorphic SVOA $\bar V_2$ acts on the space $\bar\CH_{BPS}$ of states satisfying $k_l^0=k^1_l$, and is a symmetry of the amplitudes built only from the states in such space. Note that the space $\bar\CH_{BPS}$ is different from $\CH_{BPS}$, so that there is apparently no space of states where both $\g$ and $\bar \g$ act.
\subsection{The BPS subspace as a representation of the BKM algebra}
\label{sec:BPSrep}

In this section we define the action of $\g$ on $\CH_{BPS}$. We begin by focusing on the states of nonzero momentum, which factorize into chiral and anti-chiral components.
The algebra $\g$ associated with $V_1$ is the direct sum of finite-dimensional components graded by `momentum' $k$ taking values in the even unimodular lattice $\Gamma^{1,1}\cong \ZZ\oplus \ZZ$:
\be \g=\bigoplus_{k\in \Gamma^{1,1}} \g(k)=\bigoplus_{m,w\in \ZZ}\g(m,w)\ ,
\ee where the components $\g(k)$ are isomorphic, as vector spaces, to the BRST cohomology with respect to the left-moving BRST charge $Q_l$ at momentum $k$,
\be \g(k)\cong H^1(k)\ ,\qquad\qquad  k\in \Gamma^{1,1}\ .
\ee
 There is an analogous BKM superalgebra $\bar\g$ associated with the SVOA $V_2$, which admits a similar decomposition
 \be \bar\g=\bigoplus_{k\in \Gamma^{1,1}} \bar\g(k)=\bigoplus_{m,w\in \ZZ}\bar\g(m,w)\ ,
 \ee with
 \be \bar\g(k)\cong\bar H^1(k) \ ,\qquad\qquad  k\in \Gamma^{1,1}\ .
 \ee

We work in the relative, rather than semi-relative, cohomology, so that the BPS states in $\CH_{BPS}$ admit representatives that are holomorphically factorized
\be \varrho\otimes \bar \varsigma
\ee as tensor products of representatives $\varrho$ and $\bar \varsigma$ for classes in $H^1(k_l)\cong \g(k_l)$ and $\bar H^1(k_r)$ respectively (here, the isomorphisms are to be understood as isomorphisms of vector spaces; we do not claim that an algebra structure is preserved on the tensor product).  The BPS condition $k^0_r=k^1_r$, together with the quantization of momentum $k^1$ along the circle $S^1$ of radius $R$ and the conditions $k^0_l=k^0_r$ (from the uncompactified time direction), imply that
\be k^0_l=k^0_r=k^1_r=\frac{1}{\sqrt{2}}(\frac{m}{R}+wR)\ ,\qquad k^1_l=\frac{1}{\sqrt{2}}(\frac{m}{R}-wR)\ ,\qquad m,w\in\ZZ
\ee where $m$ and $w$ are the quantized momentum and winding number along $S^1$, respectively.
This means that the left-moving momentum takes values in a lattice isomorphic to $\Gamma^{1,1}\cong \ZZ\oplus \ZZ$. The right-moving momentum takes values in a null subspace  of $\RR^{1,1}$, and depends on $m,w\in \ZZ$ as
\be k^0_r=k^1_r=\frac{1}{\sqrt{2}}(\frac{m}{R}+wR)\ .
\ee Let us focus on the space $\bar H^1(k_r)$. Suppose that $\{\bar v^a\}_{a=1\ldots,\chi^{\rm NS}}$ and $\{\bar u^{i+}\}_{i=1,\ldots,\chi^{\rm R}_+}$ are bases for the spaces $\bar V_2^{NS}(1/2)$ and $\bar V_2^{R+}(1/2)$ of states of conformal weight $1/2$ in, respectively, the NS and positive fermion number Ramond sector of the anti-holomorphic SCFT $\bar V_2$. For $k^0_r=k^1_r\neq 0$, the space of $Q_r$-closed states with ghost number $1$, picture number $-1$ and weight $0$ in the anti-holomorphic matter+ghost CFT is spanned by:
\be\label{states} \bar v ^a_{-1/2}e^{-\tilde \phi}\bar c_1e^{ik_r X_r}|0\rangle\ ,\qquad  (\bar\psi^0_{-1/2}-\bar\psi^1_{-1/2})e^{-\tilde \phi}\bar c_1e^{ik_r X_r}|0\rangle\ ,
\ee and the latter state is the only $Q_r$-exact state. It is proportional to $Q_r \bar\beta_{-1/2}e^{-\bar \phi}\bar c_1e^{ik_r X_r}|0\rangle$. In the $-1/2$ picture, the space of $Q_r$-closed states with $k^0_r=k^1_r\neq 0$ is spanned by 
\be\label{statesR} \bar u ^{i+}_{-1/2}e^{-\tilde \phi/2}\bar c_1e^{ik_r X_r}|0,-\rangle\ ,& i=1,\ldots,\chi^{{\rm R}}_+(V_2)\ ,
\ee where $|0,-\rangle$ is a Ramond ground state of the $(\bar \psi^\mu,\bar\partial X^\mu)$ SVOA with negative chirality.
This shows that all cohomology spaces $\bar H^1(k_r)$, with $k^0_r=k^1_r \neq 0$, have  dimension $\chi^{\rm NS}(V_2)+\chi^{\rm R+}(V_2)$ and are all isomorphic to each other. 
In fact, there is a distinguished isomorphism among them. One considers the vertex operator $e^{i(k'_r-k_r)X_r}$, which has conformal weight $0$ and non-singular OPE with all the states \eqref{states} and \eqref{statesR}. It is easy to see that the zero mode $(e^{i(k'_r-k_r)X_r})_0$ induces a well defined map \be\label{expmap} (e^{i(k'_r-k_r)X_r})_0:\bar H^1(k_r)\rightarrow \bar H^1(k'_r)\ ,\ee which is an isomorphism for $k_r,k'_r\neq 0$. 

With these preparations, one can then define the action of an element $x\in \g(m,w)$ on a physical state $\varrho\otimes \bar \varsigma\in \CH_{BPS}$ by
\be \delta_x(\varrho\otimes \bar \varsigma):=[x,\varrho]\otimes (e^{\frac{i}{\sqrt{2}}(\frac{m}{R}+wR)X_r})_0\bar \varsigma\ .
\ee Here, we use the fact that the left-moving factor $\varrho\in H^1(k_l)\cong \g(k_l)$ can be seen as an element of the algebra $\g$ itself. The shift in the right-moving momentum is necessary to preserve the condition $k^0_l=k^0_r=k^1_r$.

If we have two elements $x\in \g(m,w)$, $x'\in \g(m',w')$, one has
\begin{align}
(\delta_x\delta_{x'}-\delta_{x'}\delta_{x})(\varrho\otimes \bar \varsigma)&=([x,[x',\varrho]]-[x',[x,\varrho]])\otimes (e^{\frac{i}{\sqrt{2}}(\frac{m+m'}{R}+(w+w')R)X_r})_0\bar \varsigma \\&=[[x,x'],\varrho]\otimes (e^{\frac{i}{\sqrt{2}}(\frac{m+m'}{R}+(w+w')R)X_r})_0\bar \varsigma =\delta_{[x,x']}(\varrho\otimes \bar \varsigma)
\end{align} where in the second line we used the Jacobi identity. This shows that $\delta$ is indeed a representation of $\g$ -- in fact, it is the tensor product of the adjoint representation of $\g$ times a trivial (in general, not irreducible) representation of dimension $\chi^{NS}(V_2)+\chi^R_+(V_2)$. This proves the first claim in section~\ref{sec:BPSstates}. 
\subsection{Subtleties with zero momentum}\label{s:zeromom}

The procedure described in the previous subsection is problematic when either $\varrho\otimes \bar \varsigma\in \CH_{BPS}$ or $\delta_x(\varrho\otimes \bar \varsigma)$ has zero right-moving momentum.\footnote{As explained below, we are going to exclude the ghost-dilaton from the space $\CH_{BPS}$. This allows us to represent the states in $\CH_{BPS}$ as elements in the tensor product $H^1(k_l)\otimes \bar H^1(k_r)$ even at zero momentum $k_l=k_r=0$. See appendix \ref{app:semirelative} for details.} The problem is that the operator $(e^{i(k'_r-k_r)X_r})_0$ from states of momentum $k_r$ to states of momentum $k'_r$ does not map, in general, $Q_r$-closed states to $Q_r$-closed states and $Q_r$-exact states to $Q_r$-exact states.

Indeed, the vertex operator $e^{ip_rX_r}(\bar z)$, with $p^1_r=p^0_r$,  does not commute with $Q$ 
\be [Q,e^{ip_r X_r}]=p_{r\mu}(\bar c \bar\partial X^\mu+\bar\gamma  \bar\psi^\mu)e^{ip_r X_r}\ .
\ee Given this result, it is quite surprising that, at least for certain momenta, the zero mode $(e^{ip_rX_r})_0$ gives a well-defined map on the cohomology at all. Note however, that the zero mode of $p_{r\mu} \bar\partial X^\mu$, when acting on states with $k_r^0=k^1_r$, gives a term proportional to $k_r\cdot p_r=0$, since $k_r$ and $p_r$ are both null and proportional to each other. As for the second term, the only cases where it is potentially non-vanishing is when the  commutator $[Q,e^{ip_r X_r}]$ is applied to the ghost number $0$ state $\bar\beta_{-1/2}e^{-\bar \phi}\bar c_1|k_r\rangle$ or the ghost number $1$ states $\bar\psi^\mu_{-1/2}e^{-\tilde \phi}\bar c_1|k_r\rangle$ and $\bar u ^{i\mp}_{-1/2}e^{-\tilde \phi/2}\bar c_1|k_r,\pm\rangle$. In these cases, we have
\begin{align} (e^{ip_rX_r})_0Q_r\,\bar\beta_{-1/2}e^{-\bar \phi}\bar c_1|k_r\rangle&=k_{r\mu}\bar\psi^\mu_{-1/2}e^{-\tilde \phi}\bar c_1|k_r+p_r\rangle\ ,\\ Q_r(e^{ip_r X_r})_0\,\bar\beta_{-1/2}e^{-\bar \phi}\bar c_1|k_r\rangle&=(k_{r\mu}+p_{r\mu})\bar\psi^\mu_{-1/2}e^{-\tilde \phi}\bar c_1|k_r+p_r\rangle\ ,
\end{align}
as well as
\begin{align} (e^{ip_rX_r})_0Q_r\epsilon_\mu\bar\psi^\mu_{-1/2}e^{-\tilde \phi}\bar c_1|k_r \rangle&=\epsilon_\mu k^\mu_r \gamma_{-1/2}e^{-\tilde \phi}\bar c_1|k_r+p_r\rangle\ ,\\  Q_r(e^{ip_rX_r})_0\epsilon_\mu\bar\psi^\mu_{-1/2}e^{-\tilde \phi}\bar c_1|k_r\rangle&=\epsilon_\mu (k^\mu_r+p^\mu_r) \gamma_{-1/2}e^{-\tilde \phi}\bar c_1|k_r+p_r\rangle\ ,
\end{align} 
and
\begin{align} (e^{ip_rX_r})_0Q_r\bar u ^{i\mp}_{-1/2}e^{-\tilde \phi/2}\bar c_1|k_r,\pm\rangle&=k_{r\mu}\bar\gamma_0\bar\psi_0^\mu\bar u ^{i\mp}_{-1/2}e^{-\tilde \phi/2}\bar c_1|k_r+p_r,\pm\rangle\ ,\\  Q_r(e^{ip_rX_r})_0\bar u ^{i\mp}_{-1/2}e^{-\tilde \phi/2}\bar c_1|k_r,\pm\rangle&=(k_{r\mu}+p_{r\mu})\bar\gamma_0\bar\psi_0^\mu\bar u ^{i\mp}_{-1/2}e^{-\tilde \phi/2}\bar c_1|k_r+p_r,\pm\rangle\ .
\end{align}
Since $k_r$ and $k_r+p_r$ are proportional to each other, then so long as both $k_r\neq 0$ and $k_r+p_r\neq 0$, exchanging $(e^{ip_rX_r})_0$ and $Q_r$ gives just a rescaling of the resulting state. This explains why $(e^{ip_rX_r})_0$ provides an isomorphism of cohomology in these cases. Of course, these arguments only work for states in $\CH_{BPS}$, so there seems to be no natural way to extend the action of $\g$ to the whole space of physical states $\CH^{phys}$.

The problems arise when either $k_r$ or $k_r+p_r$ is zero. Recall that for $k^0_r=k^1_r=0$ and ghost number $1$ the  $Q_r$-closed states are 
\begin{align}\label{stateszero} &\vsf^a:=\bar v ^a_{-1/2}e^{-\tilde \phi}\bar c_1|0\rangle\ ,& a=1,\ldots,\chi^{\rm NS}(V_2)\\
&\vsf_-:=(\bar\psi^0_{-1/2}- \bar\psi^1_{-1/2})e^{-\tilde \phi}\bar c_1|0\rangle\ ,\\ 
&\vsf_+:=(\bar\psi^0_{-1/2}+ \bar\psi^1_{-1/2})e^{-\tilde \phi}\bar c_1|0\rangle\ ,
\end{align} in the $-1$-picture (NS sector), and \begin{align}\label{stateszeroR} &\usf^{i}_+:=\bar u ^{i-}_{-1/2}e^{-\tilde \phi/2}\bar c_1|0,+\rangle\ ,& i=1,\ldots,\chi^{{\rm R}}_-(V_2)\\&\usf^{i}_-:=\bar u ^{i+}_{-1/2}e^{-\tilde \phi/2}\bar c_1|0,-\rangle\ ,& i=1,\ldots,\chi^{{\rm R}}_+(V_2)
\end{align} in the $-1/2$-picture (Ramond sector). None of these states is $Q_r$-exact. 

Therefore we have two kind of problems:
\begin{enumerate}
	\item The  state $\epsilon_\mu\bar\psi^\mu_{-1/2}e^{-\tilde \phi}\bar c_1|0\rangle$ at zero momentum, which is closed for every choice of polarization $\epsilon_\mu$, is mapped by $(e^{ip_rX_r})_0$ to $\epsilon_\mu\bar\psi^\mu_{-1/2}e^{-\tilde \phi}\bar c_1|p_r\rangle$, which is closed only when $\epsilon_\mu$ is proportional to $p_{r\mu}$. Similarly, the $Q_r$-closed states $\upsilon^{i+}$ in the Ramond sector are mapped to the states $\bar u ^{i-}_{-1/2}e^{-\tilde \phi/2}\bar c_1|p_r,+\rangle $ that are not closed for $p^0_r=p^1_r$. Therefore, $(e^{ip_rX_r})_0$ maps physical states with zero momentum to non-physical states with momentum $p_r$.
	\item The state $k_{r\mu}\bar\psi^\mu_{-1/2}e^{-\tilde \phi}\bar c_1|k_r\rangle$ at momentum $k_r\neq 0$, which is exact, is mapped by $(e^{-ik_r X_r})_0$ to the state $k_{r\mu}\bar\psi^\mu_{-1/2}e^{-\tilde \phi}\bar c_1|0\rangle$, which is closed but not exact. Therefore, $(e^{-ik_rX_r})_0$ maps exact states with momentum $k_r$ to non-exact states with momentum $0$.
\end{enumerate}

As a consequence, $(e^{ip_r X_r})_0$ does not induce a well-defined map on cohomology in these cases.

\paragraph{Possible resolutions.} For problem 1 above, there is actually a simple (though perhaps slightly unnatural) solution. In the NS sector, we can just exclude the states $\epsilon_\mu\bar\psi^\mu_{-1/2}e^{-\tilde \phi}\bar c_1|0\rangle$  from $\CH_{BPS}$, with the exception of the combination \be\label{weird} \vsf_-=(\bar\psi^0_{-1/2}-\bar\psi^1_{-1/2})e^{-\tilde \phi}\bar c_1|0\rangle\ .\ee Similarly, in the Ramond sector, we only include the states $\usf^{i}_{-}$ and exclude the states $\usf^{i}_{+}$. Since we are already considering a particular subspace of $\CH^{phys}$, there seems to be nothing wrong with making one further restriction. Note that by acting with the algebra $\g$ on $\CH_{BPS}$ there is no way to obtain a linear combination $\epsilon_\mu\bar\psi^\mu_{-1/2}e^{-\tilde \phi}\bar c_1|0\rangle$  different from \eqref{weird}, nor a state of the form $\usf^{i}_{+}$, so the restricted space is still a representation of $\g$. In the same fashion, we also exclude the ghost dilaton (see Appendix \ref{app:semirelative}).

Problem $2$, related to the fact that the state \eqref{weird} is not exact at zero momentum, is more subtle. This state, when tensored with an appropriate left-moving counterpart,  corresponds to a soft particle. Soft particles in amplitudes couple to other particles through conserved charges. In this case, the conserved charge is $k^0-k^1$ (the $0$ picture version of the corresponding vertex operator is proportional to the current $\bar\partial X^0-\bar\partial X^1$). Therefore, $\vsf_-$ should decouple from any amplitude with states in $\CH_{BPS}$, i.e. amplitudes with one state \eqref{weird} and all the other states in $\CH_{BPS}$ are always zero. In §\ref{s:symmBPS}, we check this expectation by a direct calculation. This implies that, so long as we consider symmetries of purely BPS amplitudes, the state $\vsf_-$ is equivalent to $0$ (or to a BRST exact state). For these reasons, it seems reasonable that the action of $\g$, which is a symmetry of these purely BPS amplitudes, is only defined modulo $\vsf_-$. 

The conclusion  is that, at zero momentum, one should really consider a quotient of $\bar H^1(0)$ by the state \eqref{weird}. Altogether, the zero momentum space $\Hh^2(0,0)' $   that should be included in $\CH_{BPS}$ is
\be
\Hh^2(0,0)'=H^1(0)\otimes \bar H^1(0)'\ ,
\ee where
\be \bar H^1(0)'=\Bigl(\bigoplus_{a=1}^{\chi^{\rm NS}} \CC\vsf^a\oplus  \CC\vsf_-\oplus \bigoplus_{i=1}^{\chi^{\rm R}_+} \CC\usf^{i}_-\Bigr) /\CC\vsf_-\cong \bigoplus_{a=1}^{\chi^{\rm NS}}  \CC\vsf^a\oplus \bigoplus_{i=1}^{\chi^{\rm R}_+} \CC\usf^{i}_{-}\ .
\ee

The restricted quotient space $\bar H^1(0)'$ has dimension $\chi^{\rm NS}+\chi^{\rm R}_+$, and it is isomorphic to the cohomology at non-zero momentum, with the isomorphism given by operators $(e^{ip_r X_r})_0$.

With this definition, the action of $\g$ on $\CH_{BPS}$ is well-defined, but it remains to show that it acts in the appropriate way on BPS amplitudes.
\subsection{Symmetry of BPS amplitudes}\label{s:symmBPS}

In this subsection, we will prove \eqref{BKMsymm}.
 
Let us consider a purely BPS genus zero string amplitude
\be A_n=\int_{\CM_{0,n}} \langle \prod_{i=1}^n \CV_{\varrho_i\otimes \bar \varsigma_i}(z_i,\bar z_i)\rangle 
\ee with $n\ge 3$ insertions of vertex operators $\CV_{\varrho_i\otimes \bar \varsigma_i}$ corresponding to BRST-closed states $\varrho_i\otimes \bar \varsigma_i$ representing classes in the BPS subspace $\CH_{BPS}$, i.e. such that $k^0_r=k^1_r$ for all $i$. The picture numbers are chosen so as to give a non-zero answer, and $n-3$ vertex operators are integrated. 

For states in $\CH_{BPS}$, the integrand factorizes into a holomorphic times an antiholomorphic factor
\be \langle \prod_{i=1}^n \CV_{\varrho_i\otimes \bar \varsigma_i}(z_i,\bar z_i)\rangle =\langle \prod_{i=1}^n \CV_{\varrho}(z_i)\rangle \langle \prod_{i=1}^n \CV_{\bar \varsigma_i}(\bar z_i)\rangle \ .
\ee
Let us focus on the right-moving factor, which must be anti-holomorphic in the variables $z_i$, with possible singularities when two insertion points coincide. The BPS and physical state conditions imply that the $\CV_{\bar \varsigma_i}$ are built using only the following combinations of right-moving `spacetime' operators
\be\label{stoper} \bar\psi^0-\bar\psi^1\ ,\qquad \bar\partial X^0-\bar \partial X^1\ ,\qquad e^{ik_r X_r}\ \text{(with } k_r^0=k_r^1)\ ,
\ee multiplied by (super-)ghost and internal matter operators. Using the OPE
\be \bar\partial X^\mu(\bar z)\bar\partial X^\nu(0)=-\frac{\eta^{\mu\nu}}{{\bar z}^2}+O(1)\ ,\qquad \bar\psi^\mu(\bar z)\bar\psi^\nu(0)=\frac{\eta^{\mu\nu}}{\bar z}+O(1)
\ee
\be \bar\partial X^\mu(z)e^{ik_rX_r}(0)=\frac{ik_r^\mu}{\bar z}e^{ik_rX_r}(0)+O(1)\ ,\ee
\be e^{ik_r X_r}(\bar z)e^{ik'_r X_r}(0)=\bar z^{k_r\cdot k'_r}(\pm e^{i(k_r+k_r')X_r}(0)+O(\bar z))\ ,
\ee it is easy to see that the spacetime operators \eqref{stoper} have non-singular OPE with each other. 

Now, suppose that one of the vertex operators $\CV_{\bar \varsigma}(\bar z)$ (taken in the $0$-picture, integrated form) is just a product of spacetime operators \eqref{stoper}, with no further `internal' matter or ghost factor. An example is 
\be \CV_{\bar \varsigma}(\bar z)= :(\bar\partial X^0-\bar\partial X^1) e^{ik_r X_r}: (\bar z)d\bar z\ ,
\ee which is the $0$-picture vertex operator of the closed state  $\bar \varsigma=(\bar\psi^0_{-1/2}-\bar\psi^1_{-1/2})e^{-\tilde \phi}\bar c_1e^{ik_r X_r}|0\rangle$.

 In this case, the antiholomorphic factor $\langle \CV_{\bar \varsigma}(\bar z) \prod_{i=1}^n \CV_{\bar \varsigma_i}(\bar z_i)\rangle$ has no singularities in $\bar z$, so it must be a constant in this variable. It follows that, for these closed states, the CFT correlator $\langle \CV_{\varrho\otimes \bar \varsigma}(z,\bar z) \prod_{i=1}^n \CV_{\varrho_i\otimes \bar \varsigma_i}(z_i,\bar z_i)\rangle$ is meromorphic in the insertion position $z$. 
 
 Note that the state $\bar \varsigma=(\bar\psi^0_{-1/2}-\bar\psi^1_{-1/2})e^{-\tilde \phi}\bar c_1e^{ik_rX_r}|0\rangle$ is BRST-exact if and only if $k_r\neq 0$, and in this case the full vertex operator $\CV_{\varrho\otimes \bar \varsigma}(z,\bar z)$ is a total derivative
 \be\label{vertex} \CV_{\varrho\otimes \bar \varsigma}(z,\bar z)\propto\partial_{\bar z}(\CV_{\varrho}(z)e^{ik_r X_r} (\bar z)) \ .
 \ee  
 
Let us consider an $(n+1)$-point amplitude where the final vertex operator is of the form \eqref{vertex} for $k_r\neq 0$. 
\be A_{n+1}= \int_{\CM_{0,n}} \int_{\mathbb{P}^1} dzd\bar z\, \frac{\partial}{\partial{\bar z}}\langle \prod_{i=1}^{n} \CV_{\varrho_i\otimes \bar \varsigma_i}(z_i,\bar z_i) \CV_{\varrho}(z)e^{ik_r X_r} (\bar z)\rangle\ .\ee
Since the vertex operator \eqref{vertex} corresponds to an exact state, we expect the amplitude to vanish. We will now prove that this is the case.

  The integral in $z$ needs to be regularized by cutting a small disc around each of the other insertion points $z_1,\ldots, z_{n}$ and then taking the limit where the radius of each disc goes to zero.  Since the integrand is a total derivative of a meromorphic function in $z$, by Stokes theorem we obtain 
  \be\label{variation1} A_{n+1}=\int_{\CM_{0,n}} \sum_{j=1}^{n}\oint_{\gamma_{z_j}} dz\langle \prod_{i=1}^{n} \CV_{\varrho_i\otimes \bar \varsigma_i}(z_i,\bar z_i) \CV_{\varrho}(z)e^{ik_r X_r} (\bar z)\rangle \ .
  \ee where $\gamma_{z_j}$ is a small circle centered in $z_j$. Each contour integral picks up the residue of the correlator at the corresponding singularity. For a (single-valued) meromorphic function,  the sum over all residues vanishes, so that
  \be A_{n+1}=0\ ,
  \ee as expected. 
 
 An analogous result holds even when the state $\bar \varsigma$ has zero momentum $k_r=0$. In this case, $\bar \varsigma$ coincides with the state $\upsilon^-$ of equation \eqref{weird}. The vertex operator can still be written as a total derivative, if one allows for the fields $X_r^\mu(\bar z)$ to appear without derivatives or exponentials
 \be\label{totalder} \CV_{\varrho\otimes \bar \varsigma}(z,\bar z)\propto\partial_{\bar z}(\CV_{\varrho}(z)(X_r^0-X_r^1) (\bar z)) \ .
 \ee In general, there are some subtleties in trying to apply the previous argument in presence of the fields $X_r^\mu(\bar z)$.  Indeed, these fields might in principle lead to logarithmic singularities that spoil the single-valuedness of the correlator. However, this never happens in the case of the pure BPS amplitudes we are interested in, because the combination  $X_r^0-X_r^1$ has non-singular OPE with itself and with all the operators in \eqref{stoper}. We conclude that, while $\upsilon^-$ is \emph{not} an exact state at zero momentum $k_r=0$, a pure BPS amplitude with an insertion of the corresponding vertex operator always vanishes, as anticipated in the previous subsection.
 
 Coming back to \eqref{variation1}, the singularity at the point $z=z_i$ arises from the OPE of $\CV_{\varrho}(z)e^{ik_rX_r} (\bar z)$ with $\CV_{\varrho_i\otimes \bar \varsigma_i}(z_i,\bar z_i)$. In order to prove \eqref{BKMsymm}, we just need to show that the residue of this OPE is the action of the algebra element corresponding to $\varrho$ acting on the state $\varrho_i\otimes \bar \varsigma_i$, i.e. 
 \be \oint_{\gamma_{z_i}} dz\, \CV_{u}(z)e^{ik_rX_r} (\bar z) \CV_{\varrho_i\otimes \bar \varsigma_i}(z_i,\bar z_i) =\CV_{\delta_\varrho(\varrho_i\otimes \bar \varsigma_i)}(z_i,\bar z_i)\ .
 \ee
 The OPE of the antiholomorphic operators $e^{ik_rX_r} (\bar z)$ and $\CV_{\bar \varsigma_i}(\bar z_i)$ is non-singular and gives just a shift in the momentum of $\bar \varsigma_i$
 \be e^{ik_r X_r} (\bar z)\CV_{\bar \varsigma_i}(\bar z_i)=\CV_{(e^{ik_r X_r} )_0\bar \varsigma_i}(\bar z_i)+O(\bar z-\bar z_i)\ .
 \ee Note that the terms of order $O(\bar z-\bar z_i)$ in this OPE will not contribute to the full anti-holomorphic correlator, since, as we argued above, the correlator is constant in $\bar z$.  Let us focus on the case where $\varrho$ is in the NS sector, so that it is a representative of a  left-moving BRST cohomology class with ghost number $1$ and canonical $(-1)$-picture. The vertex operator $\CV_\varrho$ is in the integrated form and in the $0$-picture, so it corresponds to the state $b_{-1}X\varrho$ with picture and ghost numbers $0$. The latter state has conformal weight $1$ and the residue picks up the zero mode acting on the state $\varrho_i$: 
 \be \oint_{\gamma_{z_i}} dz\, \CV_{\varrho}(z) \CV_{\varrho_i}(z_i)=\CV_{(b_{-1}X\varrho)_0u_i}(z_i)\ .
 \ee Now, we recall that the Lie bracket of the (even part of the) BKM algebra $\g$ is precisely defined to be
 \be [\varrho,\varrho_i]=(b_{-1}X\varrho)_0\varrho_i\ .
 \ee We conclude that 
 \be \oint_{\gamma_{z_i}} dz\, \CV_{\varrho}(z)e^{ik_r X_r} (\bar z) \CV_{\varrho_i\otimes \bar \varsigma_i}(z_i,\bar z_i)=\CV_{[\varrho,\varrho_i]}(z_i)\CV_{(e^{ik_r X_r} )_0\bar \varsigma_i}(\bar z_i)=\CV_{\delta_\varrho(\varrho_i\otimes \bar \varsigma_i)}(z_i,\bar z_i)\ .
 \ee
A similar result holds when  $\varrho$ is in the Ramond sector, and represents a left-moving BRST cohomology classes with ghost number $1$ and canonical $(-1/2)$-picture. The vertex operator $\CV_\varrho$ is in the integrated form and corresponds to the state $b_{-1}\varrho$ of conformal weight $1$ and picture number $-1/2$, and the residue picks out the zero mode of this current acting on $\varrho_i$,
\be (b_{-1}\varrho)_0\varrho_i\ .
\ee When $\varrho_i$ is in the Ramond sector with canonical $(-1/2)$-picture, the state $(b_{-1}\varrho)_0\varrho_i$ is in the NS sector with canonical picture number $-1$, and is simply the commutator $[\varrho,\varrho_i]$. When $\varrho_i$ is in the NS sector with canonical picture number $-1$, one should move the picture changing operators so that the resulting Ramond state is $X(b_{-1}\varrho)_0\varrho_i\equiv [\varrho,\varrho_i]$ with picture number $-1/2$. This seemingly \emph{ad hoc} prescription for the placement of picture changing operators in fact arises naturally when one describes the superstring amplitude in terms of an integration over supermoduli.\footnote{In particular, the limit where $z$ coincides with another insertion point $z_i$ corresponds to a degeneration of the $(n+1)$-punctured sphere into two spheres joined through a nodal point, with the first sphere containing the two insertion points $z$ and $z_i$ and the second containing all the other punctures. The nodal point is an additional puncture in each of the two spheres, and the puncture is of Ramond type if $\varrho$ and $\varrho_i$ are one NS and one R, and of NS type otherwise. While an NS degeneration is parametrized only by one even supermodulus, a Ramond degeneration corresponds to   one even and one odd supermodulus, and the integration over the odd one corresponds to the insertion of a picture changing operator.} This concludes the proof of the second claim in section~\ref{sec:BPSstates}.

\subsection{Physical interpretation}

Let us pause for a moment and discuss the physical interpretation of the results of this section.
We have shown that the subspace $\mathcal{H}_{BPS}$ of physical states is a representation of a BKM (super-)algebra $\mathfrak{g}$. Furthermore, $\mathfrak{g}$ arises as an algebra of symmetries of the string tree level amplitudes where all insertion points correspond to states in $\CH_{BPS}$.

If the uncompactified spacetime were higher dimensional, the physical interpretation would be straightforward: $\mathfrak{g}$ is an algebra of symmetries of the S-matrix for the scattering of these BPS states, at least at tree level. In the present case, the spacetime we are considering is $0+1$-dimensional (or $1+1$ dimensional on a cylinder). Naively, it seems natural to interpret the amplitudes as transition amplitudes between an initial asymptotic state at time $t=-\infty $ and a final state at time $t=+\infty$.

However, this interpretation is problematic. Let us regard our models as theories in a  $1+1$-dimensional cylindrical spacetime. The states in $\mathcal{H}_{BPS}$, in general, represent spacetime-filling strings. The initial and final states in our transition amplitudes, therefore, represent arbitrary numbers of spacetime-filling strings. However, these states are coupled to the metric and B-field, so they cannot make sense in general: the field equations for these massless degrees of freedom require a precise number of spacetime filling strings to cancel the B-field tadpole and cosmological constant (see appendix \ref{s:Bfield} for a calculation of such a tadpole). In section \ref{sec:secondquantized} we will consider a second quantized version of this theory in a zero coupling limit. In this limit, one can neglect any backreaction and regard the massless fields as a non-dynamical fixed background; these kind of inconsistencies can be safely ignored in this case. However, when considering amplitudes, as in this section, we are implicitly assuming that the string coupling is non-zero, so that this solution is not available here.

One possible attitude is to view the amplitudes as a purely formal construction. We can build a collection of functions that are symmetric under the BKM algebra. This statement is certainly true, even if the quantities do not have a sensible spacetime interpretation. 

On the other hand, the existence of this rich collection of functions is suggestive, and hints that there should be a reasonable way to interpret our results. One possibility could be that the vertex operators are accompanied by a background charge that offsets their coupling to the massless fields. This would be analogous to matter coupled to Liouville gravity in two dimensions, where the Liouville and the matter part are coupled in such a way that the combined stress energy tensor is always traceless. A more suggestive possibility is that there exists a topological version of these string models, where only the states in $\mathcal{H}_{BPS}$ survive, and the dangerous massless fields are either not present or decouple.

While we are not going to further investigate these proposals in the present article, let us conclude by mentioning evidence that the BKM symmetry persists in BPS amplitudes at all loops. We believe this provides an additional hint that a sensible physical interpretation of these amplitudes exists. 

To start, we consider the possibility that, at least, the even subalgebra  $\g_0$ is $\g$ is a symmetry of pure BPS amplitudes at all orders in perturbation theory. The argument of the previous subsection generalizes in a straightforward way. One considers an $(n+1)$-point amplitude on a genus $g$ Riemann surface $\Sigma$, where one of the vertex operators corresponds to a BRST exact NS-NS state, and has the form \eqref{totalder}, i.e. a total derivative in the insertion position. One then integrates the position of this BRST-exact state over $\Sigma$. Technically, this is an integration over the even supermodulus parametrizing this position, i.e. an integration along the fibers of the forgetful map $\CM_{g,n+1}\to \CM_{g,n}$. Being a total derivative, this integral would vanish trivially if the integrand was non-singular. However, there are singularities when $z$ collides with one of the other insertion points, so one needs to cut out from $\Sigma$ small disks around each of these insertion points. For pure BPS amplitudes, the integrand still factorizes into left- and right-moving factors, and the right-moving factor is non-singular in $\bar z$. Therefore, the integrand is a total derivative of a meromorphic form. The integration reduces to a sum over all the residues of this meromorphic form, and this sum vanishes. The sum over residues can again be interpreted as a variation of an $n$-point amplitude with respect to the action of some element of the algebra $\g_0$, so its vanishing implies that the $n$-point amplitude is invariant.
This is a highly non-trivial result, that points toward the existence of a meaningful physical interpretation of these amplitudes.

The odd component of the algebra requires a more sophisticated formalism. In this case, the BRST-exact vertex operator is in the R-NS sector, and there is no even supermodulus that simply parametrizes the position of this operator. Equivalently,  there is no map $\CM_{g,n+1}\to \CM_{g,n}$ where one `forgets' about one Ramond puncture.\footnote{To show that such a forgetful map does not exist, it is sufficient to note that the number of Ramond punctures must always be even.} In this case, one needs to integrate over all supermoduli at once. The integrand is still a total derivative, so one gets only contributions from the boundary components of the Deligne-Mumford compactification of the supermoduli space. To prove that our result at tree level generalizes to higher genus, one needs to show that the only non-vanishing boundary contributions are the ones coming from degenerations of the Riemann surface $\Sigma$, of genus $g$ and $n+1$ punctures, into a surface $\Sigma'$ of the same genus $g$ and $n-1$ punctures and a sphere with two punctures (one of them being the exact state), joined through a double point. Furthermore, the sum over all such degenerations should vanish. In the previous case of NS-NS punctures, these kind of boundary components in the Deligne-Mumford compactification correspond exactly to the limits where the exact vertex operator collides with one of the other punctures.

 These conditions are similar (though not exactly the same) as the conditions for spacetime supersymmetry in general string perturbation theory. In that case, one can show (see for example \cite{Witten2012}) that there are only a few, particular degeneration limits that can potentially spoil spacetime supersymmetry at higher loop, and that these cases are related to either tadpoles,  spontaneous supersymmetry breaking, or to higher loop modifications of supersymmetry transformations. Checking that the full algebra $\g$ (including the odd component) is a symmetry of purely BPS amplitudes at all orders in string perturbation theory would provide strong evidence that a consistent physical interpretation of this theory does exist.

\section{Second quantization and BPS indices}\label{sec:secondquantized}

Starting from the space of physical states in our type IIA or heterotic theory on $V_1\otimes \bar V_2$, let us now define a `second quantized' Hilbert space describing an arbitrary number of strings. We will consider only the limit where the string coupling constant is zero, so that we have a theory of free strings. 

We begin by constructing our second quantized Hilbert space and we define an associated supersymmetric index which only receives contributions from BPS states. We show that the second quantized space of BPS state forms a representation for the BKM algebra $\g$ associated with the (super)VOA $V_1$, so that the index is a character for that representation. We then proceed to consider two examples in detail: a family of heterotic string compactifications on $V\otimes \bar V^{f\natural}$ (\S \ref{sec:hetex1}) and type IIA compactifications on $V\otimes \bar V^{f\natural}$ (\S \ref{sec:IIAex2}). In these cases we calculate the BPS indices and show that they correspond to known (super)denominator formulas of BKM algebras.

\subsection{A second quantized Hilbert space}\label{sec:secondHilb}

Let us consider type IIA or heterotic string theory on $V_1\otimes \bar V_2$, where $V_2\in \{V^{fE_8},V^{f\natural},F_{24}\}$. Occasionaly, we will compactify one further space-like direction on a circle $S^1$. In order to construct this second quantized space, we first introduce one operator $\eta_a$ for each physical state $a\in \Hh^2(k)$ with $k\neq 0$. The operators obey a free oscillator algebra, with (anti-)commutation relations determined by the string two point functions. The $k\neq 0$ operators can be separated into creation ($k^0>0$) or annihilation ($k^0<0$) operators. The idea is that each creation/annihilation operator $\eta_a$ creates or destroys a string in the state $a$ in the $(1+1)$- dimensional spacetime.

For the physical states with $k=0$ (i.e. $k_l=k_r=0$), the construction is slightly more complicated. First note that the $k=0$ physical states are the same in the uncompactified and in the compactified theory; for these definitions, it is convenient to work in the uncompactified theories. We separate the space of zero-momentum physical states into two orthogonal subspaces, the `propagating' and `non-propagating' states. The `propagating' states are the ones that can be obtained starting from massless $(k_l)^2=(k_r)^2=0$ states with $k\neq 0$ and then taking the $k\to 0$ limit (this limit is well defined only in the uncompactified case, which is why we prefer to give the definition in this set up). The `non-propagating' states are the states that are orthogonal to the propagating ones. The physical interpretation is clear: the propagating states are the zero modes of some local propagating degrees of freedom in the $1+1$-dimensional spacetime, corresponding for example to some massless scalars (excluding the dilaton) or spin $1/2$ fermions in spacetime. The non-propagating degrees of freedom correspond, for example, to the metric, B-field, dilaton, gravitinos and gauge vectors, that have no local propagating degrees of freedom in $1+1$-dimensions. The presence of the latter states in the physical spectrum of the superstring correspond to the possibility to deform the background, such as the geometry of the spacetime and the string coupling constant. In the construction of the second quantized space of states, we will introduce one operator $\eta_a$ for each `propagating' state $a\in \CH^2(0)$, and no operator for the `non-propagating' ones. For each of the latter, we can instead introduce a fixed non-dynamical background field: for example, in the compactified case, one can set the radius of the circle, a background B-field, or Wilson lines of gauge fields along the circle (besides the string coupling, which we always fix to zero). 

 We define a ground state of the Hilbert space to be a state in the kernel of all annihilation ($k^0<0$) operators; physically, these states should correspond to the vacuum, i.e. no strings are present. The ground states might be degenerate if there are $k=0$ `propagating' operators. The second quantized Hilbert space is then defined as the Fock space constructed by acting in all possible ways on the ground states with the creation operators.

Let us consider the theory compactified on a circle $S^1$ of radius $R$. We define a supersymmetric index
\be \mathcal{Z}^{V_1 \otimes \bar{V_2}}(\beta, b,v,R, A_k):=\Tr(e^{-\beta H} e^{2\pi i bW}e^{2\pi i vM} e^{2\pi i\sum_k A_kq_k} (-1)^\bF)\ ,
\ee where the $H$, $M$, $W$ are the total Hamiltonian, momentum and winding, the $q_k$ are charges with respect to some abelian background gauge fields (corresponding to a maximal abelian subgroup of the possibly non-abelian gauge group), and $(-1)^{\bF}$ is the spacetime fermion number. The superscript denotes the factorized worldsheet theory in which we are computing the index. The real chemical potentials $\beta$, $b$, $v$, $A_k$ admit natural physical interpretations: $\beta$ is the usual inverse temperature; $b$ is a constant background B-field; $v$ is a constant off-diagonal term in the $1+1$ dimensional metric; $A_k$ are constant background for the gauge fields in a maximal abelian torus of the gauge group, giving non-trivial Wilson lines along the circle $S^1$. 
If $\eta$ is an oscillator corresponding to momentum $k^0_l=k^0_r=E$ and $k^1_{l,r}=\frac{1}{\sqrt{2}}(\frac{m}{R}\mp wR)$, one has
\be [H,\eta]=E\eta\qquad [M,\eta]=m\eta\qquad [W,\eta]=w\eta\ .
\ee The ground states are (possibly degenerate) eigenstates for $H, M,W$
\be H|0\rangle=E_0|0\rangle\ ,\qquad M|0\rangle =m_0|0\rangle\ ,\qquad W|0\rangle =w_0|0\rangle
\ .\ee Here, $w_0$ and $m_0$ have a straightforward physical interpretation as a B-field tadpole and a `vacuum momentum', respectively, since they couple to the background B-field $b$ and to the off-diagonal metric component $v$. They arise as a 1-loop effect in superstring theory, and  can be easily computed for all the theories we are considering, see appendix \ref{s:Bfield}. The `cosmological constant' $E_0$ is usually fixed in terms of $w_0,m_0$ by spacetime supersymmetry.  

\begin{table}[h]\begin{center}
\begin{minipage}{.4\textwidth}
	\begin{tabular}{c|cc}
		Type IIA &  $w_0$ & $m_0$\\\hline
		$V^{fE_8}\otimes \bar V^{fE_8}$ & 0 &0\\
		$F_{24}\otimes \bar V^{fE_8}$ & 0 & 0\\
		$V^{fE_8}\otimes \bar V^{f\natural}_-$ & 0& 0\\
		$V^{f\natural}_{+}\otimes \bar V^{f\natural}_-$ &0&$-24$ \\
		$V^{f\natural}_{-}\otimes \bar V^{f\natural}_-$ &$ -24$ & 0\\
		$F_{24}\otimes \bar V^{f\natural}_-$ & $ 24$ & $24$\\
		$F_{24}\otimes \bar F_{24}$ &0 &0\\
	\end{tabular}
	\end{minipage}

\begin{minipage}{.4\textwidth}
\begin{tabular}{c|cc}
	Het & $w_0$ & $m_0$\\\hline
	$V\otimes \bar V^{fE_8}$ & 0 & 0\\
	$V\otimes \bar V^{f\natural}_\pm$ & $\pm (N-24)$ & $\pm N$\\
	$V\otimes \bar F_{24}$ & 0 & 0
\end{tabular}\end{minipage}
\end{center}\caption{The vacuum winding $w_0$ (i.e. the B-field tadpole) and momentum $m_0$ for the type IIA and heterotic compactifications we are interested in. The details of the calculations are in appendix \ref{s:Bfield}.}\label{tbl:tadpole}\end{table}

The theories we are considering often contain some spacetime supersymmetries $\CQ^i_\alpha$ (see Tables \ref{tbl:IIAtad} and \ref{tbl:hettad}), where $i=1,2,3,\ldots$ and $\alpha\in\{\pm\}$ denote the chirality. In this case, the index gets non-vanishing contributions only from a BPS subspace of the second quantized Hilbert space, which can be constructed as a Fock space by acting on the vacuum only with the creation operators that (anti-)commute with the supercharges. In turn, these operators correspond to single string states that are annihilated by the supercharges (BPS states). Thus we can restrict our analysis to these BPS single string states. In heterotic models, the two-dimensional supersymmetry algebra is
\be \{\CQ^i_\pm,\CQ^j_\pm\}=2\delta^{ij}(P^0_r\pm P^1_r)\ ,\qquad \{\CQ^i_+,\CQ^j_-\}=0\ ,
\ee and involves only the right-moving momentum. The same relations hold in type II theories when the supercharges arise from the NS-R sector, while for the R-NS sector we have the analogous relations with the left-moving momenta,
\be \{\CQ^i_\pm,\CQ^j_\pm\}=2\delta^{ij}(P^0_l\pm P^1_l)\ ,\qquad \{\CQ^i_+,\CQ^j_-\}=0\ .
\ee Type II supercharges coming from different sectors anticommute with each other. 

When any of these supersymmetries is present, the BPS condition is equivalent to a linear relation between the energy $E=k^0_l=k^0_r$ and the winding and momentum $m,w$ of a single string state. By linearity, the same relation must hold between the Hamiltonian $H$ and the momentum and winding operators $M$ and $W$ in the BPS subsector of the second quantized Hilbert space. For example, when negative chirality supersymmetries $\CQ^i_-$ from the right-moving Ramond sector are present, the BPS condition implies
\be H=\frac{1}{\sqrt{2}} \left(\frac{M}{R}+WR\right)\ .
\ee Note that if we require the ground states of the second quantized Hilbert space to be supersymmetric, we have the corresponding relation $E_0=\frac{1}{\sqrt{2}} \left(\frac{m_0}{R}+w_0R\right)$ of vacuum eigenvalues. By restricting the trace to the BPS subsector of the second quantized Hilbert space, we can reexpress the above index as,
\be \mathcal{Z}^{V_1 \otimes \bar{V_2}}(\beta, b,v,R, A_k)=\Tr_{\rm BPS}\Big(e^{-\frac{\beta}{\sqrt{2}} \bigl(\frac{M}{R}+WR\bigr)} e^{2\pi i bW}e^{2\pi i vM} \prod_i e^{2\pi i\sum_k A_kq_k} (-1)^\bF\Big).
\ee 

It is convenient to reorganize the real parameters $\beta$, $R$, $b$ and $v$ into two complex variables,
\be T=b+i\frac{\beta R}{2\sqrt{2}\pi}\ ,\qquad U=v+i\frac{\beta }{2\sqrt{2}\pi R}\ .
\ee Both $T$ and $U$ take values in the upper half-plane and  parametrize, respectively, the K\"ahler and complex structure of the Euclidean spacetime torus obtained upon Wick rotation of the time direction. In terms of these variables, the index becomes
\begin{eqnarray}\label{eq:index0} \mathcal{Z}^{V_1 \otimes \bar{V_2}}(T,U, A_k)&=&\Tr_{\rm BPS}\Big(e^{2\pi i TW}e^{2\pi i UM} \prod_i e^{2\pi i\sum_k A_kq_k} (-1)^\bF\Big)
\nonumber \\
&=&\Tr_{\rm BPS}\Big(p^{W}q^{M} \prod_k y_k^{q_k} (-1)^\bF\Big),
\end{eqnarray}
where we introduced the variables
\be p:=e^{2\pi iT}\ ,\qquad q:=e^{2\pi i U}\ ,\qquad y_k:=e^{2\pi iA_k}\ .
\ee 

If there are bosonic or fermionic zero modes, this definition may need to be modified. Bosonic zero modes might lead to a divergence that needs to be regularized. In the examples we consider below, massless scalars always come from the R-R sector. Since there is no physical string states representing the zero momentum of R-R fields, their contribution is automatically factored out. Fermionic zero modes $\lambda^1, 
\lambda^2\ldots$ make the index vanish (unless they carry non-trivial charges $q_k$). In all the examples we will consider, the fermionic zero modes will arise as gauginos, i.e. superpartners of the gauge fields, and thus transforming  in the adjoint representation of the spacetime gauge group. The number of neutral fermionic zero modes in this case is equal to the rank of the gauge group. In order to get a non-vanishing quantity, one can absorb these zero modes by inserting them in the trace
\be \label{eq:index1} \mathcal{Z}^{V_1 \otimes \bar{V_2}}(\beta, b,v,R, A_k):=\Tr\Big(e^{-\beta H} e^{2\pi i bW}e^{2\pi i vM} e^{2\pi i\sum_k A_kq_k} (-1)^\bF\prod_{i}\lambda^i\Big)\ .\ee
We will use the notation $\mathcal{Z}(T, U, A_k)$ throughout this section to refer to the minimal nonvanishing index for each of the theories we study: if the theory has no fermionic zero modes, then $\mathcal{Z}$ indicates \eqref{eq:index0}, while if the theory does have fermionic zero modes, we absorb them as in \eqref{eq:index1}.

There are further modifications to the index which one may want to consider. First, in type II theories, the spacetime fermions may come either from the NS-R or from the R-NS sector. If there are supercharges coming from both of these sectors, then only the ground states will contribute to the index, thus giving a constant. To get a more interesting quantity, one can consider a modified index
\be \label{eq:index2} \tilde{\mathcal{Z}}^{V_1 \otimes \bar{V_2}}(\beta, b,v,R, A_k):=\Tr\Big(e^{-\beta H} e^{2\pi i bW}e^{2\pi i vM} e^{2\pi i\sum_k A_kq_k} (-1)^{\bF_r}\prod_{i}\lambda^i\Big)\ ,\ee
where the fermion number $(-1)^\bF$ is replaced by an operator $(-1)^{\bF_r}$ anti-commuting only with, say, states in the right-moving  R sector (NS-R and R-R) and commuting with states in the right-moving NS sector (NS-NS and R-NS). Under this definition, the states contributing to the index  obey $k^0_r=k^1_r$, but not necessarily $k^0_l=k^1_l$. As above, we use $\tilde{\mathcal{Z}}$ to refer to the minimal nonvanishing (with any fermionic zero modes inserted) modified index for a given theory.

When the string theory has $V_2=V^{f\natural}_-$ as its right-moving SVOA, $\mathcal{Z}, \tilde{\mathcal{Z}}$ evaluate to the 24th powers of the superdenominator and denominator (respectively) of the BKM algebra constructed from the left-moving (S)VOA. We will derive this in detail in \S \ref{sec:hetex1} and \S \ref{sec:IIAex2}, but the basic idea is that $V^{f\natural}$ has 24 states of conformal weight 1/2 in the Ramond sector and none in the NS sector. The contribution from these 24 fermionic ground states is to produce a 24-fold multiplicity of single-string BPS states, resulting in a second-quantized Fock space that has the form of a 24th tensor power. Any of the 24 factors effectively counts contributions from the left-movers, and produces the BKM (super)denominator.

There are other options to modify the index. We could impose anti-periodic boundary conditions on the fermions around the spacelike $S^1$ circle. This corresponds to taking a different quantization for the spacelike momenta $k^1_l$ and $k^1_r$ in the NS-R and R-NS sectors, where $n$ takes values in $\frac{1}{2}+\ZZ$ instead of $\ZZ$. We could also impose anti-periodic boundary conditions only on (say) the NS-R fermions and periodic ones on the R-NS fermions. We will not consider these modifications in the rest of the text, and focus on computing the indices in equations \eqref{eq:index1} and \eqref{eq:index2}. 

\subsection{The BPS Fock space as an algebra module}

As mentioned in the previous subsection, when the string theory has the right-movers fixed to be $V_2= V^{f\natural}$, the supersymmetric indices \eqref{eq:index1}, \eqref{eq:index2} reproduce the 24th power of the denominator and superdenominator of the BKM associated to the left-movers. The factor of 24 simply comes from the right-moving ground state degeneracy. In fact, the supersymmetric index can be viewed as the graded supercharacter of a certain algebra module for the left-moving BKM superalgebra $\g$. Let us restrict for simplicity to the case where $\g$ is bosonic, i.e. with no odd components; the generalization to the case where $\g$ is a superalgebra is straightforward. Let us consider the usual triangular decomposition as $\mathfrak{g} = \mathfrak{g}^+ \oplus \mathfrak{h} \oplus \mathfrak{g}^-$ in terms of positive and negative roots and the Cartan subalgebra. As discussed below, the module describing the second quantized states is $(\bigwedge \mathfrak{g}^{-})^{\otimes 24}$, 24 copies of the graded sum of all exterior powers of the subalgebra associated to the negative roots. It is well known that for ordinary Lie algebras or Kac-Moody algebras, this is an irreducible $\mathbb{Z}_2$-graded module with highest weight the Weyl vector. It also inherits the weight space grading from $\mathfrak{g}$. In the case of the heterotic string with $V_1 = V^{\natural}$, the BKM algebra $\g$ is the Monster Lie algebra, and the module was constructed in \cite{PPV1}. We will briefly review its construction and the relation with the second quantized Hilbert space described in section \ref{sec:secondHilb}, and then comment on subtleties in the generalization to our other examples. We will focus on the construction of a single copy of $\bigwedge \mathfrak{g}^-$, since the 24 copies do not interact with one another.

The $\g$-representation $\bigwedge \mathfrak{g}^-$ can be formally constructed as follows. Recall that $\mathfrak{g}$ is graded by its root lattice $\Phi \subset \mathfrak{h}^*$, $\mathfrak{g} = \bigoplus_{\alpha \in \Phi} \mathfrak{g}_{\alpha}$, and has an invariant bilinear form $(\cdot|\cdot)$ that respects the grading. The basic idea is to built a module $A_\g$ over the Clifford algebra whose underlying vector space is $\mathfrak{g}$ endowed with $(\cdot|\cdot)$. More precisely, let us consider an algebra
of fermionic operators $\xi_a, \forall a \in \mathfrak{g}$, transforming in the adjoint representation of $\g$. The fermionic modes satisfy the Clifford algebra
\be 
\left\lbrace \xi_a, \xi_b \right\rbrace = (a|b)
\ee where the right-hand side vanishes unless, for any $a \in \mathfrak{g}_{\alpha}$, $b \in \mathfrak{g}_{-\alpha}$. Let us define a module $A_\g$ for this Clifford algebra by considering a space of ground states annihilated by all $\xi_a$ with $a\in \g^+$, and by acting in all possible ways on such ground states by $\xi_a$, $a\in \mathfrak{h}\oplus \g^-$. We can view $\mathfrak{g}$ as a subalgebra of the orthogonal algebra $so(\mathfrak{g})$ that preserves the bilinear form $\kappa$, since $\kappa$ is preserved by the adjoint action of the BKM on itself.  This Clifford module can be understood as a spinorial  representation of the orthogonal algebra $so(\mathfrak{g})$ and hence of its subalgebra $\mathfrak{g}$.

The space of ground states of $A_\g$ forms a module for the Clifford subalgebra associated with $\mathfrak{h}\subset\g$. Each of these ground states is a highest weight vector for a representation of $\g\subset so(\g)$. This immediately shows that, in general, the Clifford module $A_\g$ is not irreducible as a $\g$-representation. 
In the case of the rank-2 Monster Lie algebra, one has a twofold ground state degeneracy of opposite chirality. Indeed, for general $\g$ of even rank, the module always splits, as a representation over $so(\mathfrak{g})$, into two spinors of definite chirality. Focusing on the rank-2 case for the moment, it is natural to consider the $\g$-representation obtained by restricting to a subspace $A_\g^+$ of definite chirality, which effectively allows us to skirt vacuum degeneracy issues relating to the Cartan operators, as follows. If we take a lightcone decomposition of the Cartan as $\mathfrak{h} = \mathfrak{h}^+ \oplus \mathfrak{h}^-$ such that $\left\lbrace \xi_+, \xi_- \right\rbrace$ = 1 (and all other anticommutators zero), we may define a vacuum 
\be 
\xi_a |0 \rangle = 0, \ \ \  a \in \mathfrak{g}^+ \oplus \mathfrak{h}^+.
\ee
The Clifford module $A_\g$ is recovered by acting on the vacuum in all possible ways with the $\xi_a, a \in \mathfrak{g}^- \oplus \mathfrak{h}^-$. Next, we consider a chirality operator $(-1)^J$, distinct from the fermion parity operator used in the index definition, which leaves the vacuum invariant but acts on all $\xi_a$ with a minus sign. We project to the positive chirality subspace of our module $A_\g$, to obtain a graded vector space $A^+_\g$ isomorphic to $\bigwedge \mathfrak{g}^-$. Namely, to obtain the even powers of $\mathfrak{g}^-$, simply act on the vacuum with an even number of $\xi_{a}, a \in \mathfrak{g}^-$, and to obtain the odd powers, act first with $\xi_-$ on the vacuum, and then with an odd number of $\xi_a$, $a \in \mathfrak{g}^-$.

We stress that the set of modes $\{\xi_a\}$ are different from the set $\{\eta_a\}$ considered in section \ref{sec:secondHilb}: in that case, we exclude the modes corresponding to gravitinos, while here we include all modes. Excluding the gravitinos leads, in the rank-2 case, to a non-degenerate vacuum for the second quantized Fock space. This suggests that the second quantized space is isomorphic to the positive chirality subspace $A^+_\g$ of the $\xi_a$ Clifford module; as such, it is naturally a representation for the BKM algebra $\g$. Indeed, the operators $\eta_a$ considered in section \ref{sec:secondHilb} can be identified with the operators $\xi_-\xi_a$, that are even with respect to $(-1)^J$ and carry the same grading with respect to $\Phi\subset \mathfrak{h}^*$.
Implicitly, then, we may imagine that our index can be defined as a trace over the full Clifford module $A_\g$, with an additional insertion of $(1 + (-1)^J)/2$ in the trace that has so far been suppressed. 
 With this identification, the space-time fermion number $(-1)^{\bF}$ can be defined as an operator acting trivially on the vacuum, commuting with $\xi_-,\xi_+$ and anti-commuting with the other $\xi_a$.
The fact that we have two $\mathbb{Z}_2$ operators to employ in defining our index, $(-1)^J$ and  $(-1)^\bF$, is special to supersymmetric indices in $0+1$ dimensions. We prove in \cite{PPV1} that the resulting module is indeed a suitable highest weight representation of $\mathfrak{g}$.

A similar construction holds for the more general, higher-rank BKMs we discuss in this paper. In the construction of the second quantized Fock space in section \ref{sec:secondHilb}, we consider oscillators $\eta_a$ at zero momentum only when they correspond to the zero momentum limit of `propagating' massless degrees of freedom, while we exclude the modes that exist only at zero momentum, i.e. the two gravitinos. The space spanned by the gravitino zero modes correspond to a subspace of signature $(1,1)$ in the Cartan subalgebra of $\g$. In particular,  gravitinos can be associated  to a pair of fermionic modes $\xi_+,\xi_-$ obeying $\{\xi_\pm,\xi_{\pm}\}=0$, $\{\xi_+,\xi_-\}=1$. The other elements of the Cartan subalgebra now correspond to gaugino zero modes, which we absorb in the definition of the index. Once again, the second quantized Fock space can be identified with the positive chirality subspace $A^+_\g$ of the $\xi_a$-Clifford module $A_\g$, where the $\eta_a$ operators correspond to the generators $\xi_-\xi_a$ of the even Clifford algebra.  For a BKM of even rank $r$, the Clifford module $A_\g$ has ground state degeneracy $2^{r/2}$, where each chirality $A^\pm_\g$ of the orthogonal algebra splits as $A^\pm_\g\cong (\bigwedge \g^-)^{\oplus 2^{r/2-1}} $ into $2^{r/2 - 1}$ representations of $\mathfrak{g}$, each isomorphic (as a graded vector space) to $\bigwedge \g^-$.

The separation between the gauginos and gravitinos zero modes looks unnatural from the algebra point of view, since the adjoint action of $\mathfrak{g}$ can rotate the $\xi_+,\xi_-$ operators into any other $\xi_a$, $a\in \mathfrak{h}$. In fact, the distinction between propagating and non-propagating zero modes makes sense only in the decompactification limit, where the radius $R$ of the circle $S^1$ goes to infinity. When the rank of the algebra is greater than $2$, there are different decompactification limits, which correspond to different choices of a null direction of the Cartan subalgebra.

We therefore have the option of defining the full BPS Fock space using the full module  $A_\g$ of the Clifford algebra, including both chiralities of the spinorial rep of $so(\mathfrak{g})$. This necessitates absorbing the gravitino modes to get a nonzero index. Alternatively, we may project onto a single chirality $A^+_\g$ of the spinorial rep, as we did with the Monster. As a graded vector space, the resulting chiral representation is isomorphic to the second quantized Fock space obtained from ignoring the gravitino zero modes. This is simpler and seems natural from the physical viewpoint, in particular if one sees the superstring model as a $(1+1)$-dimensional theory on the cylinder.  The downside of privileging the gravitinos relative to the gauginos in this way is that it requires choosing a null direction in the Cartan subalgebra, viewed as a space of rank $(1, r-1)$. The resulting BKM algebra action on the Fock space module is rather unnatural.

\subsection{Heterotic strings on $V\otimes \bar V^{f\natural}$}\label{sec:hetex1}

Let us now apply the general formalism described in  \S\ref{sec:secondHilb} to some specific superstring models.
As a first example, let us consider the heterotic string on $V_L\otimes \bar V^{f\natural}_+$, where $V_L$ is a lattice VOA for one of the $24$ Niemeier lattices $L$ and contains $N$ currents. The massless content of the theory is: \footnote{Notation: each massless string state is the tensor product of a chiral and an anti-chiral physical state, which in turn are in one-to-one correspondence with, respectively, a left-moving weight $1$ state and a right-moving weight $1/2$ state is the internal CFT: $\vsf^a$, $a=1,\ldots,N$ corresponds to a current in $V$; $\bar \usf_+^i$, $i=1,\ldots, 24$ correspond to the $24$ Ramond ground states in $\bar V^{f\natural}$; $\alpha_{++}, \alpha_{--}, \bar \psi_{++},\bar\psi_{--}$ correspond to the light-cone components of $\partial X^\mu$ and $\bar\psi^\mu$ from the sigma model with target $\RR^{1,1}$. The $+$ and $-$ subscripts keep track of the action of the $SO(1,1)$ spacetime Lorentz group: each $+$ or $-$ is $\pm 1/2$ in the spin; spacetime parity exchanges $+$ and $-$.}
\begin{itemize}
	\item Graviton $\alpha_{++}\bar \psi_{++}$ and $\alpha_{--}\bar\psi_{--}$; B-field $\alpha_{++}\bar \psi_{--}-\alpha_{--}\bar \psi_{++}$; dilaton $\alpha_{++}\bar \psi_{--}+\alpha_{--}\bar \psi_{++}$. These states are non-physical, except for $k=0$.
	\item $24$ chiral gravitinos $\alpha_{++}\bar \usf^i_+$ (spin $+3/2$); $24$ anti-chiral dilatinos $\alpha_{--}\bar \usf^i_+$ (spin $-1/2$). These states are also only physical for $k=0$.
	\item $N$ gauge vectors $A^a=\vsf^a\bar \psi_{++}$, $\vsf^a\bar \psi_{--}$. These states are only physical for $k=0$.
	\item $24N$ chiral gauginos $\lambda^{ai}=\vsf^a \bar \usf^i_+$ (spin $1/2$); these states are physical for all $k$ with $k^2=0$ (they are the only massless propagating degrees of freedom).
\end{itemize} The $N$ gauge vectors $A^a$ correspond to a semi-simple Lie algebra $g$ of dimension $N$; the propagating gauginos $\lambda^{ai}$ form $24$ copies of the adjoint representation for $g$. 

The zero momentum modes of the gauginos form a Clifford algebra, and the ground states of our second quantized Hilbert space must form a module over this Clifford algebra. The full Clifford algebra is a direct sum of $24$ copies of the Clifford algebra generated by a single set (i.e. a fixed $i$, say $i=1$) of $N$ gauginos $\lambda^{ai}$, $a=1,\ldots,N$. The Clifford module of ground states, therefore, is the tensor product of $24$ copies of the Clifford module for a single set of gauginos. Thus, we can focus on one of these sets and simply take the tensor product at the end.

We choose a Cartan subalgebra $h\subset g$ and take a triangular decomposition $g=g^-\oplus h\oplus g^+$. We have an associated decomposition of the gauginos zero modes into a Cartan subspace plus the positive and negative roots subspaces. We turn on non-trivial Wilson lines $y_1,\ldots,y_r$, where $r=\rank g$ (for the theories we are considering we always have $r=24$) for the gauge vectors $A_a$ in the Cartan subalgebra, under which the gauginos in $g^-\oplus g^+$ are charged.
To get a non-vanishing supersymmetric index, one needs to absorb the zero modes of the neutral $\lambda^{ai}$, i.e. the ones in the Cartan subalgebra. The ground states of the Hilbert space live in a tensor product of a Clifford module  for the `Cartan' gauginos times a Clifford module for the $g^+\oplus g^-$ gauginos. After absorbing the zero modes, the trace over the `Cartan' Clifford module just gives a non-zero constant, which, with the appropriate normalization, can be taken to be $1$. Thus, we can just focus on the factor corresponding to the Clifford module for $g^+\oplus g^-$.  The latter module can be constructed by starting from a state (a highest weight vector) annihilated by all gauginos in $g^-$  and then acting in all possible ways with the gauginos in $g^+$. The ground state contribution to the supersymmetric index is a trace $\Tr((-1)^\bF \prod y_i^{q_i})$ over this module, where $q_i$ are the charges with respect to the background gauge fields in the Cartan torus.  The final result is the Weyl denominator of the algebra $g$
\be\label{Weylden} y^{-\rho_g}\prod_{\alpha\in \Delta_g^+}(1-y^\alpha)\ ,
\ee where $\Delta^+_g$ are the positive roots of the Lie algebra $g$ and $\rho_g$ is the charge of the lowest weight vector, which equals the Weyl vector of the Lie algebra $g$. (We implicitly use the notation that $y^\alpha$ for any vector $\alpha$ is shorthand for $\prod_k y_k^{\alpha_k}$).
 Note that there are $24$ copies of this set of gauginos, and hence one needs to take the $24$th power of this Weyl denominator in computing the index.\footnote{The ground states of the second quantized Hilbert space must form a representation of the gauge algebra $g$. It is clear what kind of representation this is: one considers the group $G$, the adjoint group of $g$, as a subgroup of the orthogonal group $SO(g)$ acting on the vector space of the algebra $g$, and preserving the Cartan Killing form. The gauginos form a vector representation for $SO(g)$. The ground states form a module for the Clifford algebra associated with the vector space underlying $g$, so they form a spinorial representation of $Spin(g)$, the universal cover of $SO(g)$. The group $G$ being a subgroup of $SO(g)$ lifts to a subgroup of $Spin(g)$, and this determines the representation of $G$ on the ground states.}

 The full supersymmetric index is given by multiplying this trace over the ground states by the contributions of the positive energy oscillators. For $k\neq 0$, the BRST quantization of the string is equivalent to the light-cone quantization, so that the classes in $\Hh^2(k_l,k_r)$ are in one-to-one correspondence with states in the internal CFT $V_L\otimes \bar V^{f\natural}_+$ satisfying the physical state conditions \be L_0-1=-(k_l)^2/2\ ,\qquad \qquad  \bar L_0-\frac{1}{2}=-(k_r)^2/2\ .\ee The BPS condition forces $(k_r)^2=0$, and since $V^{f\natural}_+$ contains no NS states with weight $1/2$, all BPS states come from the Ramond sector and are therefore spacetime fermions. This means that the corresponding operators $\eta_a$ obey a free fermion oscillator algebra. 
 
 In order to count the number of such oscillators, let us consider the `flavoured' partition function for the lattice VOA $V_L$
 $$ Z_{V_L}(\tau,\xi):=\Tr_{V_L}(\qsf^{L_0-\frac{c}{24}} e^{2\pi i \sum_k \xi_kq_k})\ .
 $$ Here, $\xi\equiv(\xi_1,\xi_2,\ldots)$ are  chemical potentials, and $q_k$ are the weights with respect to the Lie algebra generated by the zero modes of the $N$ currents in $V_L$; as the notation suggests, these are also the charges of the corresponding physical string states with respect to the background gauge fields $A=(A_1,A_2,\ldots)$. For a lattice VOA $V_L$ based on the Niemeier lattice $L$, this is given by 
 \be Z_{V_L}(\tau,\xi)=\frac{\Theta_L(\tau,\xi)}{\eta(\tau)^{24}}=\sum_{n\in \ZZ,\ell\in \ZZ^{24}} c(n,\ell)\qsf^ne^{2\pi i\sum_k \xi_k\ell_k}\ ,
 \ee where $\Theta_L$ is the `flavoured' lattice theta series
 \be \Theta_L(\tau,\xi)=\sum_{\ell\in \ZZ^{24}} \qsf^{\frac{1}{2}\sum_{j,k}\ell_j\ell_k\mu_j\cdot \mu_k} e^{2\pi i\sum_k \xi_k\ell_k}\ ,
 \ee where $\mu_1,\ldots,\mu_{24}$ is a basis of $L$. 
 
 Let us first express the ground state factor \eqref{Weylden} in terms of the Fourier coefficients $c(n,\ell)$ of the partition function $Z_{V_L}(\tau,\xi)$.  The number of currents in $V$ is
 $$ N=\sum_{\ell\in \ZZ^{24}} c(0,\ell)\ ,
 $$ and the rank of the algebra is $c(0,0)$, which is always equal to $24$ for $L$ a Niemeier lattice. The set of roots $\{\lambda\in L,\ \lambda^2=2\}$ correspond to the roots of the algebra $g$, and can accordingly be written as the disjoint union of the sets of positive or negative roots. For $\ell\in \ZZ^{24}$ such that $\lambda=\sum_i\ell_i\mu_i$ is a root, we write $\ell>0$ or $\ell<0$ depending on whether $\lambda$ is positive or negative.  The ground state contribution to the index \eqref{Weylden} can thus be written as,
 \be y^{-24\rho}\prod_{\ell>0}(1-\prod_{i=1}^{24} y_i^{\ell_i})^{24c(0,\ell)},
 \ee where for each $\ell\neq 0$, $c(0,\ell)$ is either $1$ (if the vector $\sum_i\ell_i\mu_i$ has length $2$) or $0$ (otherwise).
 
 For the contribution of positive energy string oscillators to the second-quantized index, we note that, for BPS states of winding and momentum $(w,m)$, one has  $-(k_l)^2/2=mw$. Therefore, the number of BPS single string states  of winding-momentum $(w,m)$ and charge vector $\ell\equiv (\ell_1,\ldots,\ell_{24})$ is $24c(mw,\ell)$ (the factor $24$ comes from the fact that there are $24$ Ramond ground states in $V^{f\natural}$). The positive energy condition $E=\frac{1}{\sqrt{2}}(\frac{m}{R}+wR)>0$ in the large radius regime $R>1$ is satisfied for $w> 0, m\in \ZZ$ (the only possible state with $m<0$ is $m=-1,w=1$) or for $w=0, m>0$. Therefore, the contribution from positive energy oscillators is
 \be \prod_{m>0}\prod_{\ell\in\ZZ^{24}}(1-q^m\prod_{i=1}^{24} y_i^{\ell_i})^{24c(0,\ell)}\prod_{w>0,m\in \ZZ}\prod_{\ell\in\ZZ^{24}}(1-p^wq^m\prod_{i=1}^{24} y_i^{\ell_i})^{24c(mw,\ell)}\ .
 \ee  Altogether, we find the index can be written as,
 \be\label{hetVLind} \mathcal{Z}^{V_L \otimes \bar{V}^{f\natural}_+}(T, U, A_k)= p^{N-24}q^{N}y^{-24\rho}\prod_{\substack{m,w\in\ZZ,\ell\in \ZZ^{24}\\(m,w,\ell)>0}}(1-p^wq^m\prod_{i=1}^{24} y_i^{\ell_i})^{24c(mw,\ell)},
 \ee where $(m,w,\ell)>0$ means $w\ge 0$, with $m\ge 0$ if $w=0$, and with $\ell>0$ if $m=w=0$. The winding-momentum $(w_0, m_0)$ of the ground states is computed in appendix \ref{s:Bfield} and is given by $w_0=N-24$ and $m_0=N$. This index is equal to the $24$th power  of the denominator for the BKM algebra constructed from any Niemeier lattice VOA, which is known as the fake Monster Lie algebra \cite{BorcherdsFake,BorcherdsMM}. 
 
 This algebra denominator turns out to be equal to the automorphic form $\Phi_{12}$ \cite{Gritsenko:2012qn}, whose appearance in string theory has been explored earlier \cite{phi12}. $\Phi_{12}$ enjoys 24 expansions associated to 24 ``Niemeier cusps" in moduli space corresponding to the 24 complementary constructions of the BKM. 
 The fact that there is a unique BKM algebra associated with all the $24$ VOAs $V_L$ associated with a Niemeier lattice $L$ is not a coincidence. Indeed, the (non-chiral) CFTs obtained as the product of any of the $V_L$ with the CFT of a single free boson on a circle $S^1$ are all equivalent to each other. These CFTs admit a uniform description as theories of $25$ chiral and $1$ anti-chiral free boson compactified on the unique even unimodular lattice $\Gamma^{25,1}$ of signature $(25,1)$. The Narain moduli space of this CFT is $O(25)\times O(1)\backslash O(25,1)/O(25,1,\ZZ)$, and is spanned by the radius $R$ and the Wilson lines $(A_1,\ldots, A_{24})$. The different descriptions of this theory as a product $S^1\times V_L$ correspond to the $24$ inequivalent ways of writing the lattice $\Gamma^{25,1}$ as an orthogonal sum of a Niemeier lattice and a unimodular lattice of signature $(1,1)$. These  descriptions are related to one each other by the action of the T-duality group $O(25,1,\ZZ)$, which mixes the radius and the Wilson line moduli, as well as the winding-momentum $(w,m)$ with the gauge charges $q_k$.
 
 In general, if we turn off some of the Wilson lines (i.e. set $y_i=1$ for some $i$), the index vanishes because of the zero modes of some combination of gauginos in $g^+\oplus g^-$. The only exception is when $V_L$ is the VOA associated with the Leech lattice $L=\Lambda$, where $g$ is the abelian algebra $u(1)^{\oplus 24}$ and $c(0,\ell)=0$ for all $\ell\neq 0$. In this case, we can turn off all Wilson lines, and by setting
 \be c(n)=\sum_\ell c(n,\ell),
 \ee such that $c(n)$ are the Fourier coefficients of
 \be Z_{V_\Lambda}(\tau,0)=\frac{\Theta_\Lambda(\tau,0)}{\eta(\tau)^{24}}=J(\tau)+24=\sum_{n\in \ZZ} c(n)q^n=q^{-1}+24+196884q+\ldots\ ,
 \ee
  the index reduces to 
 \be \mathcal{Z}^{V_{\Lambda} \otimes \bar{V}^{f\natural}_+}(T, U, A_k=0)= \left (q\prod_{m>0}(1-q^m)^{c(0)}\prod_{w>0,m\in \ZZ}(1-p^wq^m)^{c(mw)}\right )^{24}.
 \ee Recall the Koike-Norton-Zagier identity
 \be p^{-1} \prod_{w>0,m\in \ZZ\setminus\{0\}}(1-p^wq^m)^{c(mw)}=J(\sigma)-J(\tau)
 \ee which can be interpreted as the denominator identity for the Monster BKM algebra \cite{BorcherdsMM}. Multiplying this factor by the $w=0$ and the $m=0$ parts yields
 \bea \nn \mathcal{Z}^{V_{\Lambda} \otimes \bar{V}^{f\natural}_+}(T, U, 0)&=&\left (q\prod_{m>0}(1-q^m)^{c(0)}\right ) \left (p\prod_{w>0}(1-p^w)^{c(0)}\right )\\  \nonumber
 & & \times \, p^{-1} \prod_{w>0,m\in \ZZ\setminus\{0\}}(1-p^wq^m)^{c(mw)}\\ \nonumber
 &=&\eta(\tau)^{24}\eta(\sigma)^{24}(J(\sigma)-J(\tau))\\ 
 &=&\eta(\tau)^{24}\Theta_\Lambda(\sigma)-\eta(\sigma)^{24}\Theta_\Lambda(\tau)\ ,
 \eea which is the denominator of the fake Monster Lie algebra \cite{BorcherdsFake}. 
 
 This construction should generalize straightforwardly to all $71$ self-dual VOAs $V$ with $c=24$. When we compactify one further direction on a circle $S^1$, we expect many non-trivial equivalences between these different models, as is the case for the $24$ Niemeier lattice VOAs $V_L$.  It is known \cite{Hohn2017,Chigira2021} that there are only  $11$ isomorphism classes of holomorphic vertex algebras of central charge $26$ of the form $V_{\Gamma^{1,1}}\otimes V$, where $V_{\Gamma^{1,1}}$ is the lattice vertex algebra based on the even unimodular lattice of signature $(1,1)$ and $V$ is a self-dual VOA of central charge $24$ with currents. In particular, the $24$ VOAs $V_{\Gamma^{1,1}}\otimes V_L$  are all isomorphic to each other; the proof is similar to the one showing that the (non-chiral) CFTs on $S^1\otimes V_L$ are all isomorphic. Conjecturally, there is only one additional isomorphism class $V_{\Gamma^{1,1}}\otimes V$ where $V$ is a self-dual VOA of $c=24$ \emph{without} currents -- it is the case where $V=V^\natural$ is the Monster VOA \cite{FLM}. From each of these isomorphism classes of vertex algebras, one can define a BKM algebra using Borcherds' construction based on chiral bosonic string theory compactified on $V_{\Gamma^{1,1}}\otimes V$; see  \cite{BorcherdsFake,BorcherdsMM, HS1,HS2,Creutzig2007,Moller2021} for explicit descriptions and denominator identities.\footnote{These references describe $10$ out of $12$ distinct BKM algebras; N. Scheithauer informed us that he calculated the denominator identities for the two remaining cases (unpublished).}  It is natural to conjecture that by tensoring any self-dual VOA of $c=24$ with the bosonic (non-chiral) CFT on a circle $S^1$, one gets one of precisely $12$ distinct CFTs of left and right central charges $(25,1)$. If this is correct, then there are only $12$ inequivalent compactifications of the heterotic string of the form $(V\otimes \bar V^{f\natural})\otimes S^1$. Our construction shows that the corresponding second quantized indices $\mathcal{Z}$ are just the (24th powers of the) denominators of the $12$ associated BKM algebras. 
 
\subsection{Type IIA strings on $V\otimes \bar V^{f\natural}$}\label{sec:IIAex2}
\label{IIAVfnatural}

Now we consider the type IIA string compactified on $V\otimes  \bar V^{f\natural}_-$, where $V$ is one of the self-dual $\CN=1$ SVOAs with $c=12$.\footnote{Note that in these type IIA examples we take $V^{f\natural}_-$ as the right-moving internal SVOA, while for the heterotic examples we took $V^{f\natural}_+$. The reason is that in heterotic string theory the GSO projection restricts to the right-moving R$_+$ sector, while in the type IIA string the projection is onto the right-moving R$_-$ sector. Our choice then corresponds to having $24$ gravitinos with positive chirality in all examples.}  We assume that $V$ contains $\chi^{\rm NS}$ states  of weight $1/2$ in the NS sector, $\chi^{\rm R}_+$ states  of weight $1/2$ in the R$_+$ sector and $\chi^{\rm R}_-$ states  in the R$_-$ sector. These VOA states correspond to massless physical states in the chiral cohomology, denoted, respectively, by $\vsf^a$, $a=0,\ldots,\chi^{\rm NS}$, $\usf^i_+$, $i=1,\ldots,\chi^{\rm R}_+$, $\usf^i_-$, $i=1,\ldots,\chi^{\rm R}_-$. As discussed in \S \ref{sec:models}, we distinguish between the two cases $V=V^{f\natural}_\pm$ which have $24$ Ramond ground states in the R$_\pm$ sector. The anti-holomorphic $\bar V^{f\natural}_-$ contains $24$ Ramond ground states  in the R$_-$ sector corresponding to anti-chiral physical states $\bar \usf^i_+$, $i=1,\ldots,24$. The subscripts $+$ or $-$ on the Ramond states $\usf^i_\pm$ and $\bar \usf^i_+$  refer to the spin with respect to the $SO(1,1)$ spacetime Lorentz group; note that the type IIA GSO projection identifies the sign of the spacetime spin  with the internal fermion number in the holomorphic Ramond sector, and with the opposite of the fermion number in the anti-holomorphic Ramond sector.   Additionally, there are the left and right-moving states arising from the $\CN=(1,1)$ supersymmetric non-linear sigma model with target $\RR^{1,1}$. The massless single string states are:
\begin{itemize}
	\item Gravitons $\psi_{++}\bar \psi_{++}$ and $\psi_{--}\bar\psi_{--}$; B-field $\psi_{++}\bar \psi_{--}-\psi_{--}\bar \psi_{++}$; dilaton $\psi_{++}\bar \psi_{--}+\psi_{--}\bar \psi_{++}$. These states are non-physical, except for $k=0$.
	\item $(24+\chi^{\rm R}_+)$ chiral gravitinos $\psi_{++}\bar \usf^i_+$, $i=1,\ldots,24$, and $\usf^i_+\bar\psi_{++}$ $i=1,\ldots \chi^{\rm R}_+$ (spin $+3/2$); $\chi^{\rm R}_-$ antichiral gravitinos $\usf^i_-\bar\psi_{--}$ $i=1,\ldots \chi^{\rm R}_-$ (spin $-3/2$); $(24+\chi^{\rm R}_+)$ anti-chiral dilatinos $\psi_{--}\bar \usf^i_+$, $i=1,\ldots,24$ and  $\usf^i_+\bar\psi_{--}$ $i=1,\ldots \chi^{\rm R}_+$ (spin $-1/2$); $\chi^{\rm R}_-$ chiral dilatinos $\usf^i_-\bar\psi_{++}$ $i=1,\ldots \chi^{\rm R}_-$ (spin $+1/2$). These states are only physical for $k=0$.
	\item $\chi^{\rm NS}$ gauge vectors $A^a=\vsf^a\bar \psi_{++}$, $\vsf^a\bar \psi_{--}$ from the NS-NS sector. These states are only physical for $k=0$.
	\item  From the R-R sector, $24\chi^{\rm R}_+$ states $\usf^i_+\bar \usf^j_+$, $i=1,\ldots, \chi^{\rm R}_+$, $j=1,\ldots, 24$ corresponding to derivatives $\partial_+ \phi^{ij}$ of scalar chiral fields $\phi^{ij}$ with $\partial_-\phi^{ij}=0$, and $24\chi^{\rm R}_-$ states $\usf^i_-\bar \usf^j_+$, $i=1,\ldots, \chi^{\rm R}_-$, $j=1,\ldots, 24$, corresponding to field strengths of $U(1)$ vectors. When the theory is uncompactified, only the scalar fields have propagating degrees of freedom. When we compactify the space direction on a circle $S^1$, the equations of motion force the scalars to carry no winding along $S^1$, while the vectors can carry winding but not momentum. There are no physical string states representing the zero modes of the scalars or of the vectors -- this is standard for R-R bosons. See appendix \ref{app:0mom} for more detail. 
	\item $24\chi^{\rm NS}$ chiral gauginos $\lambda^{ai}=\vsf^a \bar \usf^i_+$ (spin $1/2$) from the NS-R sector; these states are physical for all $k$ with $k^2=0$ (massless propagating degrees of freedom).
\end{itemize}
In total the spacetime theory has $(\chi^R_-,24+\chi^{\rm R}_+)$ SUSY. When we compactify the space direction on a circle $S^1$, the corresponding supercharges obey slightly different algebras then presented previously:
\begin{itemize}
	\item The `standard' $(0,24)$ supercharges $\CQ^i_{r-}$ from the NS-R gravitinos $\psi_{++}\bar \usf^i_+$ obey the algebra
	\be \{\CQ^i_{r-},\CQ^j_{r-}\}=\delta^{ij}(P^0_r-P^1_r)\ .
	\ee
	\item The  $(0,\chi^{\rm R}_+)$ supercharges $\CQ^i_{l-}$ from the R-NS gravitinos $\usf^i_+\bar\psi_{++}$ obey the algebra
	\be \{\CQ^i_{l-},\CQ^j_{l-}\}=\delta^{ij}(P^0_l-P^1_l)\ .
	\ee Note that here only left-moving momenta appear, while in the previous equation we had only right-moving momenta. In the uncompactified theory, one has $P^0_l=P^0_r$ and $P^1_l=P^1_r$ and the algebra is the same; in the compactified theory $P^1_l$ and $P^1_r$ are independent operators.
	\item The  $(\chi^{\rm R}_-,0)$ supercharges $\CQ^i_{l+}$ from the R-NS gravitinos $\usf^i_-\bar\psi_{--}$ obey the algebra
	\be \{\CQ^i_{l+},\CQ^j_{l+}\}=\delta^{ij}(P^0_l+P^1_l)\ .
	\ee 	
\end{itemize}

Now we are going to discuss the computation of the supersymmetric indices defined in equations \eqref{eq:index1}, \eqref{eq:index2} for these theories.	
Since the time direction is always uncompactified, we have $P^0_l=P^0_r$. The spacetime index $\mathcal{Z}$ gets contributions only from states that are annihilated by all the supercharges. Since the $24$ supercharges $\CQ^i_{r-}$ are always present, this implies that  $k^1_r=k^0_r=k^0_l$. If $\chi^{\rm R}_+\neq 0$, then the states contributing to $\mathcal{Z}$ have momenta satisfying
	\be\label{kayplus}  k^1_r=k^0_r=k^0_l=k^1_l\ ,
	\ee which means that they cannot have any winding around the circle $S^1$, so that the index is independent of $T$. If $\chi^{\rm R}_-\neq 0$, we get the condition
	\be\label{kayminus} k^1_r=k^0_r=k^0_l=-k^1_l
	\ee which implies that states contributing to the index cannot carry any units of momentum along $S^1$, and $\mathcal{Z}$ is independent of $U$.
	If both $\chi^{\rm R}_+$ and $\chi^{\rm R}_-$ are not zero (this happens if $V=V^{fE_8}$) then the only states contributing are the ones with zero momentum
	\be\label{kayboth} k^0_l=k^0_r=k^1_l=k^1_r=0\ ,
	\ee  so the index is just  a number.
	
	As we will show below, the supersymmetric index $\mathcal{Z}^{V\otimes \bar V^{f\natural}_-}$ (\ref{eq:index1}) and the modified index $\tilde{\mathcal{Z}}^{V\otimes \bar V^{f\natural}_-}$ (\ref{eq:index2}) reproduce, respectively, the superdenominator and the denominator of the BKM superalgebra corresponding to $V$.  Passing from $\mathcal{Z}$ to $\tilde{\mathcal{Z}}$ amounts to inserting a `left-moving' spacetime fermion number $(-1)^{\bF_l}$ in the trace, which acts by $-1$ on the left-moving Ramond and by $+1$ on the left-moving NS sector. With this insertion, the states contributing to $\tilde{\mathcal{Z}}$ are the ones that are annihilated by the $24$ supercharges $\CQ^i_{r-}$ from the NS-R gravitinos $\psi_{++}\bar \usf^i_+$. Thus, the only condition is
	\be k^0_l=k^0_r=k^1_r\ ,
	\ee with no constraints on $k^1_l$. Indeed, all the other supercharges  do not commute with the operator $(-1)^{\bF_l}$, so there are no cancellations between states that are related by these supersymmetries. Then $\tilde{\mathcal{Z}}$ will in general be a function of both $T$ and $U$ for all algebras.  This is indeed what happens for the algebra denominators.

The only propagating massless fermions are the $24\chi^{\rm NS}$ chiral gauginos $\lambda^{ai}$, that are the superpartners of the gauge vectors $A^a$. The gauge algebra $g^{\oplus 24}$ consists of $24$ copies of a $\chi^{\rm NS}$-dimensional Lie algebra $g$. Therefore, when $\chi^{\rm NS}\neq 0$, we can consider Wilson lines corresponding to the Cartan generators of $g$. Furthermore, in order to get a non-vanishing result for the index, one needs to absorb the zero modes of the gauginos corresponding to Cartan generators, which are neutral under the Wilson lines. This situation occurs when $V=V^{fE_8}$ ($\chi^{\rm NS}=8$) and for $V=F_{24}$ ($\chi^{\rm NS}=24$).

Let us now consider a case by case analysis of the three possibilities for $V$.
\begin{enumerate}
	\item For $V=V^{f\natural}_+$ (or $V=V^{f\natural}_-$) one has $\chi^{\rm R}_+\neq 0$ and $\chi^{\rm NS}=\chi^{\rm R}_-=0$ (resp., $\chi^{\rm R}_-\neq 0$ and $\chi^{\rm NS}=\chi^{\rm R}_+=0$), so the index $\mathcal{Z}$ receives contributions only from states with $w=0$ (resp., $m=0$) and therefore is a function of $U$ only (resp., of $T$ only). This fits with the fact that the superdenominator of the Conway superalgebra associated with $V^{f\natural}$ depends on one variable only (see equation 4.30 and the surrounding discussion in \cite{Harrison:2018joy}).  The two cases $V=V^{f\natural}_+$ and $V=V^{f\natural}_-$ are related by exchanging winding and momentum along $S^1$, i.e. by T-duality along this circle; this fits with the idea that T-duality is a `left-moving parity' exchanging the left-moving R$_+$ and R$_-$ sectors while keeping the right-moving sectors fixed. The conditions \eqref{kayplus} (or \eqref{kayminus}) on momenta imply that only propagating massless states contribute to the index. The only such states are the $24^2$ massless  bosons in the R-R sector; they are either scalars carrying only momentum for $V=V^{f\natural}_+$, or vectors carrying only winding number for $V=V^{f\natural}_-$.  The index is therefore 
	\be \mathcal{Z}^{V^{f\natural}_+ \otimes \bar{V}^{f\natural}_-}(U)=\Bigl(q^{-1}\prod_{m=1}^\infty\frac{1}{(1-q^m)^{24}}\Bigr)^{24}=\left(\frac{1}{\eta^{24}(U)}\right)^{24},
	\label{ConwaydenomVplus}
	\ee
	\be \mathcal{Z}^{V^{f\natural}_- \otimes \bar{V}^{f\natural}_-}(T)=\Bigl(p^{-1}\prod_{w=1}^\infty\frac{1}{(1-p^w)^{24}}\Bigr)^{24}=\left(\frac{1}{\eta^{24}(T)}\right)^{24},
	\label{ConwaydenomVminus}
	\ee 
	that are indeed the $24$-th power of the superdenominator of Conway BKM superalgebra \cite{Harrison:2018joy}.

	The modified index $\tilde{\mathcal{Z}}$ receives contributions from both massive and massless states with $k^0_r=k^1_r$. Let $c_{NS}(n)$ and $c_R(n)$ be the multiplicities of the GSO projected NS and R states in the internal SVOA $V$ at level $L_0-\frac{1}{2}=n$ -- these multiplicities are the same for both $V^{f\natural}_{-}$ and  $V^{f\natural}_{+}$. For winding and momenta $m$, $w$, we have $24c_{NS}(mw)$ fermions from the NS-R sector and $24c_{R}(mw)$ bosons from the R-R sector. The multiplicities are non-zero only for $mw\ge 0$, and the positive energy condition $E=\frac{1}{\sqrt{2}}(\frac{m}{R}+wR)>0$ implies that $m,w\ge 0$. As discussed above, the $24c_R(0)=24^2$ massless bosons must have $w=0$ for $V=V^{f\natural}_+$ or $m=0$ for $V=V^{f\natural}_-$. The product $(-1)^\bF(-1)^{\bF_l}=(-1)^{\bF_r}$ of spacetime fermion number $(-1)^\bF$ and the `left-moving' fermion number $(-1)^{\bF_l}$ acts by a minus sign on all of these states. The final result for the modified index is therefore
	\be \tilde{\mathcal{Z}}^{V^{f\natural}_+ \otimes \bar{V}^{f\natural}_-}(T, U)=\Bigl(q^{-1}\prod_{m=1}^\infty\frac{1}{(1+q^m)^{24}}\prod_{w=1}^\infty \frac{(1-p^wq^m)^{c_{NS}(mw)}}{(1+p^wq^m)^{c_{R}(mw)}}\Bigr)^{24},
	\ee
	\be\label{modindVmVm} \tilde{\mathcal{Z}}^{V^{f\natural}_- \otimes \bar{V}^{f\natural}_-}(T, U)=\Bigl(p^{-1}\prod_{w=1}^\infty\frac{1}{(1+p^w)^{24}}\prod_{m=1}^\infty \frac{(1-p^wq^m)^{c_{NS}(mw)}}{(1+p^wq^m)^{c_{R}(mw)}}\Bigr)^{24}.
	\ee 
	These are indeed the $24$-th power of the superalgebra denominator (as computed in \cite{Harrison:2018joy}).
	\item When $V=F_{24}$, we have $\chi^{\rm NS}=24$ and $\chi^{\rm R}_+=\chi^{\rm R}_-=0$. Thus, for both the index and the modified index, the states giving non-zero contribution to the trace may carry any value of winding and momentum. To be precise, there are $8$ different string models, depending on the choice of a $\CN=1$ supercurrent for $F_{24}$ (as described in \S \ref{sec:models}.) Each choice is associated with a semi-simple Lie algebra $g$ of dimension $24$; $g$ is also the algebra of gauge symmetries in spacetime for the corresponding model. One can turn on $r$ Wilson lines for these gauge fields where $r$ is the rank of the algebra. We are left with $r$ neutral massless gauginos that need to be absorbed in order to get a non-zero result. The treatment of the massless modes is completely analogous to the case of heterotic strings with $N$ currents that we saw in the previous section. The only difference is that in this case we have massive BPS bosons, while in the heterotic case all BPS states were fermions. Let us denote by $c_{NS}(n,\lambda)$ and $c_R(n,\lambda)$ the GSO projected NS and Ramond multiplicities of states in the internal CFT $F_{24}$ with level $L_0-\frac{1}{2}=n$ and charge $\lambda$ with respect to the algebra $g$. The result for the index is then
	\begin{eqnarray}\label{F24ind1} \mathcal{Z}^{F_{24} \otimes \bar{V}^{f\natural}_-}&=&\Bigl( pqy^{-\rho} \prod_{\lambda \in \Delta^+_g} (1-y^{\lambda})^{c_{NS}(0,\lambda)} \prod_{\substack{m,w=0\\(m,w)\neq (0,0)}}^{\infty}\prod_{\lambda\in \Delta^g} \frac{(1-p^wq^my^{\lambda})^{c_{NS}(mw,\lambda)}}{(1-p^wq^my^{\lambda})^{c_{R}(mw,\lambda)}} \Bigr)^{24},
	\nonumber \\
	\end{eqnarray} and for the modified index
	\begin{eqnarray}\label{F24ind2} \tilde{\mathcal{Z}}^{F_{24} \otimes \bar{V}^{f\natural}_-}&=&\Bigl( pqy^{-\rho} \prod_{\lambda \in \Delta^+_g} (1-y^{\lambda})^{c_{NS}(0,\lambda)} \prod_{\substack{m,w=0\\(m,w)\neq (0,0)}}^{\infty}\prod_{\lambda\in \Delta^g} \frac{(1-p^wq^my^{\lambda})^{c_{NS}(mw,\lambda)}}{(1+p^wq^my^{\lambda})^{c_{R}(mw,\lambda)}} \Bigr)^{24}\ .
	\nonumber \\
	\end{eqnarray}  where, for any vector $\alpha$, $y^\alpha$  is shorthand for $\prod_k y_k^{\alpha_k}=e^{2\pi i \sum_kA_k\alpha_k}$. Again, up to the power $24$, these are, respectively, the superdenominator and denominator of the associated BKM superalgebra (see \cite{Harrison:2020wxl}).
	\item Finally, for $V=V^{fE_8}$, we have $\chi^{\rm NS}=\chi^{\rm R}_+=\chi^{\rm R}_-=8$. The $8\cdot 24$ vectors from the NS-NS sectors are the gauge bosons of an abelian $U(1)^{8\cdot 24}$ algebra. The $8\cdot 24$ massless gauginos from the NS-R sector are all neutral, so their zero modes force the indices to vanish independently of the Wilson lines, unless they are absorbed. For simplicity, we consider the case where the Wilson lines are set to zero. The theory contains supersymmetries of the form $\CQ^i_{l+}$ and $\CQ^i_{l-}$ from the R-NS sector and $\CQ^i_{r-}$ from the NS-R sector; as discussed above, this implies that the contributions to the index $\mathcal{Z}$ from states of non-zero momentum always cancel. So, the index $\mathcal{Z}^{V^{f E_8} \otimes \bar{V}^{f\natural}}$ is simply a non-zero constant that can be normalized to $1$.  The modified index is more interesting, and gives
	\be\label{VE8ind} \tilde{\mathcal{Z}}^{V^{f E_8} \otimes \bar{V}^{f\natural}}(T, U)=\Bigl(\prod_{\substack{m,w=0\\(m,w)\neq (0,0)}}^\infty \frac{(1-p^wq^m)^{c_{NS}(mw)}}{(1+p^wq^m)^{c_{R}(mw)}}\Bigr)^{24}\ ,\ee where $c_{NS}(n)=c_R(n)$ are the Fourier coefficients of the GSO projected NS and Ramond characters of $V^{fE_8}$ (that are actually equal to each other).
This is the denominator of the fake Monster superalgebra (to the power $24$) \cite{BorcherdsMM, Sch2, Sch1}. 
\end{enumerate}

\section{Path integrals and theta lifts}\label{sec:pathintegrals}

The index described in the previous section, and its variations, can be calculated using a path integral. In this section we  discuss this point of view, first reviewing the structure and modular properties of second quantized partition functions (\S \ref{sec:pfn}), and then discussing the relationship between the path integral and operatorial formalism in more detail (\S \ref{sec:pathvsop}). In \S \ref{sec:paththeta} we show that the path integral reduces to a so called theta lift in the theory of automorphic forms. We then demonstrate how to recover the (super)denominator identities using this approach (\S \ref{sec:pintsuper} and \S \ref{sec:pintdenom}).

\subsection{First and second quantized partition functions}\label{sec:pfn}
Let us first consider the relation between first and second quantized partition functions in particle physics/ quantum field theory. For a single point particle of mass $m$ in a $d$-dimensional spacetime, one defines a (first quantized) partition function by summing over worldlines  all possible ways to embed a single  (Euclidean) loop in spacetime
\begin{equation} Z_{particle}(m^2)=V_d \int \frac{d^dk}{(2\pi)^d}\int_0^{+\infty}\frac{dl}{2l}e^{-\frac{1}{2}(k^2+m^2)l}\ .
\end{equation}
Here, $l$ is the length of the loop, $\frac{1}{2}(k^2+m^2)$ is the worldline Hamiltonian, and the factor $2l$ takes into account the translation and reflection invariance of the wordline parametrization.  To obtain the (second quantized) partition function in a quantum field theory, one needs to sum over an arbitrary number $n$ of disconnected worldlines describing $n$ particles going around a  closed loop in spacetime, where on each wordline one sums over all possible particles in the theory. One should also divide by a factor $n!$ to take into account the permutations of the $n$ worldlines. The final result is
\begin{equation} Z_{QFT}=\exp\left(\sum_i Z_{particle}(m_i^2)(-1)^{\bF_i}\right)\ ,
\end{equation}
where the factor $(-1)^{\bF_i}$ takes into account that fermions running in the loop contribute with negative sign with respect to the bosons (see \cite{PolchinskiVol1}, \S7.3.).

Now let us turn to the analogue of this in string theory. In order to obtain the second quantized partition function in string theory one needs to exponentiate the (first quantized) torus partition function for a single string, with opposite signs for spacetime fermions and bosons, namely
\be \mathscr{Z}=\exp\left(\frac{1}{2}\int_{SL(2,\mathbb{Z})\backslash \mathbb{H}} \frac{d^2\tau}{\tau_2} \Tr\Big(\qsf^{L_0}\bar \qsf^{\bar L_0}e^{2\pi i \sum_kA_kq_k}(-1)^{\bF}\Big)\right)\ ,
\ee where $\tau=\tau_1+i\tau_2$ and $\qsf=e^{2\pi i\tau}$. The trace is over the GSO projected single string states and $(-1)^{\bF}$ is a space-time fermion number.  The factors of $1/2$ and $1/\tau_2$ in the exponential are needed in the torus amplitude to take into account, respectively, of the reflection and translation  automorphisms of the worldsheet torus \cite{PolchinskiVol1}. The $A_k$ are background gauge fields (if present), and $q_k$ are the corresponding charges. The Virasoro operators $L_0$ and $\bar L_0$ depend implicitly on the geometry of the spacetime and on the background B-field. The target space should be a Euclidean 2-dimensional spacetime on a torus $S^1\times S^1$, where the first $S^1$ corresponds, in Lorentzian signature, to the spacelike circle of radius $R$, while the second $S^1$ is a thermal circle, i.e. a Euclidean time circle with radius the inverse temperature $\beta$. This means that both $k^0$ and $k^1$ are quantized in the path integral. The standard quantization of momenta in the path integral will lead, on the operatorial side, to the index with fermions that are periodic along the spacelike circle $S^1$ and with $(-1)^\bF$ inserted. Non-standard quantizations of momenta in the path integral (i.e. with half-integral rather than integral momentum, either in the space or in the Euclidean time directions)  for \emph{some} of the sectors will lead, in the operatorial side, to either anti-periodic fermions around the space circle, or to modifications of the $(-1)^\bF$ factor in the trace.

The path integral representation of the index should make invariance under a suitable subgroup of $O(2,2,\ZZ)\sim SL(2,\ZZ)\times SL(2,\ZZ)$ manifest. The precise invariance subgroup depends on the periodicity conditions imposed on the fields around the spatial and Euclidean time circles. The two $SL(2,\ZZ)$ factors in $O(2,2,\ZZ)$ act by fractional linear transformations on the complex variables $T,U$ parametrizing, respectively, the complexified K\"ahler and the complex structure moduli of the spacetime torus. Thus, duality invariance implies that the index is a modular function in both variables $T$ and $U$. This invariance may  be modified in the presence of non-trivial Wilson lines, which give rise to (possibly multivariate) Jacobi forms rather than modular functions. 

The path integral might have infrared divergences in the presence of massless or tachyonic modes. These can be regularized in a way that preserves space-time diffeomorphism invariance and T-duality invariance, by cutting a suitable region in the integration over the fundamental domain $SL(2,\mathbb{Z})\backslash \mathbb{H} $. Therefore, one expects the path integral to be $O(2,2,\ZZ)$-invariant even in this case.

This is in contrast with the expected modular properties of the index $\mathcal{Z}$, as obtained from the operatorial formalism of section \ref{sec:secondquantized}. For example, suppose that our theory contains some fermionic zero modes that need to be absorbed as in eqs.\eqref{eq:index1} and \eqref{eq:index2}. In this case, one expects a non-zero modular weight depending on the number of fermions one needs to absorb. For example, consider a modular transformation $U\to \frac{aU+b}{cU+d}$ of the complex structure of the Euclidean torus. Spacetime fermions are sections of a spinorial bundle on the torus; in particular, a left-moving fermion is a section of $K^{1/2}$, a square root of the canonical bundle on the torus, and thus transforms as $dz^{1/2}$, where $z$ is a complex coordinate on the torus. Similarly, right moving fermions  transform as $d\bar z^{1/2}$. This means that if one needs to absorb $n_l$ left-moving and $n_r$ right moving fermions, the corresponding index is a modular form of weights $(n_l,n_r)$, rather than a modular invariant function.

In order to explain this mismatch, we will now discuss in more detail the relationship between the path integral $\mathscr{Z}$ and the corresponding index $\mathcal{Z}$.

\subsection{Path integral vs operatorial formalism}
\label{sec:pathvsop}

Some of the theories we are considering contain chiral massless fermions or bosons, whose path integral description is notoriously problematic. For a single chiral free boson or fermion, a standard method to overcome this issue consists in introducing additional massless degrees of freedom with the opposite chirality, so as to make the whole theory non-chiral. One can then easily compute the path integral $\mathscr{Z}$ for this non-chiral theory. Finally, the chiral partition function $\mathcal{Z}$ is obtained by extracting a suitable `holomorphic square root' of the non-chiral one $\mathscr{Z}$ -- this last step might be quite non-trivial, as discussed below.

\subsubsection*{Including spurious degrees of freedom}
Since we are interested in computing indices in free theories, a similar method can be generalized to our case. Actually, it is simpler to modify the indices computed in section \ref{sec:secondquantized}  directly in the operatorial formalism, by introducing suitable spurious degrees of freedom that make the theory non-chiral. The `non-chiral' indices we obtain this way can be compared with the path integral results we discuss in the following subsections.

Let us take a closer look at the chiral content of the superstring compactifications. Chiral massless fields can appear when the SVOA $V^{f\natural}$ is the holomorphic or anti-holomorphic factor in the internal CFT. All $24$ Ramond ground states of this SVOA have the same worldsheet fermion number, which can be taken to have either positive (for $V^{f\natural}_+$) or negative ($V^{f\natural}_-$). The GSO projection relates the fermion number of these ground states to the spacetime spin of the corresponding massless fields. Finally, the mass-shell condition, arising from the requirement that the state is BRST closed, relates the spin to the spacetime momentum. For example, for chiral fermions of spin $+1/2$ or $-1/2$, the massless Dirac equation admits a non-zero solution only for $k^0=-k^1$ or $k^0=k^1$, respectively.  As a result, we obtain massless fields that are either purely left-moving or purely right-moving in spacetime. This phenomenon does not occur for $V^{fE_8}$ or $F_{24}$, for which the Ramond sector contains the same number of states with positive or negative worldsheet fermion number.

In a path integral formulation, the mass-shell condition is not imposed on fields, so that the relationship between spacetime momentum and spin (and, in turn, with the internal fermion number) is lost. One simply sums over all possible spacetime momenta and, independently, over all possible internal states. Nevertheless, this procedure   gives the correct result if the number of internal states with positive and negative worldsheet fermion numbers is the same. Therefore it is  clear how  to modify the operatorial description so as to obtain a `non-chiral' index that can be compared with the path integral: for each (holomorphic or antiholomorphic) internal SVOA $V^{f\natural}$, containing $24$ Ramond ground states having fermion number $+ 1$ or $-1$, one needs to add $24$ `spurious' ground states with the opposite fermion number.

\subsubsection*{Het/Type IIA on $V\otimes \bar V^{f\natural}_-$}
Let us analyze how the index is modified by these spurious degrees of freedom. Let us first suppose that the theory is heterotic string theory on $V\otimes \bar V^{f\natural}_+$ or type IIA on $V\otimes \bar V^{f\natural}_-$, where $V$ is a VOA or SVOA different from $V^{f\natural}$. As described in section \ref{sec:secondquantized}, in this case, the only single string physical states relevant for the supersymmetric index are the ones with $k^0_r=k^1_r$, and correspond to products $v\otimes \bar \usf^i_+$, where $v$ is a state in $V$ and $\bar \usf^1_+,\ldots,\bar \usf^{24}_+$ are the antiholomorphic Ramond ground states in $\bar V^{f\natural}_+$ (heterotic) or $\bar V^{f\natural}_-$ (type IIA)  (the subscript $+$ in $\bar \usf^i_+$ denotes positive spacetime chirality, and not the worldsheet fermion number; notice that, according to our conventions, the GSO projection on the antiholomorphic internal factor for type IIA is the opposite than for heterotic). For each of these states, there is a creation or annihilation (depending on the sign of the energy) operator acting on the second quantized Hilbert space. Adding  $24$ further antiholomorphic Ramond ground states $\bar \usf^1_-,\ldots,\bar \usf^{24}_-$ with opposite worldsheet fermion number to $\bar V^{f\natural}$ corresponds to introducing a second copy of each creation/annihilation operator, but now with right-moving momenta satisfying $k^0_r=-k^1_r$ rather than $k^0_r=k^1_r$. This means that, for these new operators,  the relation between energy and winding-momentum changes sign and becomes $E=-\frac{1}{\sqrt{2}}(\frac{m}{R}+wR)$, so that the second copy of a creation (i.e. positive energy) operator  has the same energy and the opposite winding and momentum with respect to the first copy.  Changing the sign of $m$ and $w$ while keeping the energy fixed is equivalent to the transformation $(T,U)\mapsto (-\bar T,-\bar U)$. In turn, this is equivalent to applying complex conjugation $\overline{\mathcal{Z}(T,U)}$ to the index, because the latter is a power series in $p=e^{2\pi iT}$ and $q=e^{2\pi iU}$ with integral coefficients. 

This construction generalizes to the case where Wilson lines are present. In this case, one has \be \overline{\mathcal{Z}(T,U,A_k)}=\mathcal{Z}(-\bar T,-\bar U,-A_k)=\mathcal{Z}(-\bar T,-\bar U,A_k)\ .\ee The last equality follows because the spectrum of a self-dual VOA $V$ is invariant under charge conjugation -- in general, applying charge conjugation to a VOA $V$ leads to a different $V$-module, but since $V$ is self-dual the charge conjugate $V$-module must be isomorphic to $V$ itself.

Finally, the `non-chiral' index obtained by including both the original and the spurious degrees of freedom, therefore, is just the modulus square, \begin{equation}|\mathcal{Z}(T,U,A_k)|^2=\mathcal{Z}(T,U,A_k)\overline{\mathcal{Z}(T,U,A_k)},
\end{equation}
of the original index. A similar result was already discussed in \cite{PPV1} in the case of a heterotic string theory on $V^\natural\otimes V^{f\natural}$.

\subsubsection*{Type IIA on $V^{f\natural}_\pm \otimes \bar V^{f\natural}_-$}
The treatment is slightly more complicated for type IIA compactified on $V^{f\natural}_+\otimes \bar V^{f\natural}_-$ or $V^{f\natural}_-\otimes \bar V^{f\natural}_-$, because one needs to add spurious degrees of freedom to both the holomorphic and the antiholomorphic factor. Let us start with $V^{f\natural}_+\otimes \bar V^{f\natural}_-$. For the index $\mathcal{Z}$ (corresponding to the superdenominator of the associated BKM superalgebra), the only relevant single string physical states  are the ones satisfying $k^1_l=k^0_l=k^0_r=k^1_r$, i.e. with zero winding $w=0$ along the circle, so that the index equals $\mathcal{Z}(T,U)=(1/\eta(U)^{24})^{24}$. Let us first add $24$ spurious ground states with fermion number $-1$ to the holomorphic factor $V^{f\natural}_+$. This amounts to introducing creation/annihilation operators with momentum $-k^1_l=k^0_l=k^0_r=k^1_r$, i.e. with $m=0$, which multiply the index by a factor $(1/\eta(T)^{24})^{24}$. Similarly, the modified index $\tilde{\mathcal{Z}}$, corresponding to the denominator of the associated BKM superalgebra, gets multiplied by $\bigl(p^{-1}\prod_{w=1}^{\infty}\frac{1}{(1+p^w)^{24}}\bigr)^{24}$. Therefore, once these holomorphic spurious states are introduced, the  index $\mathcal{Z}$ becomes
\be\label{spurious1} \mathcal{Z}_{sp}^{V^{f\natural}_+\otimes \bar V^{f\natural}_-}(T,U)=\Bigl(\frac{1}{\eta(T)^{24}\eta(U)^{24}}\Bigl)^{24}\ ,
\ee and the modified index $\tilde{\mathcal{Z}}$ becomes
\begin{eqnarray}\label{spurious2}
& & \tilde{\mathcal{Z}}_{sp}^{V^{f\natural}_+\otimes \bar V^{f\natural}_-}(T,U)= 
 \\
& & \Bigl(q^{-1}\prod_{m=1}^\infty\frac{1}{(1+q^m)^{24}}\Bigr)^{24}\Bigl(p^{-1}\prod_{w=1}^\infty\frac{1}{(1+p^w)^{24}}\Bigr)^{24}\Bigl(\prod_{m,w=1}^\infty \frac{(1-p^wq^m)^{c_{NS}(mw)}}{(1+p^wq^m)^{c_{R}(mw)}}\Bigr)^{24}\ . \nonumber
\end{eqnarray}  Exactly the same result can be obtained by adding spurious states to the holomorphic factor of $V^{f\natural}_-\otimes \bar V^{f\natural}_-$. This is obvious, since the only difference between these two theories is the fermion number of the Ramond ground states.

But we are not done yet: one needs to add also the spurious states for the antiholomorphic factor $\bar V^{f\natural}$. The treatment is similar to the other $V\otimes \bar V^{f\natural}$ compactifications: the new spurious degrees of freedom just introduce a second copy of each creation and annihilation operator with the same energy but with opposite winding and momentum. Therefore, the complete non-chiral index (respectively, non-chiral modified index) is just the modulus square of $\mathcal{Z}_{sp}$ in \eqref{spurious1} (respectively, of $\tilde{\mathcal{Z}}_{sp}$ in \eqref{spurious2}).

\subsubsection*{Scalar and fermionic zero modes}

One final subtlety occurs when there are scalar or fermionic zero modes. It is well known that the path integral representation for the partition function of a free massless scalar field on a two dimensional torus is not holomorphically factorized, i.e. it is not simply the modulus square of a function depending holomorphically on the complex structure modulus $U$. Something similar can happen for a free fermion on the torus with doubly periodic boundary conditions, due to the zero modes. In Euclidean signature, the fermion is necessarily complex, and its partition function is a regularized determinant $\det' \Delta_{1/2}$, where $\Delta_{1/2}$ is the two-dimensional (spacetime) Laplace operator acting on $1/2$-differentials, and the symbol $\det'$ means that the zero eigenvalues are excluded, i.e. the zero modes have been absorbed. At first sight, there seems to be no problem with the holomorphic factorization of $\det' \Delta_{1/2}$. Indeed, if one chooses a conformal gauge $ds^2=\rho(z,\bar z)dz\,d\bar z$ for the metric of the spacetime torus, the functional determinant can be regularized in such a way that it is  the modulus square of a chiral determinant $\det' \bar \partial_{1/2}$ depending homolorphically on $U$ (in fact, it is the modulus square of an eta function). This is the standard result one obtains in two dimensional conformal field theory for the torus $1$-point function of the free fermion with periodic boundary conditions. This factorized expression for $\det' \Delta_{1/2}$, however, is not invariant under diffeomorphisms of the spacetime torus: indeed, being a $1$-point function on a torus, it transforms as a section of $K^{1/2}$, a square root of the cotangent bundle.  On the other hand, the path integral we are considering in this section is manifestly diff-invariant -- it depends on the spacetime metric only through diff-invariant quantities, such as the volume, the lengths of geodesics or the angle between them. Therefore it must correspond to a different, diff-invariant regularization of the functional determinant $\det' \Delta_{1/2}$. 

A classical result by Belavin and Knizhnik \cite{Belavin:1986cy} shows that, in general, a diff-invariant regularized determinant $\det'\Delta_n$ of the Laplacian acting on $n$-differentials on a Riemann surface $\Sigma$ fails to be holomorphically factorized. More precisely, $\det'\Delta_n$ is related to a chiral determinant $\det'\bar \partial_n$ by
\be \frac{\det'\Delta_n}{\det\langle \phi^n_a|\phi^n_b\rangle \det\langle \phi^{1-n}_a|\phi^{1-n}_b\rangle}=e^{-c_nS_L(\rho)}|\det{}' \bar \partial_n|^2\ .
\ee Here, $\{\phi^n_1,\phi^n_2,\ldots\}$ and $\{\phi^{1-n}_1,\phi^{1-n}_2,\ldots\}$ are bases of, respectively, the kernel and cokernel of $\bar \partial_n$, i.e. holomorphic $n$- and $1-n$-differentials, $S_L(\rho)$ is the Liouville action, depending on the metric $\rho(z,\bar z)$ used to define the Laplacian, $c_n$ is the conformal anomaly, and $(\det' \bar \partial_n)\,\phi^n_1\wedge\phi^n_2\wedge\ldots\wedge \phi^{1-n}_1\wedge\phi^{1-n}_2\wedge\ldots$ is a section of a holomorphic line bundle on the moduli space of Riemann surfaces. Furthermore, the scalar product $\langle \phi^n_a|\phi^n_b\rangle$ is defined by \be \langle \phi^n_a|\phi^n_b\rangle=\int_\Sigma (\rho(z,\bar z)dz\,d\bar z)^{1-n}\phi^n_a(z)\overline{\phi^n_b(z)}\ ,\ee and an analogous definition holds for the scalar product between $1-n$ differentials.

In our case, $\Sigma$ is the Euclidean spacetime torus $T^2$, which can be parametrized by the complex coordinate $z$ taking values in $\CC/(\ZZ+U\ZZ)$, with respect to which the metric is of the form $ds^2=\rho(z,\bar z)dz\,d\bar z$. The holomorphic $1$-differential $dz$ is canonically normalized $\oint_A dz=1$, $\oint_Bdz=U$, where the $A$- and $B$-cycles are identified with the spatial and the Euclidean time circle, respectively. The metric  is constant $\rho(z,\bar z)\equiv \rho$ and normalized so that the torus volume is $\Im T$. These conditions imply that
\be \int_{T^2} dz\,d\bar z=\Im U\ ,\qquad  \int_{T^2} \rho\, dz\,d\bar z=\Im T\ .
\ee There is a $1$-dimensional space of holomorphic $1/2$-differentials with doubly periodic boundary conditions, and we can take a generator to be $\phi^{1/2}(z)=(dz)^{1/2}$. Note that free chiral  fermions with standard normalization of the OPE, such as the ones used to absorb the zero modes in the indices $\mathcal{Z}$ and $\tilde{\mathcal{Z}}$ in eqs. \eqref{eq:index1} and \eqref{eq:index2}, correspond, up to a numerical ($T,U$-independent) factor, to the half-differential $(dz)^{1/2}$.   The square norm of $\phi^{1/2}$ is
\be
\langle \phi^{1/2}|\phi^{1/2}\rangle=\int_{T^2} (\rho\, dz\,d\bar z)^{1/2} (dz)^{1/2}\,(d\bar z)^{1/2}=(\Im U)^{1/2}(\Im T)^{1/2}\ .
\ee 

For a flat torus, the Liouville factor $e^{-S_L}$ admits holomorphic factorization
so that it can be reabsorbed into the definition of the chiral determinants
\be \det{}'\Delta_{1/2}=(\Im U)(\Im T)e^{c_{1/2}S_L}|\det{}' \bar \partial_{1/2}|^2=(\Im U)(\Im T)|\widetilde{\det}{}' \bar \partial_{1/2}|^2\ .
\ee An analogous result holds for scalar fields. The partition function of a free \emph{complex} scalar field is the inverse $(\det'\Delta_0)^{-1}$ of the regularized functional determinant for the scalar Laplacian. We can choose $\phi^1(z)=dz$ and $\phi^0(z)=1$ as generators of the spaces of holomorphic $1$- and $0$-differentials on the torus, respectively, so that
\be \langle \phi^{1}|\phi^{1}\rangle=\int_{T^2} dz\, d\bar z=\Im U\ ,\qquad \langle \phi^{0}|\phi^{0}\rangle=\int_{T^2} \rho dz\, d\bar z=\Im T\ .
\ee We conclude that
\be \det{}'\Delta_{0}=(\Im U)(\Im T)|\widetilde{\det}{}' \bar \partial_{0}|^2\ ,
\ee where, once again, the Liouville factor is absorbed in a redefinition of the chiral determinant.

This discussion finally suggests the correct relation between the indices $\mathcal{Z}$ and $\tilde{\mathcal{Z}}$ of section \ref{sec:secondquantized} and their path integral representation $\mathscr{Z}$ considered in this section. In particular, if the string theory we are considering contains $n_b$ massless scalars and $n_f$ massless fermions, then the path integral equals
\be\label{finalpath}\mathscr{Z}= [(\Im U)(\Im T)]^{n_f-n_b}|\mathcal{Z}_{sp}|^2\ ,
\ee where $\mathcal{Z}_{sp}$ possibly contains additional `spurious' degrees of freedom, as in equation \eqref{spurious1}. 

Note that if the massless degrees of freedom are charged with respect to some non-trivial background gauge fields, then the corresponding Laplacian is modified in such a way that the zero modes are lifted. This means that one can take the full regularized determinant $\det \Delta$, without projecting onto non-zero modes, and such a determinant automatically factorizes. As a consequence, in the presence of non-trivial Wilson lines for the gauge fields, the correct power $[(\Im U)(\Im T)]^{n_f-n_b}$ in eq.\eqref{finalpath} is given by the number $n_f$ of \emph{neutral} massless fermions minus the number $n_b$ of \emph{neutral} massless scalars.

An analogous formula to \eqref{finalpath} holds for the path integral $\tilde{\mathscr{Z}}$ corresponding to $\tilde{\mathcal{Z}}$.  The insertion of the factor $(-1)^{\bF_l}$ in the trace corresponds, in the path integral formulation, to requiring the fermions and bosons to obey periodic or antiperiodic boundary conditions along the Euclidean time circle, depending on their $(-1)^{\bF_l}$ eigenvalue. Fermions and bosons with antiperiodic boundary conditions on the torus have no zero modes, therefore the analog \eqref{finalpath} holds if  $n_f$ and $n_b$ are the number of zero modes of \emph{periodic} fermions and bosons.

\subsubsection*{Examples}
Let us consider some examples:
\begin{itemize}
    \item For heterotic strings on $V_L\otimes V^{f\natural}_+$, where $V_L$ is the lattice VOA of the Niemeier lattice $L$, the number of massless fermions is $24N$, where $N$ is the number of currents in the VOA $V_L$. Only $24^2$ of them are neutral with respect to the background gauge fields in a maximal abelian torus  of the gauge group. Therefore, $n_f=24^2$. There are no massless propagating scalars, so $n_b=0$. Thus, we expect the path integral to reproduce 
    \be \mathscr{Z}^{V_L\otimes V^{f\natural}_+}(T,U,A_k)=[(\Im U)(\Im T)]^{24^2}|\mathcal{Z}^{V_L\otimes V^{f\natural}_+}(T,U,A_k)|^2\ ,
    \ee where $\mathcal{Z}^{V_L\otimes V^{f\natural}_+}(T,U,A_k)$ is given in \eqref{hetVLind}.
    \item Consider type IIA on $V^{f\natural}_{\pm}\otimes V^{f\natural}_-$. Once the spurious states are included, the theory contains $24^2$ massless scalars from the R-R sector, and no propagating massless fermions. When we consider the index $\mathcal{Z}$, therefore, we have $n_b=24^2$ and $n_f=0$, so that the path integral will reproduce
    \be\label{piVfVf1} \mathscr{Z}(T,U)=[(\Im U)(\Im T)]^{-24^2}|\mathcal{Z}_{sp}(T,U)|^2=\frac{1}{[(\Im U)(\Im T)]^{24^2}}\left|\frac{1}{\eta(T)^{24^2}\eta(U)^{24^2}}\right|^2\ ,
    \ee where $\mathcal{Z}_{sp}$ was calculated in \eqref{spurious1}. As for the modified index $\tilde{\mathcal{Z}}$, recall that $(-1)^{\bF_l}$ acts by $-1$ on the R-NS and R-R sector and trivially on the NS-R and NS-NS sector. Therefore, all $24^2$ massless R-R scalars obey anti-periodic boundary conditions along the Euclidean time circle, and there are no zero modes. It follows that $n_b=n_f=0$ and the path integral $\tilde{\mathscr{Z}}$ factorizes as the modulus square
    \be \tilde{\mathscr{Z}}^{V^{f\natural}_\pm \otimes V^{f\natural}_-}=|\tilde{\mathcal{Z}}^{V^{f\natural}_\pm \otimes V^{f\natural}_-}_{sp}|^2
    \ee
    of the index in eq.\eqref{spurious2}.
    \item In type IIA on $F_{24}\otimes V^{f\natural}_-$ there are $24^2$ massless fermions from the NS-R sector and no massless bosons. Only $24r$ of these fermions are neutral, where $r$ is the rank of the gauge group, which depends on the choice of the $\CN=1$ supercurrent in $F_{24}$. Thus, both for the index $\mathcal{Z}$ and the modified index $\tilde{\mathcal{Z}}$, one has $n_f=24r$ and $n_b=0$, so that the path integral is expected to give
    \be \mathscr{Z}^{F_{24} \otimes V^{f\natural}_-}(T,U,A_k)=[(\Im U)(\Im T)]^{24r}|\mathcal{Z}^{F_{24} \otimes V^{f\natural}_-}(T,U,A_k)|^2
    \ee and 
    \be \tilde{\mathscr{Z}}^{F_{24} \otimes V^{f\natural}_-}(T,U,A_k)=[(\Im U)(\Im T)]^{24r}|\tilde{\mathcal{Z}}^{F_{24} \otimes V^{f\natural}_-}(T,U,A_k)|^2
    \ee where $\mathcal{Z}^{F_{24} \otimes V^{f\natural}_-}$ and $\tilde{\mathcal{Z}}^{F_{24} \otimes V^{f\natural}_-}$ are given in \eqref{F24ind1} and \eqref{F24ind2}, respectively.
    \item Finally, in type IIA on $V^{fE_8}\otimes V^{f\natural}_-$, there are $8\cdot 24$ bosons from the R-R sector and $8\cdot 24$ fermions from the NS-R sector. The gauge group is abelian and all of these states are neutral. For the index $\mathcal{Z}$, one has $n_b=n_f=8\cdot 24$, so that the path integral $\mathscr{Z}$ is holomorphically factorized (in fact, it is just a constant). For the modified index, the scalars obey anti-periodic boundary conditions, so that $n_f=8\cdot 24$ and $n_b=0$. Thus, we expect the path integral to reproduce
    \be \tilde{\mathscr{Z}}^{V^{fE_8} \otimes V^{f\natural}_-}=[(\Im U)(\Im T)]^{8\cdot 24}|\tilde{\mathcal{Z}}^{V^{fE_8} \otimes V^{f\natural}_-}|^2\ ,
    \ee where $\tilde{\mathcal{Z}}^{V^{fE_8} \otimes V^{f\natural}_-}$ is given in eq.\eqref{VE8ind}.
\end{itemize}
In \S \ref{sec:pintsuper} and \S \ref{sec:pintdenom}, we will verify some of these expectations by explicit path integral calculations.

\subsection{Path integrals and theta lifts}\label{sec:paththeta}

Let us consider  heterotic or type II superstring theory on a generic internal CFT $V_1\otimes \bar V_2$. Let us first consider the case where there are no Wilson lines. As mentioned above, the path integral representation of the second quantized index is given by
\be\label{pathint} \mathscr{Z}=\exp\left(\frac{1}{2}\int_{SL(2,\mathbb{Z})\backslash \mathbb{H}} \frac{d^2\tau}{\tau_2} \Tr\Big(\qsf^{L_0}\bar \qsf^{\bar L_0}(-1)^{\bF}\Big)\right).
\ee
   The trace in the integrand is taken over the full GSO projected ghost and matter CFT, restricted to $\ker b_0\cap \ker \bar b_0$. The latter condition corresponds to the insertion of suitable ghost vertex operators in the path integral.   The operator $(-1)^{\bF}$ is a spacetime fermion number; for example, in type II theories it acts trivially on the NS-NS and R-R sectors and by a minus sign on the NS-R and R-NS sectors. As usual, the contributions from non-zero modes of the ghosts and superghosts exactly cancel the bosonic and fermionic matter oscillators in the $1+1$ spacetime directions. Thus, the trace is the product of a term coming from a sum over winding-momenta along the two dimensional Euclidean spacetime torus, times the partition function of the internal CFT $V_1\otimes \bar V_2$.
 
 The winding-momenta take values in an even unimodular lattice $\Gamma^{2,2}$ of signature $(2,2)$, so that their contribution gives a Narain-Siegel theta function
 \be \Theta_{\Gamma^{2,2}}(\tau;T,U)=\tau_2\sum_{m_1,w_1,m_0,w_0\in \ZZ} \qsf^{\frac{k_l^2}{2}}\bar \qsf^{\frac{k_r^2}{2}}\ ,
 \ee (we included a factor $\tau_2$ to make it modular invariant and match with the mathematical literature) where 
 \begin{align}\label{momenta1}
 k_l^2&=\frac{\left|\begin{pmatrix}
 	T & 1
 	\end{pmatrix}\begin{pmatrix}
 	w_0 & w_1\\ -m_1 & m_0
 	\end{pmatrix}\begin{pmatrix}
 	-U \\ 1
 	\end{pmatrix}
 	\right|^2}{2T_2U_2}\\
 k_r^2&=k_l^2-2m_1 w_1-2 m_0 w_0\ .\label{momenta2}
 \end{align}
 The partition function of the CFT $V_1\otimes \bar V_2$ factorizes as a product $f(\tau,V_1)\overline{f(\tau,V_2)}$ of a holomorphic and an  anti-holomorphic factor. 
 
 Let us focus on type II models; in this case, the spacetime fermion number factorizes as $(-1)^{\bF}=(-1)^{\bF_l}(-1)^{\bF_r}$, where $(-1)^{\bF_{l,r}}$ acts by a minus sign on the (left- or right-moving) Ramond sector. In terms of the traces of eq. \eqref{eq:partfns}, for each given SVOA $V$, one has 
 \be\label{strace} f(\tau,V)=\frac{\phi_{NS}(\tau,V)-\phi_{\tilde{NS}}(\tau,V)}{2}-\frac{\phi_{R}(\tau,V)\pm \phi_{\tilde R}(\tau,V)}{2}=\begin{cases}
 	0 & \text{for } V=V^{fE_8}\ ,\\
 	- 24 & \text{for } V=V^{f\natural}_\pm\ , \\
 	24 & \text{for } V=F_{24}\ .
 \end{cases}
 \ee  In this formula, the first two terms yield the GSO-projected NS sector. The relative sign between the NS and the Ramond sector is due to the spacetime fermion number. The last two terms can be explained as follows. In the Ramond sector, the GSO projection should be implemented by taking just one of the two spacetime chiralities for each Ramond state of the internal SVOA $V$, depending on the eigenvalue of the internal worldsheet fermion number $(-1)^F$. However, in our Euclidean path integral, the description of chiral fermions is problematic. A way out is as follows. For $V=F_{24}$ or $V=V^{fE_8}$, one observes that $\phi_{\tilde R}(\tau,V)=0$, i.e. the number of states with positive and negative fermion number is the same for each $L_0$ eigenvalue. This means that the states in the Ramond sector come in pairs with opposite fermion number, and we can count one complex (non-chiral) spacetime fermion for each such pair, thus giving $\frac{\phi_R(\tau,V)}{2}$ as a counting function. The same reasoning applies to $V^{f\natural}_\pm$ with $L_0>1/2$: the number of states with positive and negative fermion number is the same for $L_0>1/2$, so we can count one complex fermion for each pair of states.
 On the other hand, the $24$ Ramond ground states ($L_0=1/2$) in $V^{f\natural}_\pm$, only appear for either positive (for $V^{f\natural}_+$) or negative  (for $V^{f\natural}_-$) fermion number, i.e. they would correspond to a chiral spacetime fermion. Since we are not able to describe these chiral fermions in our Euclidean path integral formulation, we simply count one complex non-chiral fermion for each of these $24$ states, irrespective of their fermion number; this gives $\frac{\phi_{R}(\tau,V^{f\natural}_\pm)\pm \phi_{\tilde R}(\tau,V^{f\natural}_\pm)}{2}$. In this way we have a slight overcounting of Ramond ground states: this is the origin of the `spurious states' discussed in section \ref{sec:pathvsop}.  
 
 By \eqref{strace}, in type IIA theories, the partition function $f(\tau,V_1)\overline{f(\tau,V_2)}$ is just a constant. Note that when either $V_1$ or $ V_2$ is $V^{fE_8}$, the partition function vanishes. 
 
 For heterotic models on $V_1\otimes \bar V_2$, we have a similar result for  $f(\tau,V_1)\overline{f(\tau,V_2)}$, where the anti-holomorphic factor $\overline{f(\tau,V_2)}$ is a constant given in in \eqref{strace},  while the holomorphic factor
 \be f(\tau,V_1)=J(\tau)+N=\qsf^{-1}+N+196884\qsf+\ldots
 \ee is the partition function of the bosonic VOA $V_1$, which equals the $J$-function up to the constant term.
 
 Thus, in all cases we are interested in, the integrand $\tau_2\Tr(\qsf^{L_0}\bar \qsf^{\bar L_0}(-1)^{\bF})$ is just the product $F(\tau)\Theta_{\Gamma^{2,2}}(\tau;T,U)$ of a theta series and a holomorphic function $F(\tau)=f(\tau,V_1)\overline{f(\tau,V_2)}$.

We conclude that the path integral will amount to performing an integral of the form 
\begin{equation}
    \int_{SL(2,\mathbb{Z})\backslash \mathbb{H}}F(\tau)\Theta_{\Gamma^{2,2}}(\tau;T,U) \frac{d^2\tau}{\tau_2^2}.
\end{equation}
This is recognized as a \emph{theta lift} in the mathematics literature. More generally one can lift a modular form $F$ on $\mathbb{H}$ to an automorphic form on $SO(m,n)$ via 
\begin{equation}
F\quad \longmapsto\quad  \Theta_F(g)=\int_{SL(2,\mathbb{Z})\backslash \mathbb{H}}F(\tau)\Theta_{\Gamma^{m,n}}(\tau;g) \frac{d^2\tau}{\tau_2^2}.
\end{equation}
 Here, $\Theta_{\Gamma^{m,n}}$ is a Siegel-Narain theta series for the lattice $\Gamma^{m,n}$, depending on $\tau\in \mathbb{H}$ and $g\in SO(m,n;\mathbb{R})$. The image $\Theta_F(g)$ of the theta lift  is then an automorphic form on 
\begin{equation}
\Gamma \backslash SO(m,n;\mathbb{R})/ (SO(m)\times SO(n)),
\end{equation}
where $\Gamma\subset SO(m,n;\mathbb{R})$ is an arithmetic subgroup (T-duality group). 

Consider the case of the lattice $\Gamma^{2,d}$, which is the level of generality that we are interested in. The case when $d>2$ corresponds to including Wilson lines. Then $\Theta_{\Gamma^{2,d}}$ transforms like a modular form of weight $d/2-1$ with respect to $SL(2,\mathbb{Z})$. Assume further that $F(\tau)$ is a weakly holomorphic modular form of weight $1-d/2$ with Fourier coefficients $c(m,n)$. Since $F$ may have singularities at the cusps the integral diverges. However,  it can be regularized by introducing a cutoff $\Im(\tau)\le t$ and computing the integral in the limit $t\to \infty$ \cite{Dixon:1990pc,HM1}. Physically, the divergence is an IR effect, due to the presence of massless or tachyonic modes (the latter only arises in  off-shell heterotic strings).  The result is of the form 
\begin{equation}
\Theta_F(g)=\log ||\Phi_F(g)||^2_{\text{Pet}}+\text{const},
\end{equation}
where $|| \cdot ||_{\text{Pet}}$ denotes the Petersson metric on the line bundle $\mathcal{L}$ of modular forms of weight $c(0,0)/2$ over $SO(2,d;\mathbb{R})/(SO(2)\times SO(d))$. Moreover, in the neighbourhood of a cusp, $\Phi_F(g)$ has an infinite product expansion. The theta lift generalizes to the case where $\Theta_\Lambda$ is the theta series of a lattice $\Lambda$ that is not unimodular, and $F(\tau)$ is a vector valued modular form for a suitable $SL(2,\ZZ)$-representation, see appendix \ref{app:vectorvaluedtheta} for details and references.

To illustrate this let us restrict again to the case of having no Wilson lines, i.e. $d=2$, as in the beginning of this section.
In this case, $F(\tau)$ is a weight 0 modular form. The simplest possibility is to let $F$ be a constant, say $c$. Then we obtain
\begin{equation}
\Theta_c(T,U)=\int_{SL(2,\mathbb{Z})\backslash \mathbb{H}} c\, \cdot \, \Theta_{\Gamma^{d,d}}(\tau;T,U) \frac{d^2\tau}{\tau_2^2}=-c\log ||\eta(T)^{2} \eta(U)^{2}||^2_{\text{Pet}} +\text{const},
\label{thetaliftofidentity}
\end{equation}
where the Petersson metric is given explicitly by 
\begin{equation}
||\eta(T)^{2} \eta(U)^{2}||^2_{\text{Pet}}=\Im(T)\Im(U)\big|\eta^{2}(T)\eta^{2}(U)\big|^2.
\end{equation}
The constant in (\ref{thetaliftofidentity}) is explicit and involves the Euler-Mascheroni constant (see, e.g, \cite{HM1}). We recognize the general structure of the path integral~(\ref{finalpath}), where we recall that the prefactors $\Im(T)\Im(U)$ arise from the bosonic and fermionic zero modes.

As mentioned in section \ref{sec:pathvsop}, the path integral we started from is manifestly invariant under spacetime diffeomorphisms. The dependence on the metric is only through the complex moduli $T$ and $U$, which in turn are defined in terms of diff-invariant quantities, such as the volume of the spacetime torus, the length of geodesics or the angle between them. The infrared regularization is also manifestly diff-invariant -- we just cut a region of integration over the moduli of the \emph{worldsheet} torus. Thus, the final result is expected to be free of gravitational anomalies. In particular, the arithmetic T-duality group $\Gamma\subset SO(2,d;\mathbb{R})$ contains a group of large spacetime diffeomorphisms, so that the absence of gravitational anomalies can be seen as the physical counterpart of the automorphy properties of the theta lift.
 
\subsection{The path integral for the superdenominator}\label{sec:pintsuper}

Let us consider a type IIA model on $V\otimes \bar V^{f\natural}_-$, where $V$ is a self-dual $\CN=1$ SVOA with $c=12$. The second quantized supersymmetric index $\mathcal{Z}$ corresponding to this model is (the $24$-th power of) the superdenominator of the BKM superalgebra associated with $V$.

In the following subsections, we will describe how the  superdenominator arises explicitly from the path integral for the type II string on $V\otimes V^{f\natural}$, with $V=F_{24}$, $V^{f\natural}$, and $V^{fE_8}$.

 \subsubsection*{The case $F_{24}\otimes \bar{V}^{f\natural}_-$}
For $V=F_{24}$ we have $\chi^{\rm R}_+=\chi^{\rm R}_-=0$. Let us  consider the case when we turn off all Wilson lines, i.e. $y_i=1$ for all $i$, so that the lattice is $\Gamma^{2,2}$. We expect the index to depend on both $T$ and $U$, the K\"ahler and complex structure moduli of the torus. 
By \eqref{strace}, we have $f(\tau,F_{24})=24$ and $f(\tau,{V}^{f\natural}_-)=-24$. The result of the theta lift is thus
\begin{equation}
\Theta_{F_{24}\otimes \bar{V}^{f\natural}}(T,U)=\int_{SL(2,\mathbb{Z})\backslash \mathbb{H}} 24\cdot (-24)\, \cdot \, \Theta_{\Gamma^{2,2}}(\tau;T,U) d\mu=24\cdot 24\log ||\eta^{2}(T)\eta^{2}(U)||^2_{\text{Pet}}.
\label{thetaliftfullGamma22}
\end{equation}

The full non-holomorphic partition function in this case is given by 
\begin{equation}
\mathscr{Z}^{F_{24}\otimes \bar{V}^{f\natural}}\big|_{y=1}=e^{\frac{1}{2}\Theta_{_{F_{24}\otimes \bar{V}^{f\natural}}}(T,U)}=\text{const}\cdot ||(\eta^{24}(T))^{24} (\eta^{24}(U))^{24}||_{\text{Pet}}^{2}.
\end{equation}

Now let us give a rough idea of the generalization to the case where the Wilson lines are turned on. Recall that $y_i=e^{2\pi i A_i}$, $i=1, \dots, r$, where $A_i$ are Wilson line moduli and $r$ is the rank of the gauge group $G$.  The integral now involves the Siegel-Narain theta function $\Theta_{\Lambda}(\tau; T, U, A)$ for a lattice $\Lambda=\Gamma^{2,2}\oplus L$ of signature $(2,2+r)$, where $L$ is the $r$-dimensional lattice of charges with respect to gauge group $G$. The partition function $F$ becomes a Jacobi form, with theta decomposition of the form $F=\sum F_\gamma \theta_\gamma$, where $\gamma$ label the cosets in $L^\vee/L$. Using the vector valued theta lift of $F$ discussed in appendix~\ref{app:vectorvaluedtheta} we arrive at
\begin{equation}
{\mathscr Z}^{F_{24}\otimes \bar{V}^{f\natural}_-}=e^{\frac{1}{2}\Theta_{_{F_{24}\otimes \bar{V}^{f\natural}}}(T,U,A)}=\text{const}\cdot ||\Phi(T, U, A)^{24}||_{\text{Pet}}^{2},
\end{equation}
where $\Phi(T, U, A)$ has the following infinite product representation
\begin{equation}
\Phi=e^{-2\pi i (T+U+(\rho, A))} \prod_{\lambda\in \Delta^+_g} (1-e^{2\pi i (\lambda, A)})^{c(0,\lambda)} \prod_{m,w=0\atop (m,w)\neq (0,0)}^\infty\prod_{\lambda\in \Delta^g} (1-e^{2\pi i( mT+ wU+ (\lambda, A))})^{c(mw, \lambda)}
\end{equation}
This matches perfectly with the index $\mathcal{Z}^{F_{24}\otimes \bar{V}^{f\natural}_-}$ obtained in section~\ref{IIAVfnatural}.

 \subsubsection*{The case $V^{fE_8} \otimes \bar{V}^{f\natural}$}
 In this case we have both $\chi^{\rm R}_+\neq 0$ and $\chi^{\rm R}_-\neq 0$, and only states with all momenta vanishing will contribute. Thus, we expect the theta lift to be independent of $T$ and $U$. Indeed, the contribution to $F(\tau)$ from the $f(\tau,V^{fE_8})$ vanishes (see eq.\eqref{strace}). Hence the partition function is just
 \begin{equation}
 \mathscr{Z}^{V^{fE_8}\otimes \bar{V}^{f\natural}}=e^{\frac{1}{2}\Theta_0}=1,
 \end{equation}
 which agrees with the result in section~\ref{IIAVfnatural}.
 
 \subsubsection*{The cases $V_{\pm}^{f\natural} \otimes \bar{V}^{f\natural}_-$}
 For $V_{\pm}^{f\natural}$ we have $\chi^R_\pm \neq 0$ (but $\chi^R_\mp=0$). Recall from section \ref{IIAVfnatural} the case $\chi^{\rm R}_+\neq0$ implies that, on-shell, there is a constraint on the winding and momenta such that there is no winding around the spacelike $S^1$. Similarly, the case $\chi^{\rm R}_-\neq 0$ implies that there is no momenta around the $S^1$.  Therefore, in the operatorial formalism, one has that $\mathcal{Z}^{V_{+}^{f\natural} \otimes \bar{V}^{f\natural}_-}$ depends only on $U$, while $\mathcal{Z}^{V_{+}^{f\natural} \otimes \bar{V}^{f\natural}_-}$ depends only on $T$. 
In the path integral \eqref{pathint}, these conditions are not imposed, so one expects $\mathscr{Z}$ to depend on both $T$ and $U$ and be the same function for both $V_{+}^{f\natural} \otimes \bar{V}^{f\natural}_-$ and $V_{-}^{f\natural} \otimes \bar{V}^{f\natural}_-$.

 The calculation is very similar to the case $F_{24} \otimes \bar{V}^{f\natural}_-$, but now \be c=f(\tau,{V}^{f\natural}_\pm)\overline{f(\tau,{V}^{f\natural}_-)}=(-24)^2\ .\ee
  The result is 
 \begin{eqnarray}
 \mathscr{Z}^{V_{\pm}^{f\natural} \otimes \bar{V}^{f\natural}_-}(T, U)&=&\text{const}\cdot ||\eta^{-24}(T)\eta^{-24}(U)||_{\text{Pet}}^{2\cdot 24}. 
 \end{eqnarray}
 
The superdenominator formula should correspond to the 24th power of the holomorphic piece of this partition function, namely $\eta^{-24}(T)$ (or $\eta^{-24}(U)$). This indeed agrees with eq.~\eqref{piVfVf1}.

 \subsection{Path integral for the denominator of Type II on  $V^{f\natural}_{\pm}\otimes \bar V^{f\natural}_-$}\label{sec:pintdenom}
 We shall now move on to compute the path integral for the modified index $\tilde{\mathcal{Z}}^{V_1\otimes \bar V^{f\natural}_-}$ for type IIA on $V_1\otimes  \bar V^{f\natural}_-$, corresponding to the denominator function of the BKM superalgebra associated with $V_1$. The modified index $\tilde{\mathcal{Z}}$ is obtained by inserting an operator $(-1)^{\bF_l}$ in the trace over the second quantized Hilbert space. The left-moving fermion number  $(-1)^{\bF_l}$ acts by $+1$ on the physical states in the left-moving NS sector (which give the even elements of the BKM superalgebra) and by $-1$ on physical string states in the left-moving R sector (corresponding to odd elements in the BKM superalgebra). In the path integral formulation of the index, this insertion corresponds to taking either integral or half integral quantization for the momenta in the time direction, depending on the $(-1)^{\bF_l}$ eigenvalue of the string running in the loop. Formally, this is equivalent to taking the standard path integral for an orbifold theory, analogous to the CHL string models. One starts from a theory whose Euclidean time circle has twice the radius, and then takes an orbifold by a symmetry $\mathfrak{z} $ acting as a half-period shift along the Euclidean time circle, together with the action of $(-1)^{\bF_l}$. As a result, we get a theory whose fields obey $(-1)^{\bF_l}$-twisted boundary conditions along the Euclidean time circle.
From a mathematical viewpoint, this procedure amounts to taking the theta lift of vector-valued (rather than scalar) modular forms in the integrand. The general structure of such integrals is recalled in appendix~\ref{app:vectorvaluedtheta}.

 Now let us apply this machinery to our setting. Let us consider the example of type IIA string theory on $V^{f\natural}_-\otimes \bar V^{f\natural}_-$.   Recall that the path integral for the superdenominator, in this example, is given by \eqref{pathint}, where the integrand is $\Theta_{\Gamma^{2,2}}(T,U,\tau)f(\tau,V^{f\natural}_-)\overline{f(\tau,V^{f\natural}_-)}$. In particular, in the CFT trace inside the integral, the spacetime oscillators cancel against the ghost and superghost contributions, and we are left with the theta series for the winding-momentum lattice on the Euclidean torus times the trace over the internal CFT.
 
 Now, in order to compute the path integral for the denominator, we have to take an orbifold by an order $2$ symmetry $\mathfrak{z}$ acting by a half-period shift along the Euclidean time circle and simultaneously by $(-1)^{\bF_l}$. This symmetry acts trivially on the spacetime oscillators and (super)ghosts, so that the same cancellations as for the superdenominator case occur. Therefore, the CFT trace in the integral of \eqref{pathint} is still $\overline{f(\tau,V^{f\natural}_-)}=-24$ times the contributions from the holomorphic internal SVOA and from the winding-momentum. We claim that the integrand has the form
 \be  \sum_{r\in \mathbb{Z}/2\mathbb{Z}}\Tr_{\mathfrak{z}^r}\Bigl(\qsf^{L_0}\bar\qsf^{\bar L_0}(-1)^{\bF}  \frac{1}{2}\sum_{s\in \mathbb{Z}/2\mathbb{Z}}\mathfrak{z}^s\Bigr)=\frac{1}{2}\sum_{r,s\in \mathbb{Z}/2\mathbb{Z}} f(r,s;\tau)\Theta^{r,s}(T,U;\tau)\overline{f(\tau,V^{f\natural}_-)}\ .
 \ee
 Here, $ \Tr_{\mathfrak{z}^r}$ denotes the trace over the $\mathfrak{z}^r$-twisted sector, and $\frac{1}{2}\sum_{s} \mathfrak{z}^s$ is the projection onto the $\mathfrak{z}$-invariant states. The antiholomorphic SVOA $\bar V^{f\natural}_-$ contributes with the usual factor $\overline{f(\tau,V^{f\natural}_-)}=-24$, since $\mathfrak{z}$ acts trivially on this theory. For each $r,s\in \ZZ/2\ZZ$, $f(r,s;\tau)$ is the contribution from the holomorphic internal SVOA $V^{f\natural}_-$, while $\Theta^{r,s}(T,U;\tau)$ comes from the sum over winding and momentum along the space-time torus. Explicitly,
 \begin{eqnarray}
f(0,0;\tau)&=&\frac{\phi_{\text{NS}}(\tau)-\phi_{\tilde{\text{NS}}}(\tau)}{2}-\frac{\phi_{\text{R}}(\tau)-\phi_{\tilde{\text{R}}}(\tau)}{2} =-24\nonumber \\
f(0,1;\tau)&=&\frac{\phi_{\text{NS}}(\tau)-\phi_{\tilde{\text{NS}}}(\tau)}{2}+\frac{\phi_{\text{R}}(\tau)-\phi_{\tilde{\text{R}}}(\tau)}{2} =24+4096\qsf+98304\qsf^2+1228800\qsf^3+\cdots \nonumber \\
f(1,0;\tau)&=& \frac{\phi_{\text{NS}}(\tau)+\phi_{\tilde{\text{NS}}}(\tau)}{2}-\frac{\phi_{\text{R}}(\tau)+\phi_{\tilde{\text{R}}}(\tau)}{2} =\qsf^{1/2}+276 \qsf^{1/2}-2048\qsf+11202\qsf^{3/2}+\cdots
\nonumber \\
f(1,1;\tau)&=&-\frac{\phi_{\text{NS}}(\tau)+\phi_{\tilde{\text{NS}}}(\tau)}{2}-\frac{\phi_{\text{R}}(\tau)+\phi_{\tilde{\text{R}}}(\tau)}{2} =-\qsf^{-1/2}-276 \qsf^{1/2}-2048\qsf-11202\qsf^{3/2}+\cdots 
\nonumber \\ 
\nonumber 
\end{eqnarray}
 where in the second line we use the fact that $\mathfrak{z}=(-1)^{\bF_l}$ on $V^{f\natural}_-$, and the third and fourth line follow from modular transformations of the second (recall that $f_{\text{NS}}$, $f_{\tilde{\text{NS}}}$, and $f_{\text{R}}$ span a $3$-dimensional representation of $SL(2,\ZZ)$, while $f_{\tilde{\text{R}}}$ is $SL(2,\ZZ)$-invariant).
 As for the theta series, we have
 \be \Theta^{r,s}(T,U;\tau)=\tau_2\sum_{m_1,w_1,m_0\in \ZZ}\sum_{w_0\in \ZZ+\frac{r}{2}} \qsf^{\frac{k_l^2}{2}}\bar \qsf^{\frac{k_r^2}{2}}(-1)^{sm_0}\ ,
 \ee where $k_l,k_r$ are given in \eqref{momenta1},\eqref{momenta2}. This follows from the fact that the winding number $w_0$ is half-integral in the $\mathfrak{z}$-twisted sector, and that a shift by half a period corresponds to multiplication by $(-1)^{m_0}$. 
 
 \bigskip
 
 The connection with vector-valued theta lifts can be easily made manifest (see appendix~\ref{app:vectorvaluedtheta} for notation).
 Let us consider the lattice $\Lambda=\Gamma^{1,1}\oplus \Gamma^{1,1}(2)$, where $\Gamma^{1,1}$ is the unique even unimodular lattice of signature $(1,1)$, and $\Gamma^{1,1}(2)$ is the same lattice with the quadratic form rescaled by $2$. This implies 
 \begin{equation}
 \Lambda^{\vee}=\Gamma^{1,1}\oplus \tfrac{1}{2}\Gamma^{1,1}(2), \qquad \Lambda^{\vee} /\Lambda\cong \tfrac{1}{2}\Gamma^{1,1}(2)/\Gamma^{1,1}(2)\cong \mathbb{Z}_2\times \mathbb{Z}_2.
 \end{equation}
 The elements $\lambda \in \tfrac{1}{2}\Gamma^{1,1}(2)$ can be parametrized by 
 \begin{equation} 
 \lambda=(m/2, n/2), \qquad \lambda^2=mn.
 \end{equation}

 The cosets $\gamma\in \Lambda^{\vee}/\Lambda$ are represented by pairs $(r/2,s/2)$ with $r,s\in \mathbb{Z}/2\mathbb{Z}$. With the lattice $\Lambda$ is associated a vector--valued representation $\rho_\Lambda$ of $SL(2,\ZZ)$, as in \eqref{rhoLambda}. We define a vector--valued modular form $F$ for $\rho_\Lambda$ via the discrete Fourier transform as follows
 \begin{equation} F_{r,s}(\tau):=\frac{1}{2}\sum_{t\in \mathbb{Z}/2\mathbb{Z}}e^{-\pi i ts}f(r,t;\tau)\overline{f(\tau,V^{f\natural}_-)}.
 \end{equation}
 This yields 
 \begin{eqnarray}
 F_{00}(\tau)&=&-24(2048\qsf+\cdots)
 \nonumber \\
 F_{01}(\tau)&=&-24(-24-2048\qsf-\cdots)
 \nonumber \\ 
 F_{10}(\tau)&=&-24(-2048\qsf-\cdots)
 \nonumber \\
 F_{11}(\tau)&=&-24(\qsf^{-1/2}+276\qsf^{1/2}+\cdots).
 \end{eqnarray}
 For later reference let us denote the Fourier expansions of these functions by 
 \begin{equation}
 F_{r,s}(\tau)=-24\sum_{n\in \frac{1}{2}\mathbb{Z}} C_{r,s}(n)\qsf^n.
 \label{FourierCoeff}
 \end{equation}
 These functions represent the components $F_\gamma(\tau)$ of the vector-valued modular form $F(\tau)$. 
 
 Now let us turn to the theta function. The shifted theta series $\theta_{\Lambda+\gamma}$ can be written as follows in terms of the shifted Narain theta series $\Theta^{r,s}$:
 \begin{equation}
 \theta_{\Lambda+\gamma}=\frac{1}{2}\sum_{t\in \mathbb{Z}/2\mathbb{Z}}e^{\pi i st}\Theta^{r,t}(T, U; \tau),
 \end{equation} where $\gamma\equiv(r/2,s/2)\in \Lambda^\vee/\Lambda$. 

The inner product in the integrand of the theta lift can now be written out explicitly as follows
\begin{equation}
\left(\overline{\Theta}_\Lambda, F\right)=\sum_{\gamma, \gamma'\in \Lambda^{\vee}/\Lambda}\theta_{\Lambda+\gamma}F_{\gamma'}\delta_{\gamma'-\gamma, 0}=\frac{1}{2}\sum_{r,s\in \mathbb{Z}/2\mathbb{Z}} f(r,s;\tau)\Theta^{r,s}(T,U;\tau)\overline{f(\tau,V^{f\natural}_-)}.
\end{equation}
Thus, the theta integral that we wish to perform is explicitly given by 
\begin{equation}
\frac{1}{2}\int_{SL(2,\mathbb{Z})\backslash \mathbb{H}} \left(\overline{\Theta}_\Lambda, F\right)d\mu.
\end{equation}
We can now invoke the general theorems of Borcherds \cite{Borcherds:1996uda} and Carnahan \cite{Carnahan} to deduce that the result of this integral is\footnote{The function $\Phi_F$ considered here is the inverse of the automorphic function $\Phi_M$ defined in theorem 13.3 of \cite{Borcherds:1996uda}. With our conventions, $\Phi_F$ can be directly identified with the path integral $\tilde{\mathscr{Z}}_{sp}$.}
\begin{equation}
\log ||\Phi_F(T, U)||_{\text{Pet}} +\text{const},
\end{equation}
where $\Phi_F(T, U)$ is an automorphic form on $O(2,2;\mathbb{R})/(O(2)\times O(2))$, with the following product representation:
\begin{equation}
\Phi_F(T,U)=
p^{-24}q^{-24}\prod_{m,w\in \mathbb{Z}\atop (\lambda, W)> 0}\prod_{t\in \ZZ/2\ZZ} (1-(-1)^tp^w  q^{m})^{24C_{0,t}\left(mw\right)}.
\end{equation}
The condition $(\lambda, W)>0$ includes all states $\lambda = (m, w)$ such that $m,w \geq 0$, and $(m,w) \neq (0,0)$.\footnote{We thank Scott Carnahan for clarifications about this point.} We may separate this into states with $w=0, m>0$ and states with $w\geq 1, m\ge 0$. Since $C_{0,0}(0)=0$ and $C_{0,1}(0)=-24$ the first set of states (with the $q^{-24}$ factor) give
\begin{equation}
\left(q^{-1}\prod_{m=1}^{\infty}(1+q^m)^{-24}\right)^{24}
\end{equation}
which we can rewrite in terms of $\eta$-functions according to 
\begin{equation}
\left(\frac{\eta(U)^{24}}{\eta(2U)^{24}}\right)^{24}.
\end{equation}
The complete function $\Phi_F$ may now be written as 
\begin{equation}
\Phi_F(T, U)=\left(q^{-1}\prod_{m=1}^{\infty}\frac{1}{(1+q^m)^{24}}\right)^{24}\Psi(T,U)^{24},
\end{equation}
where we defined
\begin{equation}
\Psi(T, U)=p^{-1}\prod_{w=1}^{\infty} \prod_{m\ge 0} \left(1-p^wq^m\right)^{C_{0,0}(mw)}\left(1+p^wq^m\right)^{C_{0,1}(mw)}.
\end{equation}
Here $C_{0,0}$ and $C_{0,1}$ are the Fourier coefficients defined in (\ref{FourierCoeff}). The function $\Psi(T,U)^{24}$ matches exactly with the modified index  $\tilde{\mathcal{Z}}^{V^{f\natural}_- \otimes \bar{V}^{f\natural}_-}$ in eq.\eqref{modindVmVm}, and its $24$-th root $\Psi(T, U)$ is the denominator formula in eq.(4.18) of \cite{Harrison:2018joy}. The factor $\left(q^{-1}\prod_{m=1}^{\infty}\frac{1}{(1+q^m)^{24}}\right)^{24}$ is the contribution of the spurious states, so that $\Phi_F$ matches with  $\tilde{\mathcal{Z}}_{sp}^{V^{f\natural}_-\otimes \bar V^{f\natural}_-}(T,U)$ in  equation~(\ref{spurious2}).

Using similar techniques one may compute the path integrals for the other cases and reproduce the corresponding denominator functions obtained in section~\ref{sec:IIAex2}.

\section{Discussion}
\label{sec:discussion}

In this paper we construct a family of distinguished type IIA and heterotic string compactifications to two spacetime dimensions. These compactifications are distinguished by the fact that there exists an action of a BKM algebra on the BPS subspsace of the Hilbert space of physical states in the models, which we illustrate in \S \ref{sec:algebras}, following the original construction for the monster BKM algebra in \cite{PPV1, PPV2}. Furthermore, we demonstrate how a spacetime supersymmetric index in our theories---computed from both a trace in the Hilbert space (\S \ref{sec:secondquantized}) and via a path integral (\S \ref{sec:pathintegrals})---furnishes an automorphic form which is (closely related to)  a denominator function for the corresponding BKM algebra. In this family we find a spacetime string--theoretic setting for the BKM algebras constructed in \cite{Sch1,Harrison:2018joy,Harrison:2020wxl}, among many others.

The compactifications we consider in this paper are also distinguished because they are highly symmetric, and the corresponding BKM algebras which arise have large finite symmetry groups, in some cases sporadic groups. For example, $V^{f\natural}$ has an action of the Conway group \cite{Duncan}, which carries over to an action on the corresponding  BKM arising from the chiral string compactification \cite{Harrison:2018joy}. The lattice VOAs $V_L$ based on the Niemeier lattices $L$ have symmetry group corresponding to (extensions of) the automorphism group of the lattice.\footnote{These are the groups appearing in the umbral moonshine phenomenon \cite{CDH2}. It would be very interesting if there was a connection between these physical models and umbral moonshine.} (For $L$ the Leech lattice, this is the group $Co_0$, while other cases have symmetry groups $M_{24}$ and $2.M_{12}$ arising from the Mathieu groups.) In light of this fact, a natural question is to explore so--called CHL models \cite{CHL} based on these compactifications, which arise from  orbifolds of string theory on $\mathcal{C} \times T^d$. These orbifolds arise by quotienting by a symmetry of the internal CFT $\mathcal C$ combined with a translation along the torus $T^d$. For a general discussion, see \cite{Persson:2017lkn}. In the monster case of \cite{PPV1, PPV2}, where $\mathcal{C}= V^\natural \otimes \bar V^{f\natural}$, considering these CHL models leads to a family of ``twisted" denominator formulas for the spacetime BKM algebra. It would be interesting to generalize this to the case of models considered in this paper.

Furthermore, these CHL models also inspired a proposed physical explanation for the genus zero property of monstrous moonshine \cite{CN}. As the Conway moonshine arising from the SVOA $V^{f\natural}$ also satisfies a genus zero property \cite{Duncan:2014eha},\footnote{To be precise, Conway moonshine is usually expressed in terms of $V^{s\natural}$, which is an SVOA isomorphic to $V^{f\natural}$, but with a different action of the Conway group. In our setup, it would be more natural to re-formulate this moonshine phenomenon in terms of $V^{f\natural}$.} it would be very interesting to offer a physical explanation along the lines of that in \cite{PPV1,PPV2} based on the constructions in this paper and their corresponding CHL models.

Besides finding explicit physical models where one can explore the physics of BPS algebras, another one of our original motivations for this work was to uncover new 2d spacetime string theoretic dualities, where one can explore connections to BKM algebras and automorphic forms. It is the case that a single automorphic form on the moduli space $\mathcal M$ of some family of string compactifications can have multiple distinct Fourier expansions at different ``cusps" of $\mathcal M$, and that the expansion at each cusp corresponds to the (super)denominator formula a unique BKM (super)algebra, as in \cite{Gritsenko:2012qn}. In a string theory context, these BKMs arise as symmetries of different perturbative duality frames of a given compactification, which can be related by discrete duality transformations. We imagine that this furnishes a physical explanation for the following observations:
\begin{enumerate}
    \item The denominator of the Conway BKM superalgebra constructed in \cite{Harrison:2018joy} arises as the Fourier expansion of automorphic form on $(\mathbb H\times \mathbb H)/O(2,2,\mathbb Z)$ at a given cusp. The authors of this paper gave evidence for the existence of a distinct BKM superalgebra, which also has an action of the Conway group, and whose denominator and superdenominator should correspond to the expansion of the same automorphic form at a different cusp of $(\mathbb H\times \mathbb H)/O(2,2,\mathbb Z)$. The construction in the present article provides a natural description of this new mysterious superalgebra. In section \ref{sec:pathintegrals}, we described the denominator of the Conway superalgebra as a path integral with `twisted' boundary conditions along the Euclidean time circle, i.e. with the fields being periodic or antiperiodic depending on their eigenvalue with respect to the left-moving spacetime fermion number $(-1)^{\bF_l}$. The expansions of this automorphic function at different cusps are related to one another by $O(2,2,\ZZ)$ T-duality transformation.  A general T-duality does not necessarily preserve these boundary conditions, but can map them to $(-1)^{\bF_l}$-twisted boundary conditions along the spacelike circle or around both circles. In particular, the path integrals obtained from these two boundary conditions are related, respectively, to the denominator and superdenominator of a different BKM superalgebra, the one obtained by taking the CHL model by $(-1)^{\bF_l}$ along the spacelike circle. 
    \item Along similar lines, one of the BKM superalgebras discussed in \cite{Harrison:2020wxl} (and originally constructed by Borcherds in \cite{BorcherdsFake,Borcherds96,BorcherdsMM}) arises from the expansion of an automorphic form for a group of the form $\Gamma \backslash SO(2,10)/SO(2) \times SO(10)$ (where $\Gamma$ is a discrete group of  automorphisms of a particular lattice.) Expanding this automorphic form at a different cusp gives rise to another BKM superalgebra, based of the SVOA $V^{fE_8}$ and originally constructed by Scheithauer in \cite{Sch1}. See example 13.7 in \cite{Borcherds:1996uda} for a description of this relation, and sections 6 and 7 of \cite{Harrison:2020wxl} for further elaboration of this point. We expect this connection arises physically from dualities of certain models we consider in this paper.
\end{enumerate}
These are just two examples where we believe string dualities acting on models of the type we study in this paper lead to nontrivial relations among different BKM algebras, though we expect there are many more.

Finally, we close with a number of additional natural questions which are raised by our analysis:
\begin{itemize}
    \item While our analysis in this paper has been limited to 2-dimensional string compactifications, we hope that BKM algebras act as symmetries of special string solutions in higher dimensions in a similar way. For example, it would be interesting to study decompactification limits of our models to higher dimensions (perhaps along similar lines as \cite{KPV}), and understand what, if any, symmetries of the BKM algebras are preserved in these limits. One could also try to construct such algebras directly from worldsheet theories of certain higher dimensional string compactifications.
    \item It would be very interesting to understand the D-brane states in the family of type II string compactifications considered here, which we imagine may furnish representations of the BKM algebras we have constructed. 
    \item The denominator of the fake monster superalgebra, arising as the modified index for type IIA strings on $V^{fE_8}\otimes \bar V^{f\natural}_-$, also appears as the genus one topological amplitude $F_1$ in in type II string theory on the Enriques Calabi-Yau threefold $X$ (the ``FHSV-model'') \cite{Klemm:2005pd}. From this point of view the modified index can be interpreted as a generating function for Gromov-Witten invariants on $X$. It would be interesting to understand more generally whether such geometric interpretations exist for other denominator formulas obtained in this paper. In view of heterotic-type II duality the perturbative BPS-states contributing to our indices should correspond to non-perturbative states in another duality frame. Mathematically, this would correspond to counting Donaldson-Thomas invariants on $K3, T^4$, or the Enriques Calabi-Yau threefold $X$. 
    
    \item Many of the (S)VOAs which appear as part of the worldsheet theory in our models have connections to large sporadic symmetry groups, arising as automorphisms that preserve some suitable subVOA (see, e.g., \cite{DuncanMC, M5, Bae:2020pvv}.) It would be interesting to understand if these structures are present in the BKM algebras which appear in the spacetime  of our models.
    
    \item In \cite{Kawaguchi2021jinvariantAB} it was observed that the denominator formula for one of the BKM superalgebras in \cite{Harrison:2020wxl} can be   obtained starting from the $SL(2,\ZZ)$-invariant $J$-function. Can this connection be explained in terms of heterotic-type II string duality?
    
    \item It would be interesting to determine if any of our models arise as special points in the moduli space of any of the 2d string models considered in \cite{Sen:1996na}, where certain dual pairs of  2d superstring compactifications were identified.
    
    \item As discussed in more detail in section \ref{sec:algebras}, it would be satisfying to study a twisted version of our models, so that our space of BPS states becomes a space of physical states in a topological string theory, perhaps with a natural action of the BKM algebras.
\end{itemize}

\section*{Acknowledgements} It is a pleasure to thank Scott Carnahan and Nils Scheithauer for very helpful email correspondences. 

The work of S.M.H. is supported by the National Science and Engineering Council of Canada, an FRQNT new university researchers start-up grant, and the Canada Research Chairs program.
NMP is supported by the grant NSF PHY-1911298, and the Sivian Fund. D.P. was supported by the Swedish Research Council (Vetenskapsr\aa det), grant nr. 2018-04760.

\appendix
\section{Anomalies and tadpoles}\label{s:Bfield}
In this section, we discuss  the tadpole for the B-field which may arise when compactifying string theory to two dimensions. This was first computed in \cite{Lerche:1987sg,Lerche:1987qk,Vafa:1995fj}; see also \cite{Sethi:1996es} and \cite{Sen:1996na}. We will adapt their methods to the theories of interest to us in this paper.

As a consistency check, we also verify that gravitational anomalies cancel in all the models we are interested in. 
In the second quantized theory considered in section \ref{sec:secondquantized}, where one further spatial direction is compactified on a circle $S^1$, the B-field tadpole can be interpreted as a vacuum winding number $w_0$ along $S^1$. In this case, we argue that the vacuum might also carry non-trivial momentum $m_0$ along $S^1$, and we compute $m_0$ for the models we are interested in.

\subsection{Type II models}

Let us compute the B-field tadpole $w_0$ in the type IIA models we are interested in. As argued in \cite{Vafa:1995fj}, the relevant contribution to the tadpole is a 1-loop string amplitude with the insertion of one B-field vertex operator at zero-momentum, as well as the suitable ghost and superghost insertions to make the amplitude non-vanishing. The only non-zero contribution to this 1-point function comes from path integral with either (even,odd) or (odd,even) spin structure.\footnote{A reminder about spin structures: the odd spin structure means fully periodic boundary conditions for fermions on the torus. In the Hamiltonian formalism, this is the Ramond sector with the insertion of worldsheet fermion number in the trace, which we denote by $\tilde R$; it is $SL(2,\ZZ)$-invariant. The even spin structures have some anti-periodic boundary condition for fermions. In the Hamiltonian formalism, these are NS sector with or without fermion number insertion, or the Ramond sector without fermion number. We denote these sectors by $\tilde{NS}$, NS and $R$, respectively; the three sectors are permuted by $SL(2,\ZZ)$ transformations. } 

According to \cite{Vafa:1995fj}, the (even,odd) and the (odd,even) spin structures give the same contribution, so they just compute the (even,odd) case and put a factor $2$ in front of the final result. This step assumes that the internal CFTs are left-right symmetric, which is not necessarily true in our case; therefore, with respect the the formulas in \cite{Vafa:1995fj}, we drop the factor $2$ and consider the explicit sum of the (even,odd) and the (odd,even) spin structures. The computation of the (even,odd) case is completely analogous to \cite{Vafa:1995fj}; after the usual cancellations between ghost, superghosts and light-cone oscillators, plus some further non-trivial steps, the final result is that one has to take the coefficient of the $\qsf^0$-term of the  combination
\be-\frac{1}{48} (Z_{NS,\tilde R}(\tau)-Z_{\tilde{NS},\tilde{R}}(\tau)-Z_{R,\tilde R}(\tau))_{\qsf^0}\ .
\ee where $Z_{X,Y}$ denotes the partition functions of the internal CFT in the $X,Y$ sector. The relative signs arise  because the central charge of the internal CFT is not a multiple of $24$, so that $Z_{NS,\tilde R}(\tau+1)=-Z_{\tilde{NS},\tilde{R}}(\tau)$.  Note that $Z_{NS,\tilde R}(\tau)$ and $-Z_{\tilde{NS},\tilde{R}}(\tau)$ only differ for the sign of the half-integral powers of $\qsf$, while the integral powers of $\qsf$ (in particular, the $\qsf^0$ term) are the same. Thus, we can set $n_{NS,\tilde{R}}=(Z_{NS,\tilde R}(\tau))_{\qsf^0}=(-Z_{\tilde{NS},\tilde{R}}(\tau))_{\qsf^0}$ and $n_{R,\tilde{R}}=(Z_{R,\tilde{R}})_{\qsf^0}$, and get the formula
\be -\frac{1}{48}(2n_{NS,\tilde R}-n_{R,\tilde R})\ ,
\ee which is, up to a factor $2$, eq.(2.6) in \cite{Sen:1996na}.  The sum  $Z_{NS,\tilde R}(\tau)-Z_{\tilde{NS},\tilde{R}}(\tau)$ effectively gives twice the GSO projected partition function in the NS sector, so $n_{NS,\tilde{R}}$ is also the $\qsf^0$ term in the GSO projected partition function.\footnote{This explains Sen's comments in \cite{Sen:1996na} that the $n_{NS,\tilde R}$ count states that are GSO projected on the left- but not on the right-moving side, while $n_{R,\tilde R}$ counts states that are not GSO projected either on the left or on the right.}

The (odd,even) contribution is completely analogous (it is given by the $\bar \qsf^0$ term in the analogous combination of partition functions), so that the total contribution to the tadpole is
\be w_0=-\frac{1}{48}(2n_{NS,\tilde R}-n_{R,\tilde R}+2n_{\tilde R,NS}-n_{\tilde R,R})\ .
\ee   For left-right symmetric internal CFTs, one has $n_{NS,\tilde R}=n_{\tilde R,NS}$ and $n_{R,\tilde R}=n_{\tilde R,R}$, so that one reobtains formula (2.6) in \cite{Sen:1996na}. When the internal CFT is holomorphically factorized and of the form $V_1\otimes \bar V_2$, then one has $n_{X,Y}=n_X \bar n_Y$, where $n_X$ refers to the left-moving SVOA $V_1$ and $n_Y$ to the right-moving SVOA $\bar V_2$, so that the tadpole is
\be w_0=-\frac{1}{48}[(2n_{NS}-n_{R})\bar n_{\tilde R}+n_{\tilde R}(2\bar n_{NS}+\bar n_{R})]\ .
\ee In the table below, we report $n_{NS}=-n_{\tilde{NS}}=\chi^{\rm NS}$, $n_R=\chi^{\rm R}_++\chi^{\rm R}_-$, $n_{\tilde R}=\chi^{\rm R}_+-\chi^{\rm R}_-$ and $2n_{NS}-n_R=2\chi^{\rm NS}-\chi^{\rm R}_+-\chi^{\rm R}_-$ for the SVOAs we are interested in. The quantities $\chi^{\rm NS},\chi^{\rm R}_\pm$ are the same as presented in \S \ref{sec:VOAs}.
\begin{center}
	\begin{tabular}{ccccccc}
		SVOA $V$& $\chi^{\rm NS}(V)$ & $\chi^{\rm R}_+(V)$ & $\chi^{\rm R}_-(V)$  & $n_R$ & $n_{\tilde R}$ & $2n_{NS}-n_R$\\
		$V^f_{E_8}$            & 8 & 8 & 8 &  16 & 0 & 0\\
		$V^{f\natural}_{-}$ & 0 &  0 & 24&  24 & $- 24$ & $-24$\\
		$V^{f\natural}_{+}$ & 0 & 24 & 0&  24 & $+ 24$ & $-24$\\
		$F_{24}$                 & 24 & 0 & 0 &  0 & 0 & 48.
	\end{tabular}
\end{center}
Note that for $V^{fE_8}$ one has $n_{\tilde R}=2n_{NS}-n_R=0$, so that when either $V_1$ or $V_2$ is $V^{fE_8}$ there is no tadpole. More generally, the tadpole is always an integer, so it can be canceled by a suitable number of spacetime filling strings. 

In terms of $\chi^{\rm NS}$, $\chi^{\rm R}_+$, $\chi^{\rm R}_-$ and their right-moving counterparts $\bar \chi^{\rm NS}$, $\overline \chi^{\rm R}_+$, $\overline \chi^{\rm R}_-$ the tadpole is
\be\label{Btadpole} w_0=\frac{1}{48}\left[(2\chi^{\rm NS}-\chi^{\rm R}_+-\chi^{\rm R}_-)(\overline\chi^{\rm R}_--\overline{\chi}^{\rm R}_+)+( \chi^{\rm R}_--\chi^{\rm R}_+)(2\overline \chi^{\rm NS}-\overline\chi^{\rm R}_+-\overline\chi^{\rm R}_-)\right].
\ee
 The values of $w_0$ for the theories we are interested in are given in table \ref{tbl:tadpole} in the main text.

Let us now check that the gravitational anomaly always cancels. The chiral massless content of these theories, as a function of $\chi^{\rm NS}$, $\chi^{\rm R}_+$, $\chi^{\rm R}_-$, $\bar \chi^{\rm NS}$, $\overline \chi^{\rm R}_+$, $\overline \chi^{\rm R}_-$, is (the NS-NS sector does not contain chiral bosons):
\begin{center}
	\begin{tabular}{c|ccc|c}
		Spin & NS-R & R-NS & R-R & Total\\\hline
		+3/2 & $\overline \chi^{\rm R}_-$ & $\chi^{\rm R}_+$ & 0 & $\chi^{\rm R}_++\overline \chi^{\rm R}_-$ \\
		-3/2 & $\overline \chi^{\rm R}_+$ & $\chi^{\rm R}_-$ & 0 & $\chi^{\rm R}_-+\overline \chi^{\rm R}_+$ \\
		+1/2 (dilatinos)& $\overline \chi^{\rm R}_+$ & $\chi^{\rm R}_-$ & 0 & $ \chi^{\rm R}_-+\overline\chi^{\rm R}_+$\\	
 	-1/2 (dilatinos) & $\overline \chi^{\rm R}_-$ & $\chi^{\rm R}_+$ &  0& $ \chi^{\rm R}_++\overline\chi^{\rm R}_-$\\		
 			+1/2 (gauginos) & $\chi^{\rm NS}\overline \chi^{\rm R}_-$ & $\chi^{\rm R}_+\overline\chi^{\rm NS}$ & 0 & $ \chi^{\rm NS}\overline \chi^{\rm R}_-+\chi^{\rm R}_+\overline\chi^{\rm NS}$\\	
 	-1/2 (gauginos) & $\chi^{\rm NS}\overline \chi^{\rm R}_+$ & $ \chi^{\rm R}_-\overline\chi^{\rm NS}$ &  0& $\chi^{\rm NS}\overline \chi^{\rm R}_+ +\chi^{\rm R}_-\overline\chi^{\rm NS}$\\		 	
  chiral +1 & 0 & 0 & $\chi^{\rm R}_+\overline \chi^{\rm R}_-$ & $\chi^{\rm R}_+\overline \chi^{\rm R}_-$ \\ 
anti-chiral -1 & 0 & 0 & $\chi^{\rm R}_-\overline \chi^{\rm R}_+$ & $\chi^{\rm R}_-\overline \chi^{\rm R}_+$ 
\end{tabular}
\end{center}
The contribution to the gravitational anomaly polynomial of each of these massless chiral fields is \cite{AlvarezGaume:1983ig}
\footnote{Here, we consider Majorana-Weyl spin $1/2$ and $3/2$ fields and real chiral and anti-chiral bosons. The contributions of the spin $1/2$ and $3/2$ fields are half of the ones calculated in \cite{AlvarezGaume:1983ig}, because in that case complex Weyl fermions were considered.} 
\begin{itemize}
	\item Gravitinos (spin $\pm \frac{3}{2}$): $\pm \frac{23}{48}p_1$;
		\item  Chiral fermions (spin $\pm \frac{1}{2}$): $\mp \frac{1}{48}p_1$;
		\item Chiral bosons (spin $\pm 1$): $\mp \frac{1}{24}p_1$.
\end{itemize} Here,
\be p_1=\frac{1}{16\pi^2}\Tr(R\wedge R)\ ,
\ee is the first Pontryagin class, and $R$ denotes Riemann tensor.
Therefore, the total gravitational anomaly is (note that each gravitino of spin $\pm \frac{3}{2}$ comes with a dilatino of spin $\mp \frac{1}{2}$, so it is easier to consider these fields together, contributing $\pm \frac{24}{48}p_1$ each)
\begin{align} &(\chi^{\rm R}_++\overline \chi^{\rm R}_--\chi^{\rm R}_--\overline \chi^{\rm R}_+)\frac{24}{48}p_1+(\chi^{\rm NS}\overline \chi^{\rm R}_++ \chi^{\rm R}_-\overline\chi^{\rm NS}-\chi^{\rm NS}\overline \chi^{\rm R}_-- \chi^{\rm R}_+\overline\chi^{\rm NS})\frac{p_1}{48}+(\chi^{\rm R}_-\overline \chi^{\rm R}_+ -\chi^{\rm R}_+\overline \chi^{\rm R}_-)\frac{p_1}{24}\notag \\
&=\frac{p_1}{48}[(\chi^{\rm NS}+\chi^{\rm R}_-+\chi^{\rm R}_+-24)(\overline \chi^{\rm R}_+-\overline\chi^{\rm R}_-)+(\chi^{\rm R}_--\chi^{\rm R}_+)(\overline\chi^{\rm NS}+\overline \chi^{\rm R}_++\overline \chi^{\rm R}_--24)]=0
\end{align}
where the last equality follows because all internal SVOAs satisfy $\chi^{\rm NS}+\chi^{\rm R}_-+\chi^{\rm R}_+=24$. Therefore, there is no gravitational anomaly.

Finally, when the theory is further compactified on a circle of radius $R$, there might be a non-zero vacuum momentum $\frac{2\pi m_0}{R}$ \cite{Ganor:1996xg,Dasgupta:1996yh}.\footnote{The vacuum momentum can arise in type IIB theories, while it is usually forbidden in type IIA theories, because it violates spacetime parity. In our models, while we impose a type IIA GSO projection, the final theory in general is not invariant under spacetime parity, due to the fact that the internal CFT is not necessarily left-right symmetric; therefore, there is no reason to expect $m_0$ to be $0$.}
This momentum is related to the B-field tadpole $w_0$ of the T-dual theory. A similar phenomenon arises in two dimensional conformal field theory on a cylinder, where $L_0-\bar L_0$ along the circle might have a non-zero vacuum level. More precisely, $L_0-\bar L_0$ is the generator of translations when the radius is rescaled so that $R=2\pi$, so that $m_0$ is precisely the vacuum value of $L_0-\bar L_0$. This analysis shows that a massless chiral boson of spin $\pm 1$ contributes $\mp \frac{1}{24}$ to $m_0$, while a chiral fermion of spin $\pm \frac{1}{2}$ with periodic boundary conditions contributes $\pm \frac{1}{24}$ (with antiperiodic boundary conditions, the contributions is $\mp \frac{1}{48}$). For this calculation, we take into account only propagating degrees of freedom, so from the previous description of massless chiral fields we exclude all gravitinos of spin $\pm \frac{3}{2}$ and the same number of dilatinos of spin $\mp \frac{1}{2}$. Therefore, for periodic fermions, one has
\be m_0=\frac{1}{24}[(\chi^{\rm NS}\overline \chi^{\rm R}_-+\chi^{\rm R}_+\overline\chi^{\rm NS}+\chi^{\rm R}_-\overline \chi^{\rm R}_+)- (\chi^{\rm NS}\overline \chi^{\rm R}_++\chi^{\rm R}_-\overline\chi^{\rm NS}+\chi^{\rm R}_+\overline \chi^{\rm R}_-)]\ .
\ee This expression can be obtained from the B-field tadpole \eqref{Btadpole} by exchanging $\chi^R_-$ and $\chi^R_+$, as expected from T-duality
\be m_0=\frac{1}{48}[(2\chi^{\rm NS}-\chi^{\rm R}_+-\chi^{\rm R}_-)(\overline \chi^{\rm R}_--\overline \chi^{\rm R}_+)+( \chi^{\rm R}_+-\chi^{\rm R}_-)(2\overline\chi^{\rm NS}-\overline \chi^{\rm R}_+-\overline \chi^{\rm R}_-)]\ .
\ee The values of $m_0$ for the theories we are relevant for our construction are listed in table \ref{tbl:tadpole}.

\subsection{Heterotic models}
In the heterotic string case, the B-field tadpole is given by \cite{Lerche:1987sg,Lerche:1987qk}
\be w_0=\frac{1}{24}(Z_{\tilde R}(\tau)E_2(\tau))_{\qsf^0}\ ,
\ee where $Z_{\tilde R}(\tau)$ is the partition function of the internal CFT in the $\tilde R$ sector, and $E_2(\tau)=1-24\qsf+O(\qsf^2)$ is the Eisenstein series of weight $2$.

For a holomorphically factorized theory $V_1\otimes \bar V_2$ one has $Z_{\tilde R}(\tau)=\bar n_{\tilde R}Z_{V_1}(\tau)$, where $\bar n_{\tilde R}=\bar\chi^{\rm R}_+-\bar\chi^{\rm R}_-$, $Z_{V_1}(\tau)=N+J(\tau)$ is the partition function of the internal bosonic VOA $V_1$, $N$ is the number currents in $V_1$, and
\be J(\tau)=\qsf^{-1}+0+196884\qsf+\ldots \ ,
\ee is the modular invariant $J$-function with vanishing constant term. Thus, in this case, the B-field tadpole is equal to (see \cite{PPV1} for the case of $V_1= V^\natural$)
\begin{align} w_0&=\frac{1}{24}(\bar\chi^{\rm R}_+-\bar\chi^{\rm R}_-)(Z_{V_1}(\tau)E_2(\tau))_{\qsf^0}\nonumber\\&=\frac{1}{24}(\bar\chi^{\rm R}_+-\bar\chi^{\rm R}_-)[(\qsf^{-1}+N+\ldots)(1-24\qsf+\ldots)]_{\qsf^0}\\&=\frac{1}{24}(\bar\chi^{\rm R}_+-\bar\chi^{\rm R}_-)(N-24)\ .\nonumber
\end{align}  We summarize the result for the heterotic theories of interest to us in table \ref{tbl:tadpole} in the main text. Again, the tadpole is always an integer, so that it can be canceled by a number of spacetime-filling string with suitable orientation.

Let us verify that the gravitational anomaly cancels in this case as well. Let $\overline \chi^{\rm R}_+$ and $\overline \chi^{\rm R}_-$ be the number of weight $1/2$ states in the R$_+$ and R$_-$ sector of the SVOA $\bar V_2$, so that $\bar n_{\tilde R}=\overline \chi^{\rm R}_+-\overline \chi^{\rm R}_-$. Then, the chiral massless fermions are:
\begin{itemize}
	\item $\overline \chi^{\rm R}_+$ gravitinos with spin $+3/2$ and $\overline \chi^{\rm R}_-$ with spin $-3/2$;
	\item $\overline \chi^{\rm R}_+$ dilatinos and $N\overline \chi^{\rm R}_-$ gauginos of spin $-1/2$, and $\chi^{\rm R}_-+N\overline \chi^{\rm R}_+$ of spin $+1/2$.
\end{itemize}
Furthermore, in heterotic string, every unit of B-field tadpole contributes $p_1/2$ to the gravitational anomaly. The total anomaly, therefore, is
\be (\overline \chi^{\rm R}_+-\overline \chi^{\rm R}_-)\frac{24}{48}p_1+N(\overline \chi^{\rm R}_--\overline \chi^{\rm R}_+)\frac{p_1}{48}+\frac{(N-24)(\overline \chi^{\rm R}_+-\overline \chi^{\rm R}_-)}{24}\frac{p_1}{2}=0\ .
\ee When one inserts the spacetime-filling  strings to cancel the B-field tadpole, the additional degrees of freedom must be such that the gravitational anomaly still vanishes. Each spacetime filling string contributes $\frac{c_R-c_L}{24}p_1$ to the gravitational anomaly\footnote{The normalization and sign can be deduced by comparison with the case of a single chiral Majorana-Weyl fermion or a chiral real scalar.}. One has to take the static gauge for the string, so that $c_L$ and $c_R$ are the left- and right-moving central charges of the internal CFT. Depending on the orientation of the string, we might have either $(c_L,c_R)=(24,12)$ if the supersymmetric sector is right-moving, or $(c_L,c_R)=(12,24)$ in the other case. Matching the anomaly contribution of the B-field and of the spacetime filling strings, one gets that the number of spacetime filling strings must be 
$ |w_0|= \frac{|(N-24)(\overline \chi^{\rm R}_+-\overline \chi^{\rm R}_-)|}{24}\ ,$
with the supersymmetric sector being left-moving for $w_0>0$ and right-moving if $w_0<0$.

As in the type II case, when the theory is compactified on a circle of radius $R$, there might be a vacuum momentum $\frac{2\pi m_0}{R}$. In this case, only the chiral gauginos contribute, and for periodic boundary conditions we obtain
\be m_0=\frac{N(\overline \chi^{\rm R}_+-\overline \chi^{\rm R}_-)}{24}\ .
\ee
See table \ref{tbl:tadpole} for the relevant values of the tadpole $w_0$ and momentum $m_0$ in the heterotic models we are interested in.

\section{Semirelative cohomology}\label{app:semirelative}

When combining left and right-moving sectors of a closed string theory, the condition that the corresponding states are annihilated by $b_0$ and $\bar{b}_0$ is in general too strong. One need only require that the state is annihilated by the combination $b_0 - \bar{b}_0$. We can follow the treatment of \cite{ZW} to discuss the relevant distinctions. We denote the total left and right-moving BRST operator by $Q = Q_l + Q_r$. The three varieties of cohomology, all graded by ghost number $N$\footnote{We suppress the additional grading by picture number and momenta.}, relevant for this discussion are 
\begin{enumerate}
\item Absolute cohomology $\mathsf{H} = \oplus_N \mathsf{H}^N$ of $Q$-closed states $|\Psi \rangle$ modulo states of the form $Q | \Lambda \rangle$. \\
\item Relative cohomology  $\mathsf{H}_R = \oplus_N \mathsf{H}_R^N$ of $Q$-closed states $| \Psi \rangle$ that satisfy $b_0 |\Psi \rangle = \bar{b}_0 | \Psi \rangle = 0$, modulo states of the form $Q |\Lambda \rangle$ which satisfy $b_0 |\Lambda \rangle = \bar{b}_0 | \Lambda \rangle = 0$. \\
\item Semirelative cohomology $\mathsf{H}_{S} = \oplus_N \mathsf{H}_{S}^N$ of $Q$-closed states $| \Psi \rangle$ that satisfy $(b_0 - \bar{b}_0) |\Psi \rangle = 0$, modulo states of the form $Q |\Lambda \rangle$ which satisfy $(b_0 - \bar{b}_0) |\Lambda \rangle = 0$. 
\end{enumerate}

At each total ghost number $N$, the relative and absolute cohomologies factorize between left and right movers: 
$\mathsf{H}^N = \oplus_{n+ \bar n=N} H^n_l \otimes \bar H^{\bar n}_r$, but the semirelative cohomology does not. The cohomologies fit into the exact sequences 

\begin{equation}
\xymatrix{
  \ldots \mathsf{H}^N_R \ar[r]^-{i'} & \mathsf{H}^N_S \ar[r]^-{b^+_0} & \mathsf{H}^{N-1}_R \ar[r]^-{\left\lbrace Q, c^+_0 \right\rbrace} & \mathsf{H}^{N+1}_R \ar[r]^-{i'} & \mathsf{H}^{N+1}_S \ldots \\
}
\end{equation}

\begin{equation}
\xymatrix{
  \ldots \mathsf{H}^N_S \ar[r]^-{i''} & \mathsf{H}^N \ar[r]^-{b^-_0} & \mathsf{H}^{N-1}_S \ar[r]^-{\left\lbrace Q, c^-_0 \right\rbrace} & \mathsf{H}^{N+1}_S \ar[r]^-{i''} & \mathsf{H}^{N+1} \ldots \\
}
\end{equation}
where $b^{\pm}_0 = b_0 \pm \bar{b}_0$, and similarly  $c^{\pm}_0 = {1 \over 2}(c_0 \pm \bar{c}_0)$. The maps $i', i''$ are the maps that forget the relevant conditions involving the $b$-ghosts, and the other maps are given by multiplication by the indicated operators.

For the $k \neq 0$ states, \cite{LZ2} proved that for a chiral critical superstring at picture number $-1$ in the NS sector (to which we may always specialize canonically at nonzero momentum), the cohomology vanishes for all ghost numbers not equal to 1, so for a full (chiral and anti-chiral) superstring the only nonvanishing relative cohomology group is at ghost number 2. Then the exact sequence collapses to isomorphisms between the following relative and semirelative cohomologies:
\begin{align*}
0& \rightarrow \mathsf H^2_R \rightarrow \mathsf H^2_S \rightarrow 0 \\
0& \rightarrow \mathsf H^3_S \rightarrow \mathsf H^2_R \rightarrow 0.
\end{align*}
The first map is the forgetful inclusion and the second map produces a state in $\mathsf H^3_S$ from one in $\mathsf H^2_R$ by acting with $c^+_0$, which effectively inverts $b^+_0$ at nonzero momentum. 
Similar arguments apply for the theory in the Ramond sector at the canonical picture numbers $-1/2, -3/2$. Hence, for the states of nonzero momentum, we therefore lose nothing by considering the relative cohomology, which is reassuring since the cohomology at nonzero momentum correctly reproduces the physical states one expect from a lightcone gauge calculation.

\subsection{Zero-momentum states}
To complete our understanding of the physical string states, we would like to understand what kind of zero-momentum states we can obtain (at any ghost number), and whether this has any interesting impact on the semirelative cohomology. For this analysis and all subsequent analyses, we will only consider the canonical pictures where all positive modes of worldsheet fields annihilate the vacuum:$-1$ in NS and $-1/2, -3/2$ in R, though a priori one could consider more general pictures.

We can start more simply in a chiral superstring model and identify the $k=0$ states there. We focus on the SVOA $V^{f\natural}_+$; the generalizations to the other self-dual $c=12$ SVOAs is trivial. These states would contribute to the relative cohomology factors.  $V^{f\natural}_+$ has 24 weight 1/2 states in the R$_+$ sector which we denote by $u^{i+},\, i=1,\ldots, 24$, and 0 such states in the NS, R$_-$ sectors. Consequently, there are 2 states of ghost number 1
\begin{align}
\psi^{\mu}_{-1/2}e^{-\phi}c_1 |0\rangle, \ \ \ \mu=\pm,
\end{align} 
1 state of ghost number 2,
\begin{equation}
\gamma_{-1/2}e^{-\phi}c_1 |0 \rangle,
\end{equation}
and 1 state of ghost number 0,
\begin{equation}
\beta_{-1/2}e^{-\phi}c_1 | 0 \rangle,
\end{equation}
which can each be combined with their antiholomorphic counterparts to get states in relative cohomology. 

There are also zero-momentum states in the Ramond sector (chirally, from the R$_+$ sector), in this case at all ghost numbers. Again, we start with the states that appear in relative cohomology by writing down the chiral states of interest (which may then be combined with antichiral states). At picture number $-1/2$, there are 24 states of ghost number 1:
\begin{equation}
u^{i+}_{-1/2}e^{-\phi/2}c_1 |0, +\rangle,
\end{equation} acting on the ground state of zero momentum and positive spacetime chirality (which in this 2d model is tied to the fermion number). Additionally, we get 24 states at each positive ghost number $\geq 2$: 
\begin{equation}
(\gamma_0)^n u^{i+}_{-1/2}e^{-\phi/2}c_1 |0, +\rangle, \ \ \  n \geq 1.
\end{equation} 
Correspondingly, there are states at picture number $-3/2$ for all ghost numbers $1-m$ with $m \geq 0$:
\begin{equation}
(\beta_0)^m u^{i+}_{-1/2}e^{-3\phi/2}c_1 |0, -\rangle, \ \ \  m \geq 1.
\end{equation}

In the relative cohomology then, we get new NS-NS states of total ghost numbers 0, 1, 2, 3, 4 and new R-R states at picture number $-1/2$ at all ghost numbers 2 and higher, etc. 
Note that in the literature, the extra R-sector conditions ker$\beta_0 = 0= $ker $\bar{\beta}_0$ (the analogue of the relative cohomology conditions in the NS sector) and $G_0 = \bar{G}_0 =0$ are usually imposed although the former conditions, like their NS-sector counterpart, may be too strong in general.
\subsection{Semirelative cohomology at zero momentum}

With the chiral zero-momentum states in hand, let us ask if we can understand the semirelative cohomology using the exact sequences.
We start with the NS-NS sector at the canonical picture number, since it turns out that the picture-changing operator no longer has an inverse at zero-momentum and therefore we must treat that sector with more care.

The zeroth semirelative cohomology is isomorphic to the zeroth relative cohomology since the long exact sequence truncates due to the vanishing of negatively moded cohomology groups: $0 \rightarrow \mathsf H^0_R \rightarrow \mathsf H^0_S \rightarrow 0$. These are just 1-dimensional groups generated by the vacuum (which is constructed by acting on the $SL(2, \mathbb{C}) \times SL(2, \mathbb{C})$-invariant vacuum $|0 \rangle$\footnote{In the non-chiral setting we use the shorthand $|0\rangle$ for $|0,0\rangle$ when we think no confusion can arise.} with various ghosts): $\beta_{-1/2}\bar{\beta}_{-1/2}e^{-\phi} e^{-\bar{\phi}}c_1 \bar{c}_1 | 0 \rangle$.

Given the nonvanishing relative cohomology representatives we can construct at zero-momentum, the long exact sequence splits into the following exact sequences (plus the zeroth cohomology sequence above):
\begin{equation}
\xymatrix{
  0  \ar[r] & \mathsf{H}^5_S \ar[r]^-{b^+_0} & \mathsf{H}^4_R \ar[r]& 0 \\ 
  0 \ar[r]& \mathsf{H}^1_R \ar[r]^-{i} & \mathsf{H}^1_S \ar[r]^-{b^+_0} & \mathsf H^0_R   \ar[r]^-{\left\lbrace Q, c^+_0 \right\rbrace} & \mathsf H^2_R \ar[r]^-{i}& \mathsf H^2_S \ar[r]^-{b^+_0}& \mathsf H^1_R \ar[r]^-{\left\lbrace Q, c^+_0 \right\rbrace}& \mathsf H^3_R \ar[r]^-{i}& \mathsf H^3_S \ar[r]^-{b^+_0}& \mathsf H^2_R 
  \\  \ar[r]^-{ \left\lbrace Q, c^+_0 \right\rbrace}&  \mathsf H^4_R \ar[r]^-{i}& \mathsf H^4_S \ar[r]^-{b^+_0}& \mathsf H^3_R \ar[r]& 0.
}
\end{equation}

From this we can calculate the rest of the semirelative cohomology groups. For convenience, let's define the shorthand  $|\Omega, \bar{\Omega} \rangle := e^{-\phi} e^{-\bar{\phi}}c_1 \bar{c}_1 | 0 \rangle$ such that $\beta_{-1/2}\bar{\beta}_{-1/2}|\Omega, \bar{\Omega} \rangle$ generates the zeroth (semi)relative cohomology groups at picture number $-1$, and denote $|\Omega \rangle := e^{-\phi} c_1 |0 \rangle$ (similarly for $|\bar{\Omega}\rangle)$ the chiral counterparts at ghost numbers (1,0) (respectively, (0,1)). 

Then at ghost number 1 there are 4 states: $\psi^{\mu}_{-1/2}\bar{\beta}_{-1/2} |\Omega, \bar{\Omega} \rangle, \bar{\psi}^{\mu}_{-1/2} \beta_{-1/2}|\Omega, \bar{\Omega} \rangle$. These are identical in both relative and semirelative cohomology.

At ghost number 2, we have the representatives 
\begin{align*}
\left\lbrace \psi^{\mu}_{-1/2}\bar{\psi}^{\nu}_{-1/2}|\Omega, \bar{\Omega} \rangle, \gamma_{-1/2}\bar{\beta}_{-1/2}|\Omega, \bar{\Omega} \rangle, \bar{\gamma}_{-1/2} \beta_{-1/2} |\Omega, \bar{\Omega} \rangle \right\rbrace  &\in \mathsf H^2_R \\
\left\lbrace \psi^{\mu}_{-1/2}\bar{\psi}^{\nu}_{-1/2}|\Omega, \bar{\Omega} \rangle,  (\gamma_{-1/2}\bar{\beta}_{-1/2}-\bar{\gamma}_{-1/2} \beta_{-1/2}) |\Omega, \bar{\Omega} \rangle   \right\rbrace  &\in \mathsf H^2_S 
\end{align*} The second state in $\mathsf H^2_S$ is called the ghost dilaton \cite{BZ} and is nontrivial only in the semirelative, but not the relative, cohomology. As mentioned in the main text, this state is excluded to construct the BKM algebra on BPS states. 

It is also simple to write down representatives for the remaining relative cohomology classes at ghost numbers 3 and 4:
\begin{align*}
\left\lbrace \bar{\gamma}_{-1/2}\psi^{\mu}_{-1/2}|\Omega, \bar{\Omega}\rangle, \gamma_{-1/2}\bar{\psi}^{\mu}_{-1/2} |\Omega, \bar{\Omega}\rangle \right\rbrace  &\in \mathsf H^3_R \\
\left\lbrace \gamma_{-1/2}\bar{\gamma}_{-1/2}|\Omega, \bar{\Omega}\rangle   \right\rbrace  &\in \mathsf H^4_R.
\end{align*} 
From these, we use the exact sequence to obtain the remaining semirelative cohomology classes at ghost numbers 3, 4, and 5: 

\begin{align*}
\left\lbrace c^+_0\psi^{\mu}_{-1/2}\bar{\psi}^{\nu}_{-1/2}|\Omega, \bar{\Omega} \rangle,  c^+_0(\gamma_{-1/2}\bar{\beta}_{-1/2}-\bar{\gamma}_{-1/2} \beta_{-1/2}) |\Omega, \bar{\Omega} \rangle  \right\rbrace  &\in \mathsf H^3_S \\
\left\lbrace c^+_0 \bar{\gamma}_{-1/2}\psi^{\mu}_{-1/2}|\Omega, \bar{\Omega}\rangle, c^+_0\gamma_{-1/2}\bar{\psi}^{\mu}_{-1/2} |\Omega, \bar{\Omega} \rangle  \right\rbrace  &\in \mathsf H^4_S \\
\left\lbrace c^+_0 \gamma_{-1/2}\bar{\gamma}_{-1/2}|\Omega, \bar{\Omega}\rangle  \right\rbrace  &\in \mathsf H^5_S.
\end{align*} 
We focus on the ghost number 2 cohomology (or ghost number 1 for chiral superstrings) in the main text, but record these additional zero-momentum states for completeness.

To deal properly with the Ramond sector, we can largely follow the analysis of \cite{BeZ}. 
Now, for states of nonzero momentum, chiral relative cohomologies and closed string semirelative cohomologies at all picture numbers are isomorphic to one another (with a canonical isomorphism given by an invertible picture-changing operator) but this is not strictly true for the closed string semirelative cohomology at zero momentum. This is essentially because $b^-_0$ does not commute with the picture-changing operators; it does turn out that zero-momentum RR states are inequivalent at different picture numbers (though nonzero momentum physical states are equivalent in all pictures for all sectors). For the NS-R sectors, we have the options of being at picture numbers $(-1, -1/2), (-1, -3/2)$ which do turn out to be equivalent to one another. The computations presented in \cite{BeZ} essentially apply to this case as well.

For example, if we take the R$_+$ sector on the holomorphic side in the $-1/2$ picture and the NS sector on the antiholomorphic side in the $-1$ picture the cohomology (given by listing representatives) is \footnote{Our ghost number convention differs slightly from \cite{BeZ}.}
\begin{align*}
\mathsf H^0_{S} &= \left\lbrace \bar{\beta}_{-1/2}|-{1 \over 2}, \bar{\Omega}\rangle^i \right\rbrace \\
\mathsf H^1_{S} &=  \left\lbrace \bar{\psi}^{\mu}_{-1/2}|-{1 \over 2}, \bar{\Omega}\rangle^i , \  \bar{\beta}_{-1/2}\gamma_0|-{1 \over 2}, \bar{\Omega}\rangle^i \right\rbrace \\
\mathsf H^2_{S}&= \left\lbrace \left(\bar{\gamma}_{-1/2} - \gamma_0^2 \bar{\beta}_{-1/2} \right)|-{1 \over 2}, \bar{\Omega}\rangle^i , \ \bar{\psi}^{\mu}_{-1/2} \gamma_0 |-{1 \over 2}, \bar{\Omega}\rangle^i  \right\rbrace \\
\mathsf H^3_{S}&= \left\lbrace \gamma_0\left(\bar{\gamma}_{-1/2} - \gamma_0^2 \bar{\beta}_{-1/2} \right)|-{1 \over 2}, \bar{\Omega}\rangle^i  \right\rbrace,
\end{align*} 
where we have defined the state $|-{1 \over 2 }\rangle^i:= u^{i+}_{-1/2}e^{-\phi/2}c_1|0, + \rangle$. 
Isomorphic (but more complicated-looking) expressions can be obtained for the $(-3/2, -1)$ picture.

In the RR sector, there are subtle differences among the various picture numbers, which in turn is related to distinctions between finite and infinite cohomologies arising from the possibility of allowing arbitrary number of superghost zero modes. In particular, in the mixed picture numbers, $(\gamma_0 \bar{\beta}_0)^n$, respectively $(\bar{\gamma}_0 \beta_0)^n$, do not annihilate the vacuum for arbitrary nonnegative $n$. In the diagonal $(-1/2, -1/2)$ and $(-3/2, -3/2)$ pictures, we do not have the possibility of such combinations. In \cite{BeZ}, it is proven that the diagonal $-3/2$ picture cohomology is in fact isomorphic to the finite $(-3/2, -1/2)$ cohomology, while the diagonal $-1/2$ picture cohomology is isomorphic to the infinite $(-3/2, -1/2)$ cohomology. For simplicity, in the main text, we focus on the analogue of the relative cohomology condition in the R-sector, namely: ker$\beta_0 =$ ker $\bar{\beta}_0=0$, as we do not believe these subtleties impact the construction of the BKM algebra.

A more refined analysis can be undertaken using the results of \cite{BeZ}. When $V_2 = V^{f\natural}$, for example, one finds a nonvanishing contribution in the $(-3/2, -3/2)$ picture, where there is a nontrivial $\mathsf H^{2}_{S}$ generated by the state given by $c_0^+$ acting on the vacuum. This can be excluded from considerations of the BKM algebra action in the manner of the ghost dilaton.

\section{Zero momentum R-R sector}\label{app:0mom}
In this section, we describe in detail the massless R-R spectrum of type IIA superstring theory compactified on $V^{f\natural}\otimes \bar V^{f\natural}$. These results are needed in the calculation of the spacetime indices $\mathcal{Z}$ and $\tilde{\mathcal{Z}}$ in section \ref{sec:secondquantized}. 

Both the holomorphic and antiholomorphic SVOAs have 24 Ramond ground states, whose corresponding chiral and anti-chiral physical states we denote by $\usf^i_{\alpha}$ and $\bar \usf^j_{\beta}$  where $i,j=1,\ldots,24$ and $\alpha,\beta \in \{+,-\}$ are Dirac spinor indices in two dimensional spacetime.  There are two inequivalent choices for the internal fermion number in each of these SVOAs, leading to two inequivalent  GSO projections (up to parity). In particular, for $V^{f\natural}_+\otimes \bar V^{f\natural}_-$, the spinors $\usf^i_{\alpha}$ and $\bar \usf^j_{\beta}$ have all the same spacetime spin  $\alpha=\beta=+$. For $V^{f\natural}_-\otimes \bar V^{f\natural}_-$, the 24 holomorphic $\usf^i_{\alpha}$ have spin $\alpha=-$ and the 24 anti-homolorphic $\bar \usf^j_{\beta}$ have spin $\beta=+$. 

In the R-R sector one has $24^2=576$ ground states $\usf^i_{\alpha} \bar \usf^j_{\beta}$  where $i,j=1,\ldots,24$. Since these are actually Majorana-Weyl spinors, only one of these Dirac components is non-zero, for example $\bar \usf^j_+$ and either $\usf^i_+$ or $\usf^i_-$. Now, we impose the physical state condition. For $k\neq 0$, BRST closedness implies that, for all $i,j=1,\ldots,24$, $\usf^i_\alpha$ and  $\bar \usf^j_{\beta}$ should satisfy the massless Dirac equations
\be \begin{pmatrix}
	k^0_l+k^1_l & 0\\ 0 & -k^0_l+k^1_l 
\end{pmatrix} \begin{pmatrix} \usf^i_+\\\usf^i_-\end{pmatrix} = 0\ ,\ee 
\be \begin{pmatrix}
	k^0_r+k^1_r & 0\\ 0 & -k^0_r+k^1_r 
\end{pmatrix} \begin{pmatrix} \bar \usf^j_+\\\bar \usf^j_-\end{pmatrix} = 0\ .\ee

Since only one of the two components $\usf^i_+$ and $\usf^i_-$ is non-zero, the first equation either eliminates that component or it is trivially satisfied, depending on whether $k_l^0=k_l^1$ or $k_l^0=-k_l^1$. An analogous argument holds for the anti-holomorphic $\bar \usf^i_\beta$.

In the uncompactified theory, one has  $k^\mu_l=k^\mu_r$, so that physical states exist only if the $\usf^i_\alpha$ and $\usf^j_\beta$ have the same non-zero component, say $\alpha=\beta=+$; this corresponds to type IIA on $V^{f\natural}_+\otimes \bar V^{f\natural}_-$. In this case, the Dirac equations are satisfied if and only if $k^0=-k^1$. From the spacetime point of view, this means that there are chiral massless propagating degrees of freedom of spin $1$. We recall that, since the states corresponding to R-R fields obey physical state conditions that are of first order in momenta, they are naturally interpreted as quanta of field strengths rather than potentials. In the case of type IIA on $V^{f\natural}_+\otimes \bar V^{f\natural}_-$, this means that the fields corresponding to $\usf^i_+\bar \usf^j_+$ are the derivative $\partial_+ \phi^{ij}$ of $576$ massless chiral scalar fields $\phi^{ij}$, $i,j=1,\ldots,24$. 

On the other hand, for type IIA on $V^{f\natural}_-\otimes \bar V^{f\natural}_-$, where $\usf^i_\alpha$ and $\usf^j_\beta$ have non-zero components of opposite spin ($\alpha=-$, $\beta=+$), there cannot be any physical propagating degrees of freedom from the massless R-R fields. In this case, the  corresponding spacetime fields would be the field strengths of massless vector fields $C_\mu^{ij}$, which must be the gauge bosons of an abelian gauge symmetry. The fact that there are no physical states is consistent with the fact that massless gauge bosons in two dimensions carry no propagating degrees of freedom.

At zero momentum, all $576$ states $\usf^i_\alpha\bar \usf^j_\beta$ are physical. For $V^{f\natural}_+\otimes \bar V^{f\natural}_-$, these are just the zero modes of the scalars $\phi^{ij}$. For $V^{f\natural}_-\otimes \bar V^{f\natural}_-$, these zero modes are the fluxes of the field strengths in the two dimensional spacetime.

If the spatial direction is compactified on a circle $S^1$, for $V^{f\natural}_+\otimes \bar V^{f\natural}_-$ the physical states carry zero winding ($k^1_l=k^1_r$) but possibly non-zero momentum along $S^1$; for the $V^{f\natural}_-\otimes \bar V^{f\natural}_-$, the physical states carry zero momentum ($k^1_l=-k^1_r$) but possibly non-zero winding along the circle.

\section{Theta lifts of vector valued modular forms}
 \label{app:vectorvaluedtheta}
 The theta lift of a vector valued modular form $F(\tau)$ is of the form \cite{Borcherds:1996uda}
 \begin{equation}
 \Theta_F:=\int_{SL(2,\mathbb{Z})\backslash \mathbb{H}} \left(\overline{\Theta}_\Lambda(\tau), F(\tau)\right)d\mu,
 \end{equation}
 where $\tau=\tau_1+i\tau_2\in \mathbb{H}$, $\Lambda$ is an even lattice of signature $(2,d)$, $F$ is a weight $k=1-d/2$ vector-valued modular form and $\overline{\Theta}$ is a weight $-k$ Siegel-Narain theta series for $\Lambda$. 
  In particular, if $\Lambda^\vee$ denotes the dual lattice of $\Lambda$, then we can write $F$ as follows
 \begin{equation}
 F(\tau)=\sum_{\gamma \in \Lambda^{\vee}/\Lambda}F_\gamma(\tau) e_\gamma,
 \end{equation}
 where the components $F_\gamma$ are modular forms for a congruence subgroup $\Gamma\subseteq SL(2,\ZZ)$, and $F$ transforms in the metaplectic representation $\rho_\Lambda$ of $\widetilde{SL(2,\mathbb{Z})}$ on the group ring $\mathbb{C}[\Lambda^{\vee}/\Lambda]$. In practice this means that the components $F_\gamma$ are in one-to-one correspondence with the cosets $\Lambda^{\vee}/\Lambda$ and under the action of the modular group we have
 \begin{eqnarray}\label{rhoLambda}
 F_{\gamma}(\tau+1)&=&e^{\pi i \gamma^2}F_\gamma(\tau)
 \nonumber \\
 F_{\gamma}(-1/\tau)&=&\frac{i^{(2-d)/2}}{\sqrt{|\Lambda /\Lambda^{\vee}|}}\sum_{\delta \in \Lambda /\Lambda^{\vee}} e^{-2\pi i (\gamma, \delta)}F_\delta(\tau).
 \end{eqnarray} The Siegel theta series is similarly written as
 \begin{equation}
 \Theta_\Lambda=\sum_{{\gamma \in \Lambda^{\vee}/\Lambda}}\theta_{\Lambda+\gamma} e_\gamma, 
 \end{equation}
 where $\theta_{\Lambda+\gamma}$ is a shifted theta series. In addition the integrand involves the scalar product $(\, , \, )$ defined by 
 \begin{equation}
 (e_\gamma, e_{\gamma'})=\delta_{\gamma+\gamma', 0}, \qquad \gamma \in \Lambda^{\vee}/\Lambda.
 \end{equation} Furthermore, we define complex conjugation to act by $\overline{e_\gamma}=e_{-\gamma}$ on the elements of $\mathbb{C}[\Lambda^{\vee}/\Lambda]$. With these definitions, the integrand $\left(\overline{\Theta}_\Lambda(\tau), F(\tau)\right)$ is $SL(2,\ZZ)$-invariant, and it makes sense to integrate it on the upper half-plane $\mathbb{H}$ modulo $SL(2,\ZZ)$-transformations. The theta lift produces an automorphic form on the Grassmannian $G(2,d)$ of positive definite $2$-planes in $\mathbb{R}^{2,d}$, for a discrete group $O(\Lambda,F)$ (a subgroup of the the group $O(\Lambda)$ of automorphisms of the lattice $\Lambda$).

\end{document}